\pdfoutput=1
\documentclass[cernpreprint, atlasdraft=false, texlive=2023, UKenglish, orcidlogo, texmf, orcidlogo]{atlasdoc}
 
\usepackage{atlaspackage}
\usepackage{atlasbiblatex}
 
\usepackage{atlasphysics}

\graphicspath{{logos/}{figures/}}
 
\usepackage{ANA-HDBS-2019-12-PAPER-defs}

\AtlasTitle{A search for new resonances in multiple final states with a high transverse momentum $Z$ boson in $\sqrt{s}=13$\,TeV $pp$ collisions with the ATLAS detector}

\AtlasAbstract{
A generic search for resonances is performed with events containing a $Z$ boson with transverse momentum greater than 100\,GeV, decaying into $e^+e^-$ or $\mu^+\mu^-$. The analysed data collected with the ATLAS detector in proton--proton collisions at a centre-of-mass energy of 13\,TeV at the Large Hadron Collider correspond to an integrated luminosity of 139\,fb$^{-1}$.
Two invariant mass distributions are examined for a localised excess relative to the expected Standard Model background in six independent event categories (and their inclusive sum) to increase the sensitivity.
No significant excess is observed. Exclusion limits at 95\% confidence level are derived for two cases: a model-independent interpretation of Gaussian-shaped resonances with the mass width between 3\% and 10\% of the resonance mass, and a specific heavy vector triplet model with the decay mode $W'\to ZW \to \ell\ell qq$.
}

\AtlasRefCode{HDBS-2019-12}
 
\PreprintIdNumber{CERN-EP-2022-180}

\AtlasJournalRef{JHEP 06 (2023) 036}
\AtlasDOI{10.1007/JHEP06(2023)036}

\hypersetup{pdftitle={ATLAS document},pdfauthor={The ATLAS Collaboration}}
 
\begin{document}
 
\maketitle
 
\tableofcontents

\section{Introduction}
 
One of the key components of the Large Hadron Collider (LHC) physics programme is the search for new resonances predicted by hypotheses for physics beyond the Standard Model (BSM)~\cite{PhysRevD.30.1470,Haber:1984rc,Altarelli:1989ff,Arkani_Hamed_1998,Randall_1999,Davoudiasl_2000,Giudice_2008,Low_2011,Delgado_2017}.
A large number of searches have been performed in proton--proton ($pp$) collisions at the LHC. Most searches are optimised for some specific final states motivated by a particular BSM model, or a class of models, and are therefore model-dependent, e.g.\ those associated with a $Z$ boson~\cite{ATLAS:2021ldb,HDBS-2018-13,HDBS-2018-10,ATLAS:2021edm,HIGG-2018-09,SUSY-2018-21,CMS-EXO-19-003,EXOT-2017-24,CMS-HIG-18-023,CMS-HIG-18-012,CMS-EXO-19-007,CMS-B2G-19-006,CMS-B2G-17-013,CMS-EXO-17-005,CMS-B2G-16-023,CMS-B2G-17-004,HIGG-2016-28,HIGG-2016-19,EXOT-2016-34,EXOT-2016-29,EXOT-2016-11,EXOT-2016-10}. No new physics has been found so far by these model-specific searches. To expand the search program, generic searches can be sensitive to models that do not have the same characteristics of those already searched for and give more ability for reinterpretation for theories yet to be proposed.
 
\begin{figure}[htbp]
\centering
\subfloat[]{\includegraphics[width=0.325\columnwidth]{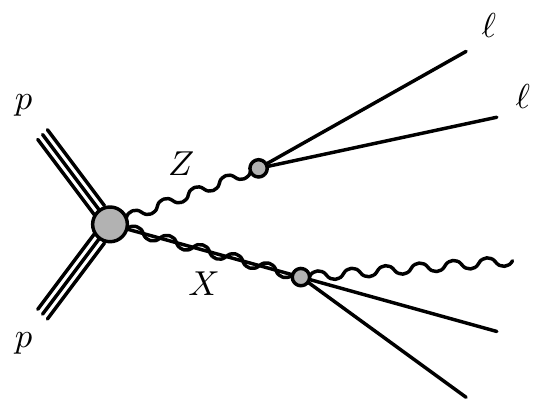} \label{fig:diagram-a}}\hspace{10mm}
\subfloat[]{\includegraphics[width=0.325\columnwidth]{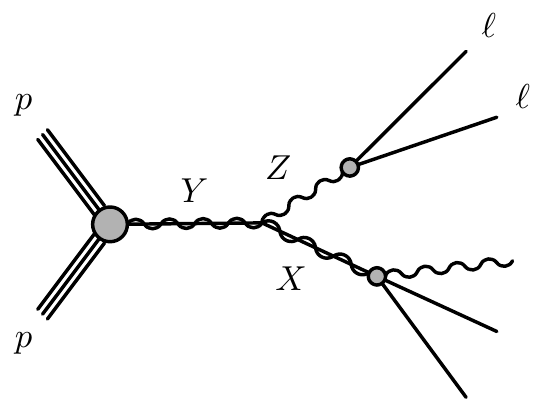} \label{fig:diagram-b}}
\caption{Schematic diagrams for \protect\subref{fig:diagram-a} $pp\to ZX$ and \protect\subref{fig:diagram-b} $pp\to Y\to ZX$ where $Y$ is a new resonance to be searched for while $X$ can be a new resonance or SM particle.}
\label{fig:diagram}
\end{figure}

This article presents a generic resonance search with both model-independent and model-dependent interpretations in events with a leptonically decaying $Z$ boson with high transverse momentum, $p_\mathrm{T}$.
It is generic since the search is performed in multiple final states and it is sensitive to new particles in various production and decay modes.
It is also model-independent since minimal features of BSM physics are assumed.
Specifically, BSM processes are assumed to produce a resonance decaying into high-transverse-momentum objects, in association with, or including, a $Z$ boson, whereas the SM background is assumed to be smoothly falling over the observables of interest.
Two generic cases are considered, as shown schematically in Figure~\ref{fig:diagram}: $pp \to ZX$, where the new resonance, represented as $X$, decays to a final state of SM particles, and $pp \to Y \to ZX$, where the resonance appears in production as an intermediate state $Y$ that subsequently decays into a $Z$ boson and SM particles, possibly via another intermediate resonance $X$.
To probe a variety of possible BSM processes in a single analysis,
events are separated into six exclusive categories according to the identity of the leading-$p_\mathrm{T}$ object in $X$.
The leading-$p_\mathrm{T}$ object can be an electron, a muon, a photon, a small-$R$ jet reconstructed with a radius parameter of $R = 0.4$ containing a $b$-hadron, a small-$R$ jet without a $b$-hadron, or a large-$R$ jet with $R=1$.
Invariant mass distributions of all the visible objects in the final state,  either excluding the leptons from the $Z$ boson decay ($m_X$) or including the leptons from the $Z$ decay ($m_{ZX}$) in each of the six event categories, are defined as observables of interest. The distributions are examined with the BumpHunter algorithm~\cite{0809.3781,choudalakis2011hypothesis} for local excesses above a data-driven estimate of the smoothly falling SM background. Depending on the event categories and mass observables, the search is performed so as to be sensitive to a large mass range from around 200\,\GeV\ up to 6\,\TeV.
 
Generic searches with minimal beyond-the-SM signal assumptions were performed by the D0 Collaboration~\cite{hep-ex/0006011,hep-ex/0011067,hep-ex/0011071,1108.5362} and the CDF Collaboration~\cite{0712.1311,0809.3781} at the Tevatron, and by the H1 Collaboration~\cite{hep-ex/0408044,0901.0507} at HERA. At the LHC, a general search was performed by each of the ATLAS~\cite{EXOT-2016-38} and CMS~\cite{CMS-EXO-19-008} collaborations, using 3.2~fb$^{-1}$ and 35.9~fb$^{-1}$ of data at a centre-of-mass energy ($\sqrt{s}$) of 13~\TeV, respectively.
ATLAS also performed searches in final states containing two-leptons as well as in three- and four-leptons~\cite{EXOT-2018-08,EXOT-2019-36}.
However, final states including a $Z$ resonance were not explicitly considered in general searches although decays through weak bosons are often possible in BSM models.
Requiring a high-$p_\mathrm{T}$ $Z$ boson may reduce the background enough to reveal signals which are not visible in an inclusive sample and also explores a signature not explicitly considered in previous general searches. Furthermore, a leptonically decaying $Z$ boson is a fully reconstructable final state with straightforward ways to reduce the multijet background which is challenging to model.
 
The analysis uses the full data sample from $\sqrt{s}=13$\,\TeV\ $pp$ collisions recorded by the ATLAS detector in Run~2 at the LHC, corresponding to an integrated luminosity of 139\,fb$^{-1}$.
The results are interpreted in a model-independent way in terms of Gaussian-shaped signals with relative mass widths of 3\%, 5\%, and 10\% for all the event categories. Furthermore, the results from the dominant leading large-$R$-jet event category are interpreted in a Heavy Vector Triplet (HVT) model~\cite{deBlas:2012qp,1402.4431} in order to demonstrate the sensitivity of the analysis.


\section{ATLAS detector}
The ATLAS detector~\cite{PERF-2007-01} covers nearly the entire solid angle around the collision point.\footnote{The ATLAS experiment uses a right-handed coordinate system with its origin
at the nominal interaction point (IP) in the centre of the detector
and the $z$-axis along the beam pipe. The $x$-axis points from the
IP to the centre of the LHC ring, and the $y$-axis points
upwards. Cylindrical coordinates $(r,\phi)$ are used in the
transverse plane. The polar angle ($\theta$) is measured from the positive $z$-axis and the azimuthal angle ($\phi$) is measured from the positive $x$-axis in the transverse plane. The pseudorapidity is defined as $\eta = -\ln \tan(\theta/2)$. Angular distance is
measured in units of $\Delta R \equiv \sqrt{(\Delta\eta)^{2} +
(\Delta\phi)^{2}}$.}
It consists of an inner tracking detector (ID) surrounded by a thin superconducting solenoid providing a 2\,T axial magnetic field, electromagnetic (EM) and hadronic calorimeters, and a muon spectrometer (MS).
 
The ID  provides charged-particle tracking in the range $|\eta|<2.5$. It consists of silicon pixel, silicon microstrip, and transition radiation tracking detectors.
The high-granularity silicon pixel detector covers the vertex region and typically provides four measurements per track, the first hit normally being in the insertable $B$-layer~\cite{ATLAS-TDR-2010-19,PIX-2018-001}. It is followed by the silicon microstrip tracker, which usually provides eight measurements per track. These silicon detectors are complemented by the transition radiation tracker, which enables radially extended track reconstruction up to $|\eta|=2.0$ and contributes to electron identification.
 
Lead/liquid-argon (LAr) sampling calorimeters provide EM energy measurements with high granularity. A steel/scintillator-tile hadronic calorimeter covers the central pseudorapidity range ($|\eta|<1.7$). The end-cap and forward regions are instrumented with LAr calorimeters for both the EM and hadronic energy measurements up to $|\eta|=4.9$. For $|\eta|<2.5$, the EM calorimeter is divided into three layers longitudinal in shower depth, which are finely segmented in $\eta$ and $\phi$, particularly in the first layer. This segmentation allows measurements of the lateral and longitudinal shower profile, and the calculation of shower shapes~\cite{EGAM-2018-01} used for particle identification and background rejection. The longitudinal segmentation of the EM calorimeter is also exploited to calibrate the energy response to electron and photon candidates~\cite{EGAM-2018-01}.
 
The MS comprises separate trigger and high-precision tracking chambers measuring the deflection of muons in a magnetic field generated by superconducting air-core toroids. The precision chamber system covers the region $|\eta|<2.7$ with three layers of monitored drift tubes, complemented by cathode-strip chambers in the forward region, where the background is highest. The muon trigger system covers the range $|\eta|<2.4$ with resistive-plate chambers in the barrel, and thin-gap chambers in the end-cap regions.
 
A two-level trigger system~\cite{TRIG-2016-01} was used during the 2015--2018 data-taking period.
The first-level trigger (L1) is implemented in hardware and uses a subset of the detector information. This is followed by a software-based high-level trigger, which runs algorithms similar to those in the offline reconstruction software, reducing the event rate to approximately 1\,kHz from the maximum L1 rate of 100\,kHz.
 
An extensive software suite~\cite{ATL-SOFT-PUB-2021-001} is used in data simulation, in the reconstruction and analysis of real and simulated data, in detector operations, and in the trigger and data acquisition systems of the experiment.


\section{Data and simulated event samples}
The analysed data comprise the full ATLAS $pp$ collisions at $\sqrt{s}=13$\,\TeV\ collected from 2015 to 2018, corresponding to an integrated luminosity of 139\,fb$^{-1}$ after data quality requirements~\cite{DAPR-2018-01} were imposed to ensure that all the subdetectors were operating normally.
 
Candidate signal events are selected using triggers~\cite{TRIG-2018-05,TRIG-2018-01} that require either a single isolated lepton or a pair of isolated leptons, targeting leptonic decays of the $Z$ bosons.
Selecting $Z$ bosons with $\pT$ greater than 100\,\GeV, decaying into $e^+e^-$ or $\mu^+\mu^-$,
reduces the SM background contribution and provides a clean sample with high trigger efficiencies.
For the single-lepton triggers, the minimum $p_\mathrm{T}$ thresholds vary between 24\,\GeV\ and 26\,\GeV\ for electrons and between 20\,\GeV\ and 26\,\GeV\ for muons, depending on the data-taking period. For the dilepton triggers, the minimum $p_\mathrm{T}$ thresholds vary between 12\,\GeV\ and 17\,\GeV\ for electrons and between 10\,\GeV\ and 14\,\GeV\ for muons.
These triggers are complemented with jet triggers~\cite{TRIG-2012-01} with minimum $p_\mathrm{T}$ thresholds of 360\,\GeV\ for events with relatively small angular separation between two electron candidates (merged-$e^+e^-$ candidates, discussed in Section~\ref{sec:recon}).
The overall trigger efficiency per event as measured in simulated \Zjets events is about 96\% at low $\pTZ$ down to 100\,\GeV\
and increases to above 99\% for $p_\mathrm{T}^Z\gtrsim 600$\,\GeV\ in the muon channel, and it is above 99\% for the entire considered $p_\mathrm{T}^Z$ range in the electron channel.
 
The SM background originates mainly from production of \Zjets, dibosons ($ZZ$, $ZW$, $Z\gamma$), top-quark pairs ($t\bar{t}$), and single top quarks.
The combination of the SM background processes is expected to yield smoothly falling spectra in the considered mass observables.
A data-driven method using a functional fit is employed to model the contributions from the SM background in the invariant mass distributions.
However, simulated Monte Carlo (MC) samples are used to optimise the analysis, to select fit functions for both observables, and to determine the corresponding fitting ranges.
 
The production of a $Z$ boson in association with jets was modelled by the \Sherpa[2.2.1]~\cite{10.21468/SciPostPhys.7.3.034} generator with the \NNPDF[3.0nnlo] parton distribution function (PDF) set~\cite{Ball:2014uwa}. Diagrams with up to two additional parton emissions were simulated at next-to-leading-order (NLO) accuracy in QCD, and with three or four additional parton emissions at leading-order (LO) accuracy~\cite{Hoeche:2012yf}. Matrix elements were merged with the \Sherpa parton shower using the \MEPSatNLO formalism~\cite{Hoeche:2012yf}. An alternative sample was generated with \POWHEGBOX[v1]~\cite{Nason:2004rx,Frixione:2007vw,Alioli:2010xd,Alioli_2008} at NLO accuracy in QCD, and it was interfaced to \Pythia[8.186]~\cite{Sjostrand:2014zea} for the modelling of the parton shower, hadronisation, and underlying event, with a set of tuned parameters called to the AZNLO tune~\cite{STDM-2012-23}. The \CT[10nlo] PDF set~\cite{Lai_2010} was used for the hard-scattering processes, whereas the \CTEQ[6L1] PDF set~\cite{Pumplin:2002vw} was used for the parton shower.
 
Diboson production with leptonic final states was simulated using \Sherpa[2.2.2]. The matrix elements were calculated using the \NNPDF[3.0nnlo] PDF set. Processes with up to one additional parton emission were calculated at NLO accuracy in QCD,
while processes with two or three parton emissions were modelled at LO accuracy.
The various jet-multiplicity final states were merged using the \MEPSatNLO formalism.
Diboson production with semileptonic decays was modelled using \Sherpa[2.2.1] and the same PDF set.
 
The production of $t\bar{t}$ and the single top quark associated with a $W$ boson ($Wt$) were modelled using \POWHEGBOX[v2]~\cite{Frixione:2007nw} interfaced to \Pythia[8.210] with the A14 parton shower tune~\cite{ATL-PHYS-PUB-2014-021}. The matrix elements were calculated at NLO precision in QCD using the \NNPDF[3.0nlo] PDF set and assuming a top-quark mass of 172.5\,\GeV. The $t\bar{t}$ production cross section is normalised to the prediction from the \TOPpp[2.0] program at next-to-next-to-leading order (NNLO) in perturbative QCD, including soft-gluon resummation calculated to next-to-next-to-leading-logarithm (NNLL) precision~\cite{Czakon:2011xx}.
 
Contributions from multi-jet and \Wjets processes, which cannot be estimated reliably from simulations, are found to be negligible when using data-driven techniques similar to those applied in Ref.~\cite{STDM-2018-14}.
 
For the model-dependent interpretation, HVT signal samples with the coupling strength to SM gauge bosons, $g_{V} $, set to 1, referred to as model A, and $g_{V} = 3$, referred to as model B, are used. Both of them have the same decay channel, $W^{\prime} \to ZW \to \ell\ell qq$.
The sample was generated using \MGNLO~\cite{Alwall:2014hca} interfaced to \Pythia[8.186] with the \NNPDF[2.3lo] PDF set~\cite{carrazza2013parton}. Thirteen signal mass points were simulated, ranging from 500\,\GeV\ to 6\,\TeV. The reconstructed $W^\prime$ resonance mass can be modelled by a double-sided Crystal Ball function~\cite{Oreglia_slac,HIGG-2019-14}.  Acceptance times efficiency, ${\cal A}\times \epsilon$, values of up to 50\% are found for these signal models in the dominant leading large-$R$-jet category.
 
All background and signal samples were processed through a \GEANT-based simulation~\cite{Agostinelli:2002hh,SOFT-2010-01} of the ATLAS detector.
The effect of multiple interactions in the same and neighbouring bunch crossings (pile-up) was modelled by overlaying the simulated hard-scattering event with inelastic $pp$ collision events generated with \Pythia[8.186] using the \NNPDF[2.3lo] PDF set and the A3 tune~\cite{ATL-PHYS-PUB-2016-017}.
The MC simulated events are weighted to match the distribution of the number of reconstructed pile-up vertices observed in different data-taking periods.
Scale factors are applied to the simulated events to account for any remaining differences between observed and simulated data.


\section{Object selection}
\label{sec:recon}
 
Selected events are required to have at least one primary vertex~\cite{ATL-PHYS-PUB-2015-026} with at least two associated tracks, each with $p_\mathrm{T}>500$\,\MeV. In each event, only the primary vertex with the highest sum of squared transverse momenta of contributing tracks, hereafter denoted by $\sum p^2_\mathrm{T}$, is considered.
 
Muons are reconstructed either by combining ID and MS tracks that have consistent trajectories and curvatures or as stand-alone MS tracks~\cite{PERF-2015-10,MUON-2018-03}. The muon candidates used in this analysis are required to have $p_\mathrm{T}>25$\,\GeV\ and $|\eta|<2.7$. They are also required to pass the `Medium' selection for $p_\mathrm{T}>25$\,\GeV\ and the `HighPt' selection for  $p_\mathrm{T}>300$\,\GeV. The efficiency for identifying a single muon is about 99\% for $p_\mathrm{T}<300$\,\GeV\ and around 80\% for $p_\mathrm{T}>300$\,\GeV.
 
Electrons are reconstructed from clusters of energy deposits in the EM calorimeter that match a track reconstructed in the ID. Reconstructed electrons are identified using likelihood identification criteria~\cite{PERF-2017-01,EGAM-2018-01}. The electron candidates are required to have  $p_\mathrm{T}>25$\,\GeV. They are also required to pass the `MediumLH' selection and be within $|\eta|<2.47$, excluding the transition region between the barrel and end-caps in the LAr calorimeter ($1.37<|\eta|<1.52$).
 
To ensure that lepton candidates originate from the interaction point, the likelihood-based electrons (muons) must satisfy $|d_0|/\sigma_{d_0}< 5\,(3)$, and both lepton types must satisfy $|z_0\sin\theta|<0.5$~mm.
Here, $d_0$ is the transverse impact parameter of the lepton measured relative to the primary vertex, $\sigma_{d_0}$ is the uncertainty in $d_0$, and $z_0$ is the longitudinal distance between the position of the primary vertex and the position of the track at the point of closest approach.
In addition, electron and muon candidates are required to be isolated from other tracks and calorimetric activity by applying $p_\mathrm{T}$- and $\eta$-dependent isolation criteria. For both the electron and muon candidates, the calorimeter isolation is based on the sum of the transverse energies of topological clusters~\cite{PERF-2014-07} within a cone of size $\Delta R=0.2$ around each lepton. The track isolation is computed by summing the transverse momenta of selected tracks within a cone of $p_\mathrm{T}^\ell$-dependent size around the lepton track. The efficiency of the isolation requirements is 95\% for muons from $Z$ bosons with $p_\mathrm{T}>100$\,\GeV. For electrons, it is 90\% but decreases for $Z$ bosons with $p_\mathrm{T} >450$\,\GeV\ as the electron signatures begin to overlap.
 
If a $Z \to e^+e^-$ decay has a sufficiently large Lorentz boost, the $e^+e^-$ angular separation becomes small enough for the $e^+e^-$ pair to merge and fail the electron isolation requirements.
The merged $e^+e^-$ pairs are then reconstructed as jets, and
a boosted decision tree (BDT) algorithm~\cite{breiman2017classification,ROE2005577} is used to separate merged-$e^+e^-$ candidates from hadronic jets.
This BDT identification is only applied when there is no $Z$ boson candidate from either an $e^+e^-$ pair or a $\mu^+\mu^-$ pair in the event.
The input variables for the BDT include the jet's calorimeter- and ID-track-related properties and kinematic information. More details are given in Appendix~\ref{app:merged-ee}. The combined efficiency of the resolved and merged identification for an $e^+e^-$ pair forming a $Z$ boson candidate is around 80\% with weak $p_\mathrm{T}$ dependence.
 
Photon candidates are reconstructed from dynamic, variable-size topological clusters of cells with significant energy in the EM calorimeter and, possibly, matching tracks reconstructed in the inner detector~\cite{PERF-2017-02}.
The photon candidates are classified as `converted' if two tracks forming a conversion vertex, or one track with the signature of an electron track but without hits in the innermost pixel layer, are matched to the cluster; otherwise, they are labelled as `unconverted'.
They are identified by using variables that characterise the lateral and longitudinal shower development in the EM calorimeter and the energy fraction leaking into the hadronic calorimeter. Photons used in this analysis are required to pass the `Tight' selection~\cite{DAPR-2018-01,EGAM-2018-01} for transverse energy $E_\mathrm{T}>25$\,\GeV\ and be within $|\eta|<2.37$, excluding the same transition region between the barrel and end-caps in the LAr calorimeter as for electrons. The selection has an efficiency of 82\%--92\% (75\%--92\%) for unconverted (converted) photon candidates depending on $E_\mathrm{T}$ and $\eta$. The fractions of the unconverted and converted photon candidates are about 68\% and 32\%, respectively.
The photon isolation requirement is based on the amount of transverse energy inside a cone of size $\Delta R=0.4$. The efficiency of the isolation requirement is 75\%--90\% (${\sim}70$\%) for unconverted (converted) photon candidates.
 
Small-$R$ jets are reconstructed using the particle-flow algorithm~\cite{PERF-2015-09}, which considers both tracker and calorimeter information,
using the anti-$k_t$ algorithm~\cite{Cacciari:2008gp} with a radius parameter of $R=0.4$ implemented in the FastJet package~\cite{Cacciari:2011ma}.
The four-momenta of these small-$R$ jets are calculated as the sum of the four-momenta of their constituents.
Jets are corrected using $p_\mathrm{T}$- and $\eta$-dependent scale factors to account for the calorimeter's non-compensating response, signal losses due to noise-threshold effects, energy lost in passive material, and contributions from pile-up interactions, as described in Ref.~\cite{PERF-2016-04}.
 
The jets are required to have $p_\mathrm{T}>30$\,\GeV\ and $|\eta|<2.4$. To suppress jets that originate from pile-up, a jet-vertex tagger using the `Tight' working point~\cite{PERF-2014-03} is applied to jets with $p_\mathrm{T}<60$\,\GeV\ and $|\eta|<2.4$.
To avoid double counting, jets of any transverse momentum are discarded if they are within a cone of size $\Delta R = 0.2$ around an electron or a photon, or if they have fewer than three associated tracks and are within a cone of size $\Delta R = 0.2$ around a muon. However, jets discarded due to an overlap with a reconstructed electron are still considered when testing for a merged-$e^+e^-$ candidate.
If a remaining jet with three or more associated tracks is within a cone of size $\Delta R =0.4$ around a muon candidate, or the separation between an electron or a photon and any jet is within $0.2 < \Delta R < 0.4$, the corresponding muon, electron, or photon candidate is rejected.
 
Small-$R$ jets containing a $b$-hadron are identified using the DL1r $b$-tagging algorithm~\cite{FTAG-2018-01}
at the 77\% $b$-tagging efficiency operating point. The misidentification rate for jets which originate from a light quark or a gluon is less than 1\%, while it is approximately 20\% for $c$-jets, as measured in simulated $t\bar{t}$ events.
 
Large-$R$ jets are reconstructed from topological clusters~\cite{PERF-2014-07} in the calorimeter, using the anti-$k_t$ algorithm with a radius parameter $R=1.0$. To reduce contributions to the jet's transverse momentum and mass that arise from pile-up, a trimming procedure~\cite{Krohn:2009th} is applied to remove $R=0.2$ subjets with less than 5\% of the original $p_\mathrm{T}$ of the jet. Corrections are applied to restore the jet's energy to that of jets reconstructed at the particle-level energy scale~\cite{JETM-2018-06}. Jet candidates are required to have $p_\mathrm{T}$ above 200\,\GeV\ and $|\eta|<2.0$. The energy resolution for large-$R$ jets with $p_\mathrm{T}$ between  200\,\GeV\ and 1\,\TeV\ is $\sim$7\%~\cite{JETM-2018-06}. The jet's mass is computed as a weighted combination of calorimetric mass and track-assisted mass~\cite{ATLAS-CONF-2020-022}. To avoid double counting, the small-$R$ jets are discarded if they are within a cone of size $\Delta R=1.0$ around the large-$R$ jets. The large-$R$ jets are discarded if they are within a cone of size $\Delta R=1.0$ around an electron or a photon.


\section{Event selection and classification}
\label{sec:selection}
 
The $Z$ boson candidates are selected by requiring two oppositely charged, same-flavour leptons ($e^+e^-$ or $\mu^+\mu^-$). Both leptons must satisfy the minimal quality criteria discussed in Section~\ref{sec:recon}. If no $Z$ boson candidate can be formed from a pair of two isolated leptons,  merged-$e^+e^-$ candidates are checked. The $Z$ boson candidates are required to have an invariant mass between 66 and 116\,\GeV.
If there are multiple $Z$ boson candidates, the one with the mass closest to the $Z$ boson mass is chosen.
To suppress the large SM background contributions at low $p_\mathrm{T}$ and to be more sensitive to high $p_\mathrm{T}$ final states expected from new heavy particles, the selected $Z$ boson candidate is also required to have $p_\mathrm{T}^Z>100$\,\GeV. The choice of this $p_\mathrm{T}^Z$ threshold is a compromise between suppressing the SM background contribution and keeping the data event yields high enough for data-driven background estimations in all search categories.
 
For a given event, all objects satisfying the selections described in Section~\ref{sec:recon}, except the lepton pair from the $Z$ boson decay, are assumed to belong to the recoil system ($X$).
The objects in $X$ are ordered in $p_\mathrm{T}$.
The leading-$p_\mathrm{T}$ object type is used to assign an event to one of six independent event categories: a small-$R$ (non-$b$) jet (LeadJ), a small-$R$ $b$-jet (LeadB), a large-$R$ jet (LeadFatJ), a photon (LeadP), an electron (LeadE) or a muon (LeadM). An inclusive category is formed by summing the six exclusive categories. This category can be useful when the signal is distributed over several considered event categories. The data yields in the six categories are shown in Table~\ref{tab:eventyield}. The invariant mass of the recoil system, $m_X$, and the invariant mass of the recoil system and the $Z$ boson, $m_{ZX}$, are the observables of the search.
 
\begin{table}[htbp]
\begin{center}
\caption{Data yields of the six event categories in the $Z\to e^+e^-$ and $\mu^+\mu^-$ decay channels. The merged-$e^+e^-$ events are included in the $e^+e^-$ channel, increasing the event yield, mainly in the leading large-$R$-jet category, by 0.6\%.}
\label{tab:eventyield}
\begin{tabular}{crrrrrr}
\hline
Category & LeadJ & LeadB & LeadFatJ & LeadP & LeadE & LeadM \\\hline
$e^+e^-$    & 979\,074 & 77\,625 & 181\,561 & 2\,601 & 565 & 530 \\\hline
$\mu^+\mu^-$    & 1\,307\,187 & 99\,927 &  228\,986 & 3\,418 & 790 & 766\\\hline
\end{tabular}
\end{center}
\end{table}


\section{Search for narrow resonances} \label{sec:search}
 
\subsection{Background modelling}\label{sec:bkgmodel}
 
The SM background is modelled by a fit to the binned distribution of $m_X$ or $m_{ZX}$ after combining the $e^+e^-$ and $\mu^+\mu^-$ channels in each event category.
MC samples are used in selecting the varying bin size for each mass distribution, ${\cal M}$ ($m_X$ or $m_{ZX}$).
The bin widths for a mass spectrum are chosen to approximate the mass resolution, which broadens with increasing mass.
Two functional forms are considered for the estimation of the background and are defined as:
\begin{equation}
f_1(x)=p_0\left(\mathrm{e}^{-p_1x}+p_2\mathrm{e}^{-(p_1+p_3)x}+p_4\mathrm{e}^{-(p_1+p_3+p_5)x}+\cdots\right)\label{eq:form1}
\end{equation}
or
\begin{equation}
f_2(x)=p_0(1-x)^{p_1}x^{p_2+p_3\ln\!x+p_4\ln^2\!x+\cdots}\label{eq:form2}
\end{equation}
where $x= ({\cal M}-{\cal M}_\mathrm{min})/({\cal M}_\mathrm{max}-{\cal M}_\mathrm{min})$ in  Eq.~(\ref{eq:form1}) and $x = {\cal M}/\sqrt{s}$ in Eq.~(\ref{eq:form2}), $p_0$ is a free normalisation factor and $p_i$ are a number of other free parameters controlling the shape of a mass distribution, and ${\cal M}_\mathrm{min}$ and ${\cal M}_\mathrm{max}$ are the lower and upper fit boundaries of the distribution. For a given mass bin, the integral of the function is fit to the selected number of events in the bin.
The number of free parameters depends on the mass spectrum and event category.
Mass spectra that cover a larger range usually have higher yields and need more parameters than those with a smaller range.
 
The choice of functional form, number of free parameters, and fit range is based on two criteria for the fits initially performed on the simulated SM mass spectrum. The statistics of the MC samples is category dependent and varies typically from a few times up to an order of magnitude larger than that of the data sample.
The global $\chi^2$ $p$-value of the background-only fit has to be greater than 0.05 for the modelling to be good enough over the considered mass range, and the significance of the spurious-signal yields over the fit range must typically be below 20\%--50\% to ensure a discovery will not be claimed falsely.
The spurious signal is evaluated by fitting the simulated background's mass spectrum with signal-plus-background models.
The statistical uncertainty of the mass spectrum corresponds to that of data.
The signal significance is defined as the ratio of the absolute value of the signal yield to its fitted uncertainty.
The upper fit boundary is dictated by the number of events in the high mass tail of the mass spectrum. In most cases, the lower fit boundary is a direct consequence of the event selection. However, in some cases, it has to be increased slightly to satisfy the requirements of a small spurious-signal significance and a global $\chi^2$ $p$-value larger than 0.05 for the background fit.
The spurious signal for the model-independent interpretation is evaluated by fitting the simulated background mass spectrum with a Gaussian-shaped signal function together with the selected background functional form.
The detector resolution, with relative values typically around 5\% for the mass observables, is also taken into account for the Gaussian-shaped signals with intrinsic relative width values of 3\%, 5\%, and 10\%. It widens the reconstructed peak from the intrinsic width.
For a model-dependent interpretation with an HVT signal, the spurious signal is evaluated with a double-sided Crystal Ball function whose parameter values are determined from the HVT MC sample.
 
\subsection{Search strategy}\label{sec:strategy}
 
\begin{figure}[htbp]
\centering
\includegraphics[width=\columnwidth]{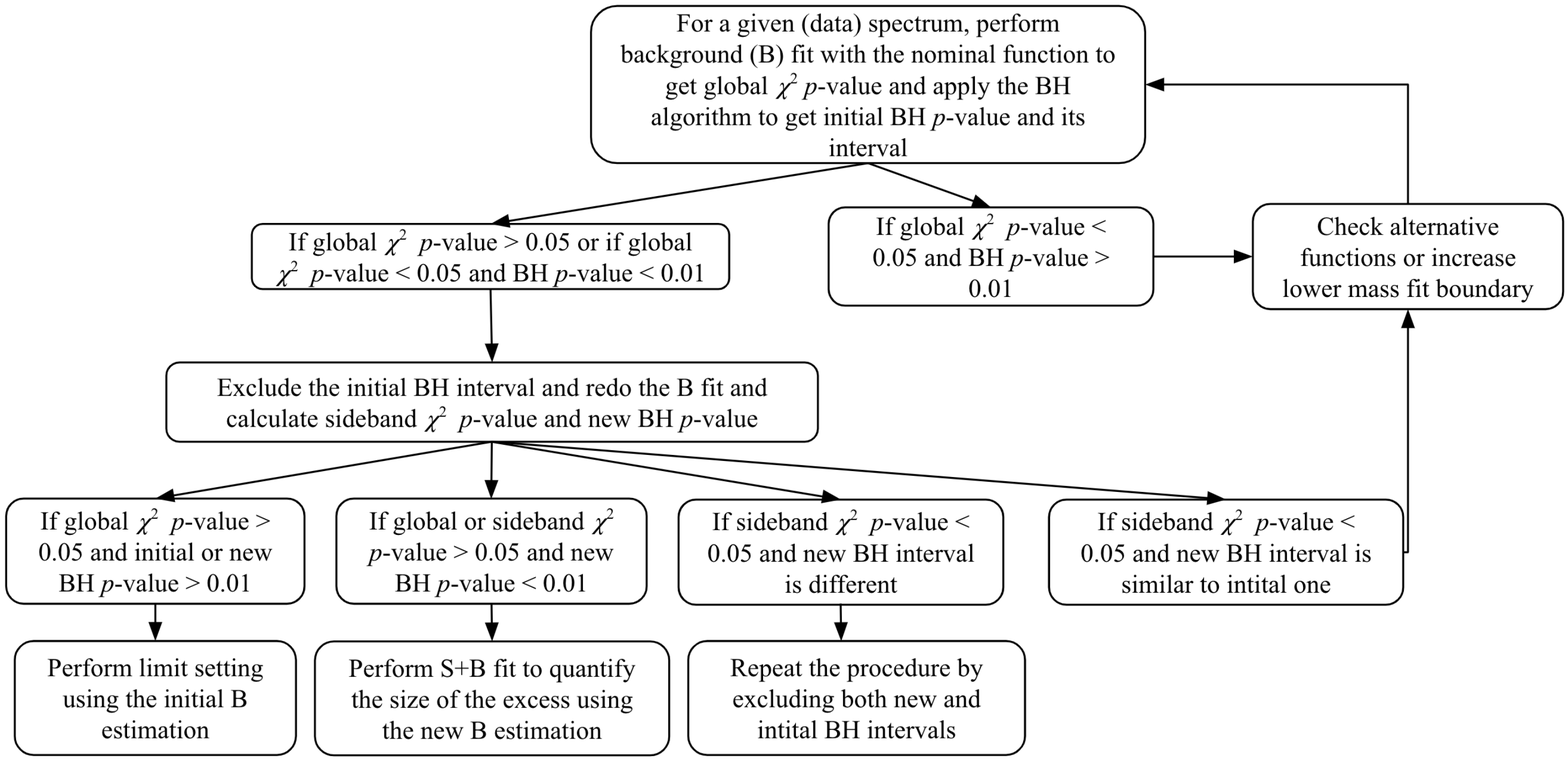}
\caption{Workflow for the search strategy}
\label{fig:search_strategy}
\end{figure}

The strategy applied in this search is diagrammed in Figure~\ref{fig:search_strategy}.
The mass spectrum in data in each category is fitted with a functional form within the fit range. Both the functional form and the fit range are chosen for each mass spectrum, as discussed in Section~\ref{sec:bkgmodel}. To search for local excesses over the background, the BumpHunter (BH) algorithm is then applied to the mass spectrum within a signal-sensitive mass range, which lies within the fit range (see below).
 
The BH algorithm calculates the significance of any excess found in continuous mass intervals of varying size at various locations in the spectrum.
The size of the search window varies from a minimum of two mass bins up to half of the bins in the sensitive mass range. For each interval in the scan, the BH algorithm computes the significance of the difference between the data and the background estimate. The interval where the data deviates most significantly from the background estimate is defined as the set of bins that have the smallest probability of arising from a Poisson background fluctuation. The probability of random fluctuations in the background-only hypothesis to create an excess at least as significant as the one observed anywhere in the spectrum defines the BH $p$-value. It is determined by performing a series of pseudo-experiments drawn from the background estimate, with the look-elsewhere effect considered.
 
The signal-sensitive mass range for each signal depends on the assumed shape and width. The sensitive range is derived using the procedures described in Appendix~\ref{app:signal-range}. A signal at a given mass point  within the fit range is injected on top of a set of 100 pseudo-experiments generated by fluctuating the fitted background mass spectrum within a Poisson distribution. The injected number of signal events  has a signal significance of $5\sigma$, calculated over the full mass spectrum using a definition recommended in Ref.~\cite{ATL-PHYS-PUB-2020-025}.
A mass point is considered to be sensitive if the signal is correctly located with a BH $p$-value below 0.01 in more than 50\% of the pseudo-experiments. In general, the procedure is more sensitive for narrow signals,  so the signal-sensitive range corresponding to the 3\% width value is used in the BH search and it is wider than the signal-sensitive mass ranges for other width values.
For Gaussian-shaped signals, the largest considered width is 10\% because wider signals are often absorbed into the estimated background.
 
To ensure that the global fitting method is sensitive to potential signal resonances and the background
functional form is appropriate, an alternative signal injection test has been performed and it was found that the injected signal can be recovered by the procedure.
 
The search strategy depends on the global $\chi^2$ $p$-value and the BH $p$-value.
If the global $\chi^2$ $p$-value is above the threshold of 0.05 and the BH $p$-value is above the threshold of 0.01, it indicates that this functional form can model the SM background for the examined mass spectrum and there is no significant excess above the background.
In this case, upper limits are set on the signal models within a limit-sensitive mass range which is derived for each signal using a procedure, described in Appendix~\ref{app:limit-range}, based on the coverage of the expected exclusion limit.
 
If the global $\chi^2$ $p$-value is below 0.05 and the BH $p$-value is above 0.01, it indicates that the initial functional form validated on the simulated SM mass spectrum does not fit the corresponding data spectrum. Alternative functional forms are then tested against the nominal one. Two alternative functional forms are considered: the same functional form as the nominal one but with more free parameters and the other functional form defined in Eq.~(\ref{eq:form1}) or~(\ref{eq:form2}). In the first case, the test is performed using an $F$-test~\cite{ATL-PHYS-PUB-2020-028}:
\begin{equation*}
F=\frac{\left(\chi^2_\textrm{nom}-\chi^2_\textrm{alt}\right)/(n_\textrm{alt}-n_\textrm{nom})}{\chi^2_\textrm{alt}/(n-n_\textrm{alt})}
\end{equation*}
where $\chi^2_\textrm{nom}$ and $\chi^2_\textrm{alt}$ are the $\chi^2$ values for fits to the spectrum using, respectively, the nominal or alternative model, $n_\textrm{nom}$ and $n_\textrm{alt}$ are the number of free parameters in each model, and $n$ is the number of bins used for the $\chi^2$ computation. The value of $F$ is larger if the alternative model provides a significantly better fit to the spectrum than the nominal model.
The alternative model is retained if the $p$-value of the $F$-test satisfies $p(F)>0.05$.
If the test is not successful, the other functional form is considered with up to six (eight) free parameters for $f_2(x)$ ($f_1(x)$). If both functional forms fail to describe the background, the lower fit boundary is raised until the global $\chi^2$ $p$-value is above the threshold of 0.05.
 
If the BH $p$-value is below 0.01 for any global $\chi^2$ $p$-value, a new global fit is performed while excluding the BH interval corresponding to the most significant excess relative to the background estimated by the fit in the full fit range. This gives the sideband $\chi^2$ $p$-value. The BH algorithm is then applied using the newly fitted background estimate to get a new BH $p$-value. If the new BH interval is at the same mass location as the initial one, the new BH $p$-value is expected to be smaller than the initial one
because more events from the excess may be absorbed in the background fit in the initial step. If the new BH interval is at a different location, the procedure is repeated while excluding both intervals.
If the global $\chi^2$ $p$-value is above 0.05, the excess is to be further quantified by performing a signal-plus-background fit to the mass spectrum.
On the other hand, if the global $\chi^2$ $p$-value is smaller than 0.05, the global fit is to be improved using alternative models described above and the procedure is then repeated.
 
\subsection{BumpHunter search results}
 
The functional forms given in Section~\ref{sec:bkgmodel} are applied to mass spectra in data for the six exclusive event categories and the inclusive one to get the initial data-driven background estimate. The upper fit boundaries for some of the event categories have to be lowered from their initial MC-based values because there are fewer high-mass data events than expected. The lower fit boundaries are driven by the kinematic selection except for the $m_{ZX}$ spectra in the leading large-$R$-jet and inclusive categories, where the search strategy requires them to be raised a little.
The fit range and nominal functional form for each mass spectrum are shown in Table~\ref{tab:bkgfit}.
\begin{table}[htbp]
\begin{center}
\caption{A list of mass spectra, event categories and their corresponding fit ranges, functional forms, numbers of free parameters and global $\chi^2$ $p$-values from background-only fits. The fit range values are rounded to the nearest 5~\GeV.}
\label{tab:bkgfit}
\begin{tabular}{clcccc}
\hline
Spectrum & Category &  Fit range [\GeV]  & Function & $N_\mathrm{parameters}$ & $\chi^2$ $p$-value \\\hline
$m_X$  & LeadJ & [350, 5270] & $f_1(x)$ & 6 & 0.68\\
$m_X$ & LeadB & [350, 3340]  & $f_1(x)$ & 6 & 0.69 \\
$m_X$ & LeadFatJ &  [630, 5670]  & $f_2(x)$ & 4 & 0.58\\
$m_X$ & LeadP &  [150, 1800]  & $f_1(x)$ & 4 & 0.98\\
$m_X$ & LeadE & [150, 1515]  & $f_1(x)$ & 4 & 0.67\\
$m_X$ & LeadM & [150, 1230]  & $f_1(x)$ & 4 & 0.12\\
$m_X$ & Inclusive & [550, 6455]  & $f_2(x)$ & 6 & 0.58\\\hline
 
$m_{ZX}$ & LeadJ & [730, 6050] & $f_1(x)$ & 6 & 0.63\\
$m_{ZX}$ & LeadB & [580, 4185]  & $f_1(x)$ & 6 & 0.52 \\
$m_{ZX}$ & LeadFatJ & [830, 6650]  & $f_1(x)$ & 4 & 0.05\\
$m_{ZX}$ & LeadP & [280, 2200] & $f_1(x)$ & 4 & 0.84\\
$m_{ZX}$ & LeadE & [325, 2185]  & $f_1(x)$ & 4 & 0.85\\
$m_{ZX}$ & LeadM & [290, 1900]  &$f_2(x)$ & 4 & 0.11\\
$m_{ZX}$ & Inclusive & [800, 6830] & $f_1(x)$ & 6 & 0.11\\\hline
\end{tabular}
\end{center}
\end{table}
 
\begin{table}[htbp]
\begin{center}
\caption{A list of mass spectra, event categories and their corresponding signal-sensitive mass ranges. The initial BH $p$-value is obtained by using the background derived from the background-only fit in the full fit range, whereas the new BH $p$-value uses the background derived from the background-only fit in the range excluding the initial BH interval.}
\label{tab:bhfit}
\begin{tabular}{clccc}
\hline
Spectrum & Category &  Sensitive range [\GeV]  & Initial BH $p$-value & New BH $p$-value  \\\hline
$m_X$  & LeadJ & [350, 5270] & 0.69 & 0.30\\
$m_X$ & LeadB & [350, 2800]  & 0.91 & 0.39 \\
$m_X$ & LeadFatJ &  [630, 5670]  & 0.79 & 0.61\\
$m_X$ & LeadP &  [150, 1580]  & 0.94 & 0.69\\
$m_X$ & LeadE & [150, 1300]  & 0.92 & 0.65\\
$m_X$ & LeadM & [150, 1025]  & 0.48 & 0.08\\
$m_X$ & Inclusive & [675, 6450]  & 0.71 & 0.30\\\hline
 
$m_{ZX}$ & LeadJ & [860, 6050]  & 0.76 & 0.27\\
$m_{ZX}$ & LeadB & [580, 3615]  & 0.95 & 0.72 \\
$m_{ZX}$ & LeadFatJ & [830, 6650]  & 0.46 & 0.10\\
$m_{ZX}$ & LeadP & [280, 2200] & 0.69 & 0.45\\
$m_{ZX}$ & LeadE & [325, 1905]  & 0.94 & 0.85\\
$m_{ZX}$ & LeadM & [290, 1650]  & 0.60 & 0.40\\
$m_{ZX}$ & Inclusive & [800, 6830] & 0.74 & 0.29\\\hline
\end{tabular}
\end{center}
\end{table}
 
\begin{figure}[htbp]
\centering
\subfloat[]{\includegraphics[scale=0.33]{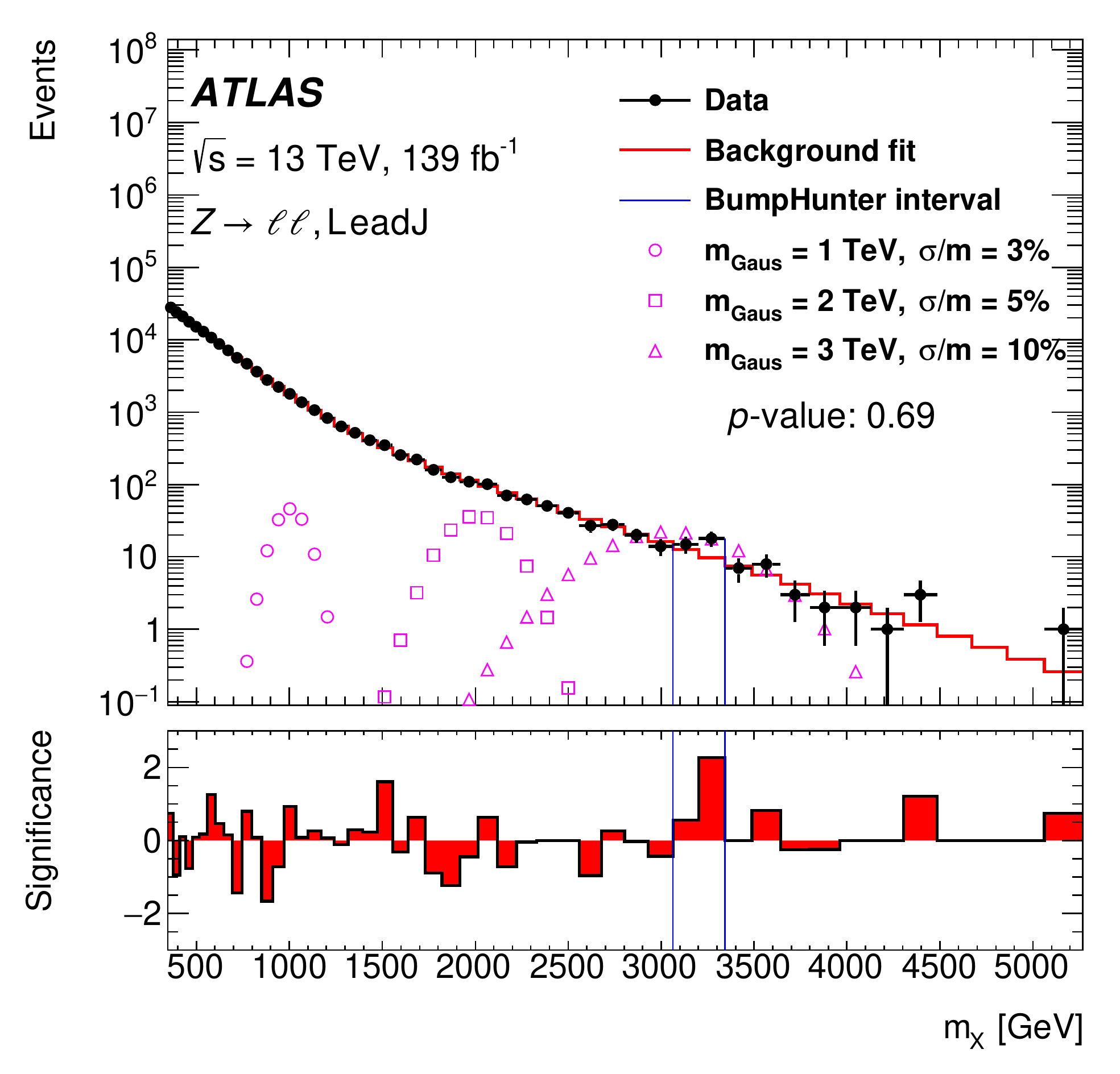} \label{fig:bh-mx-a}}
\subfloat[]{\includegraphics[scale=0.33]{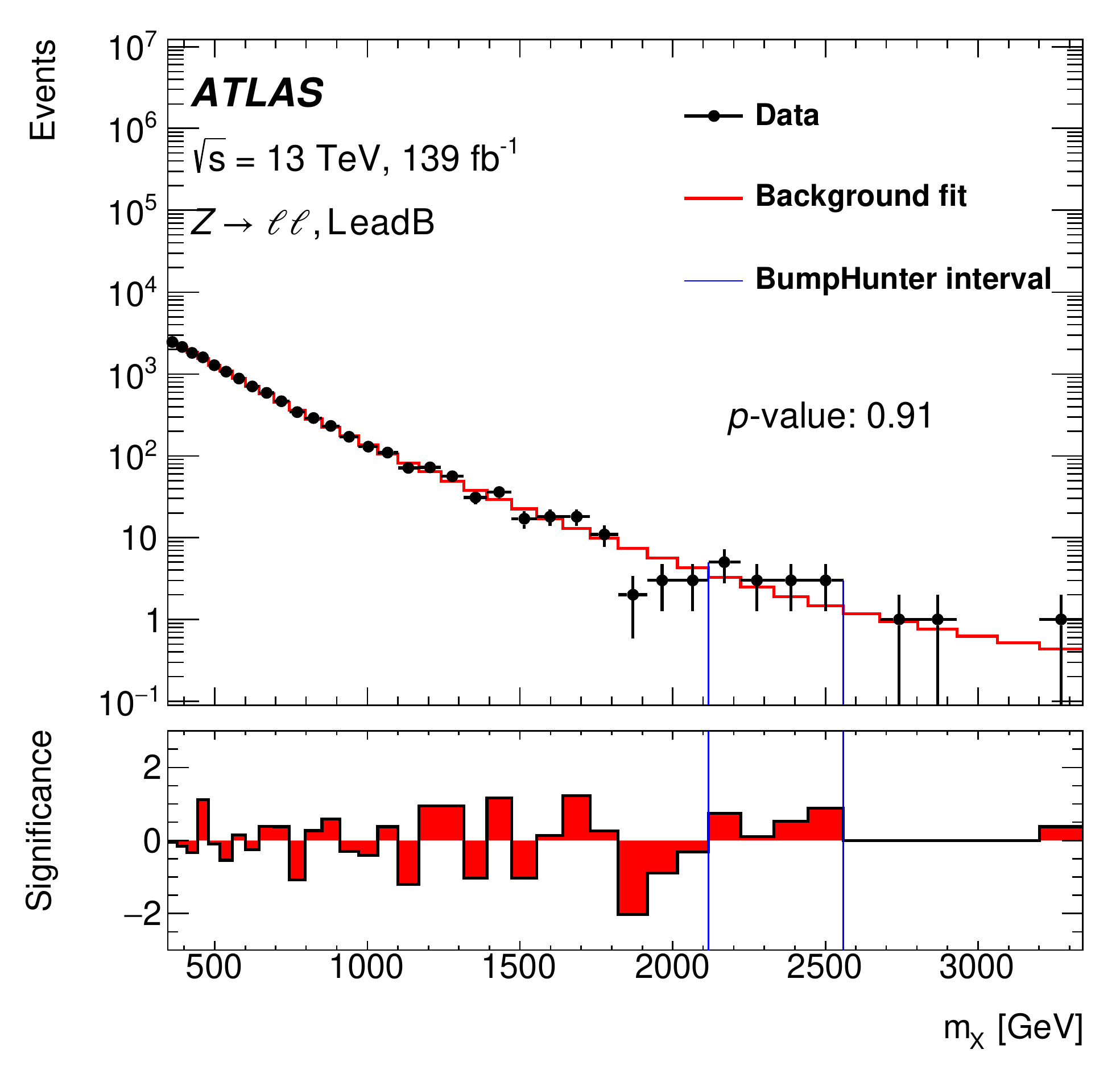} \label{fig:bh-mx-b}}
\vspace{-.5\baselineskip}
\hfill
\subfloat[]{\includegraphics[scale=0.33]{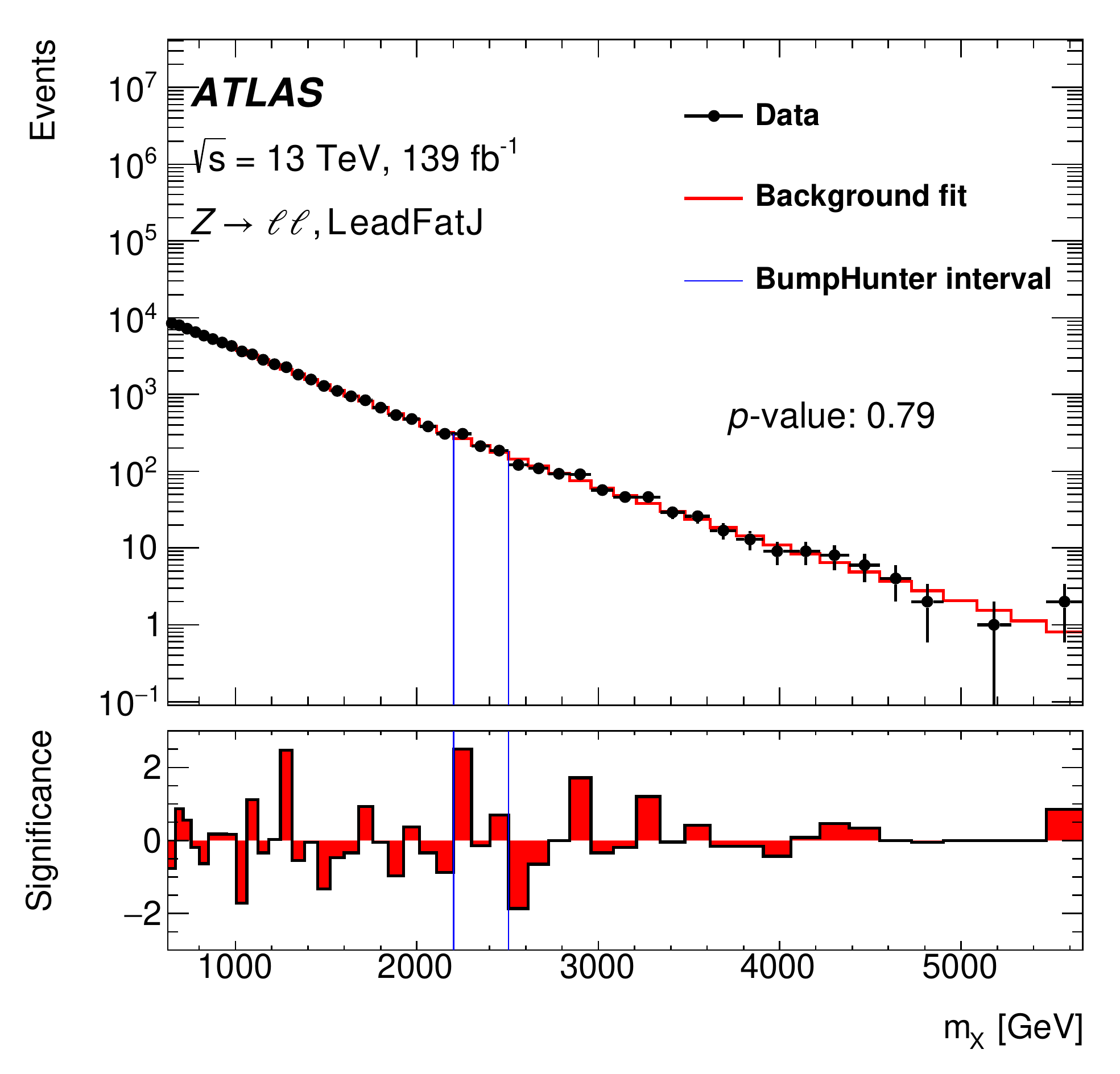} \label{fig:bh-mx-c}}
\subfloat[]{\includegraphics[scale=0.33]{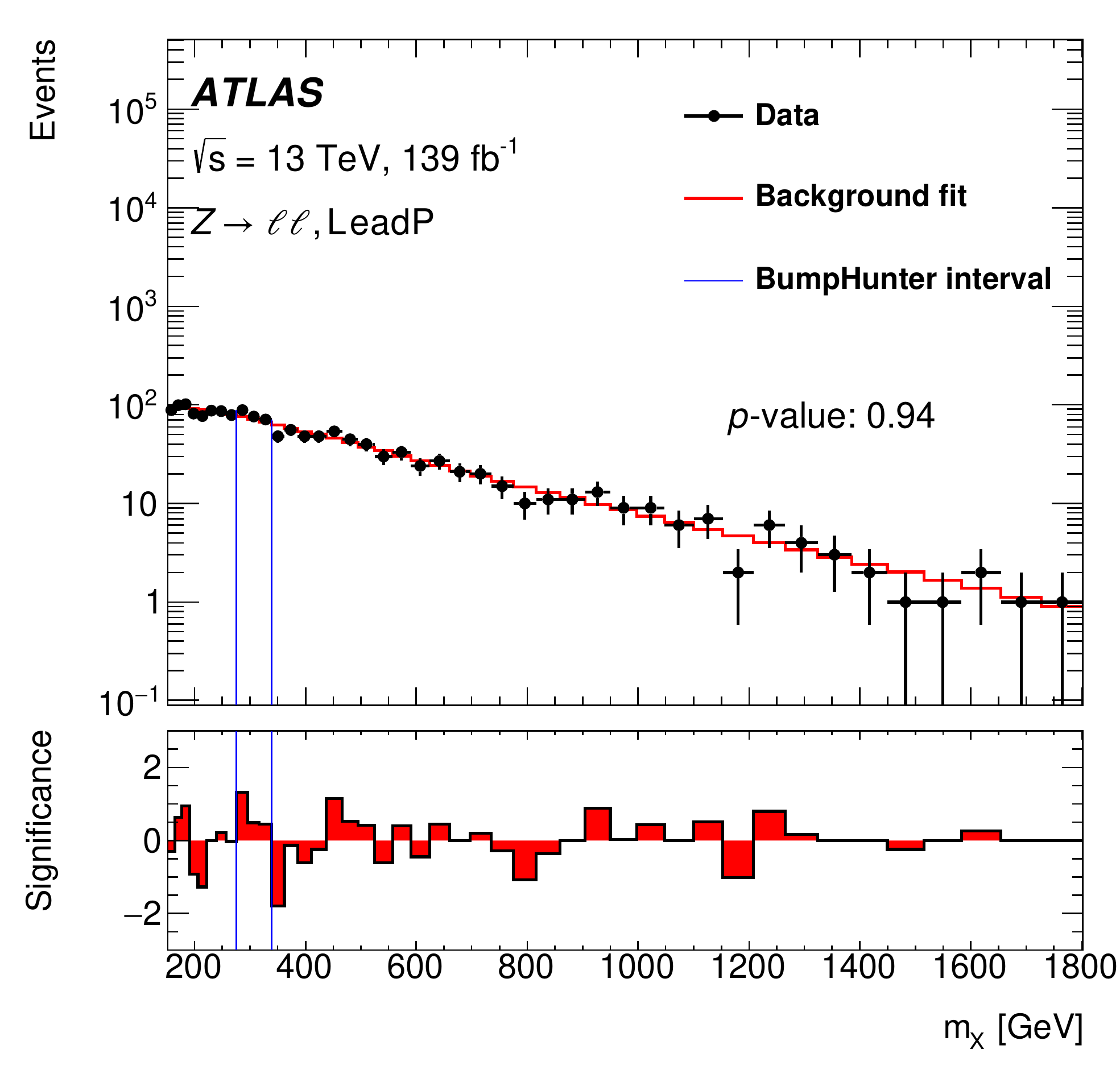} \label{fig:bh-mx-d}}
\vspace{-.5\baselineskip}
\hfill
\subfloat[]{\includegraphics[scale=0.33]{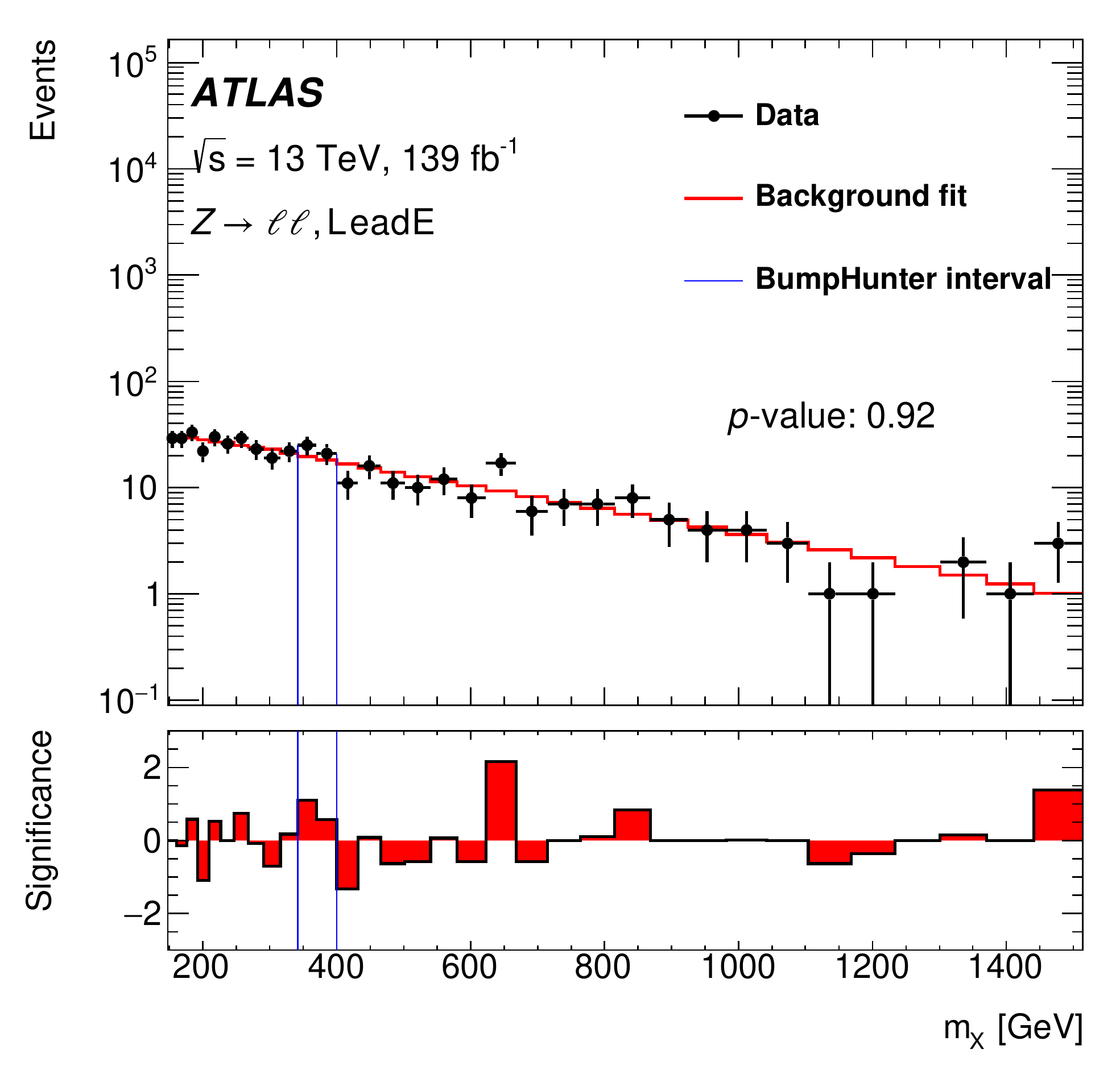} \label{fig:bh-mx-e}}
\subfloat[]{\includegraphics[scale=0.33]{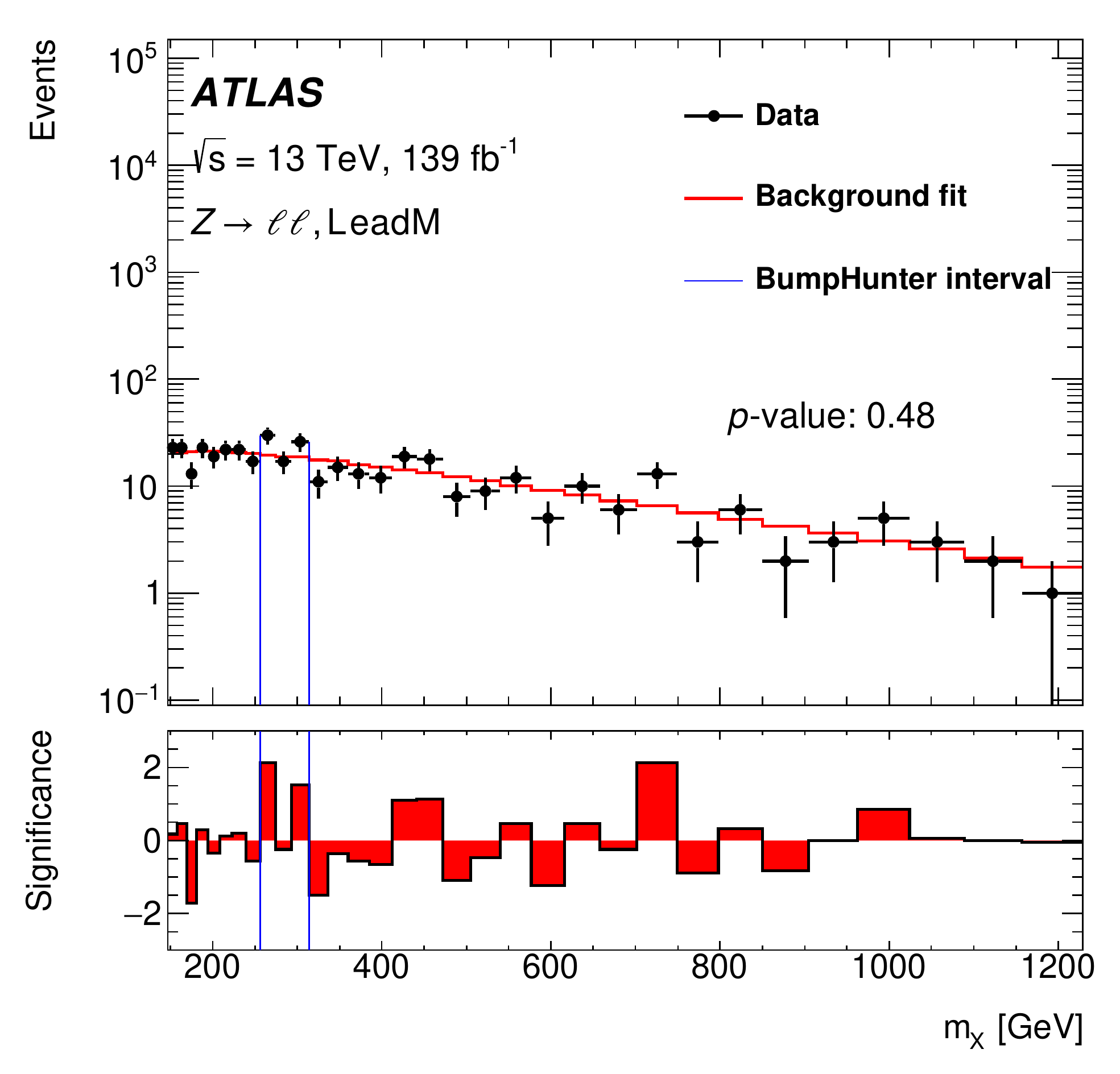} \label{fig:bh-mx-f}}
\vspace{-.5\baselineskip}
\caption{The $m_X$ spectra in the six exclusive categories: \protect\subref{fig:bh-mx-a} leading small-$R$ jet, where Gaussian-shaped signals at mass values around 1, 2, and 3\,\TeV\ with relative widths of 3\%, 5\% and 10\% corresponding to a cross section of 1\,fb each are also shown, \protect\subref{fig:bh-mx-b} leading $b$-jet, \protect\subref{fig:bh-mx-c} leading large-$R$-jet, \protect\subref{fig:bh-mx-d} leading photon, \protect\subref{fig:bh-mx-e} leading electron, and \protect\subref{fig:bh-mx-f} leading muon. The vertical lines indicate the most discrepant interval identified by the BH test from the initial step and the $p$-values correspond to those shown in the fourth column in Table~\ref{tab:bhfit}.}
\label{fig:bh-mx}
\end{figure}

\begin{figure}[htbp]
\centering
\subfloat[]{\includegraphics[scale=0.33]{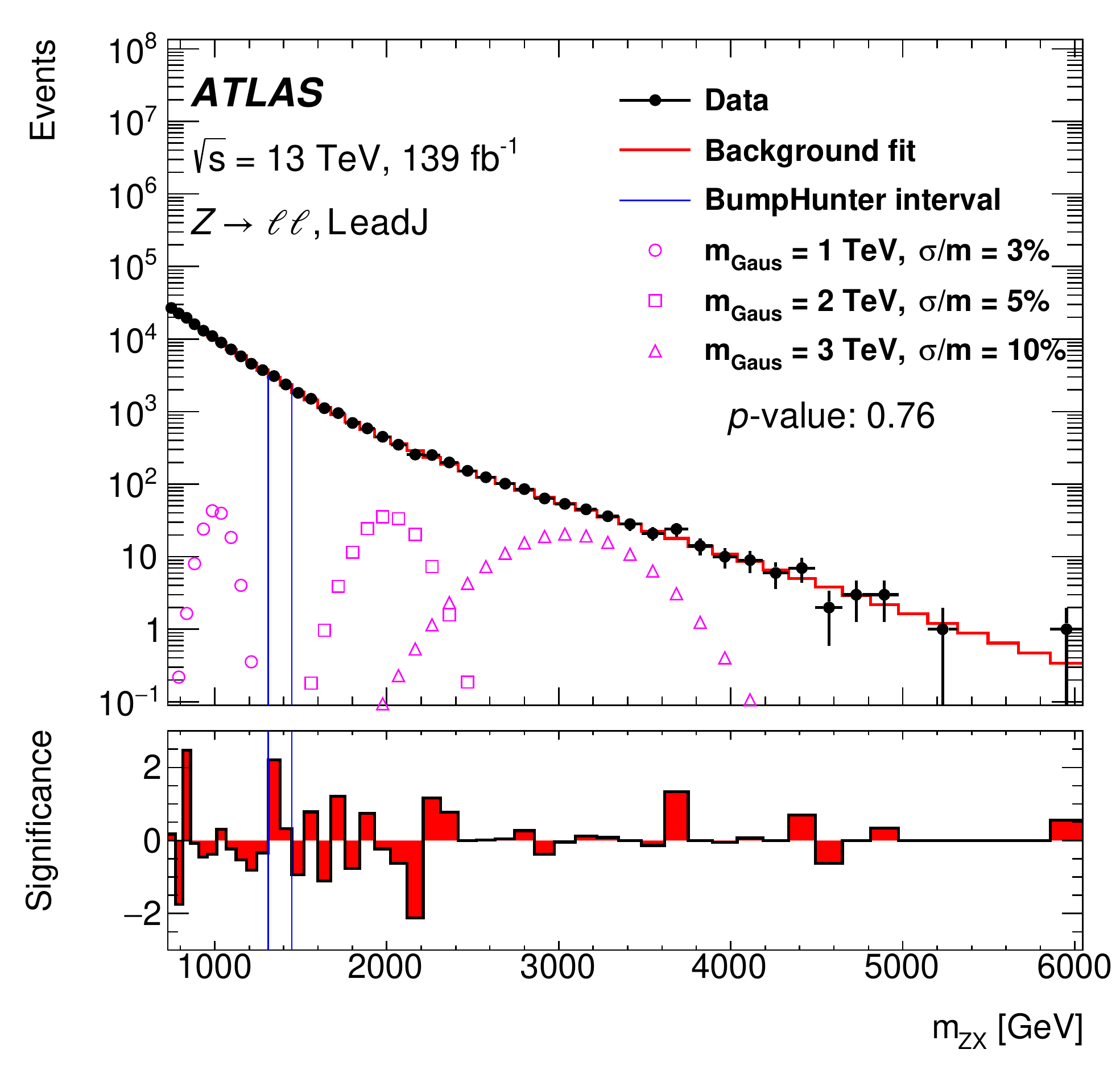} \label{fig:bh-mzx-a}}
\subfloat[]{\includegraphics[scale=0.33]{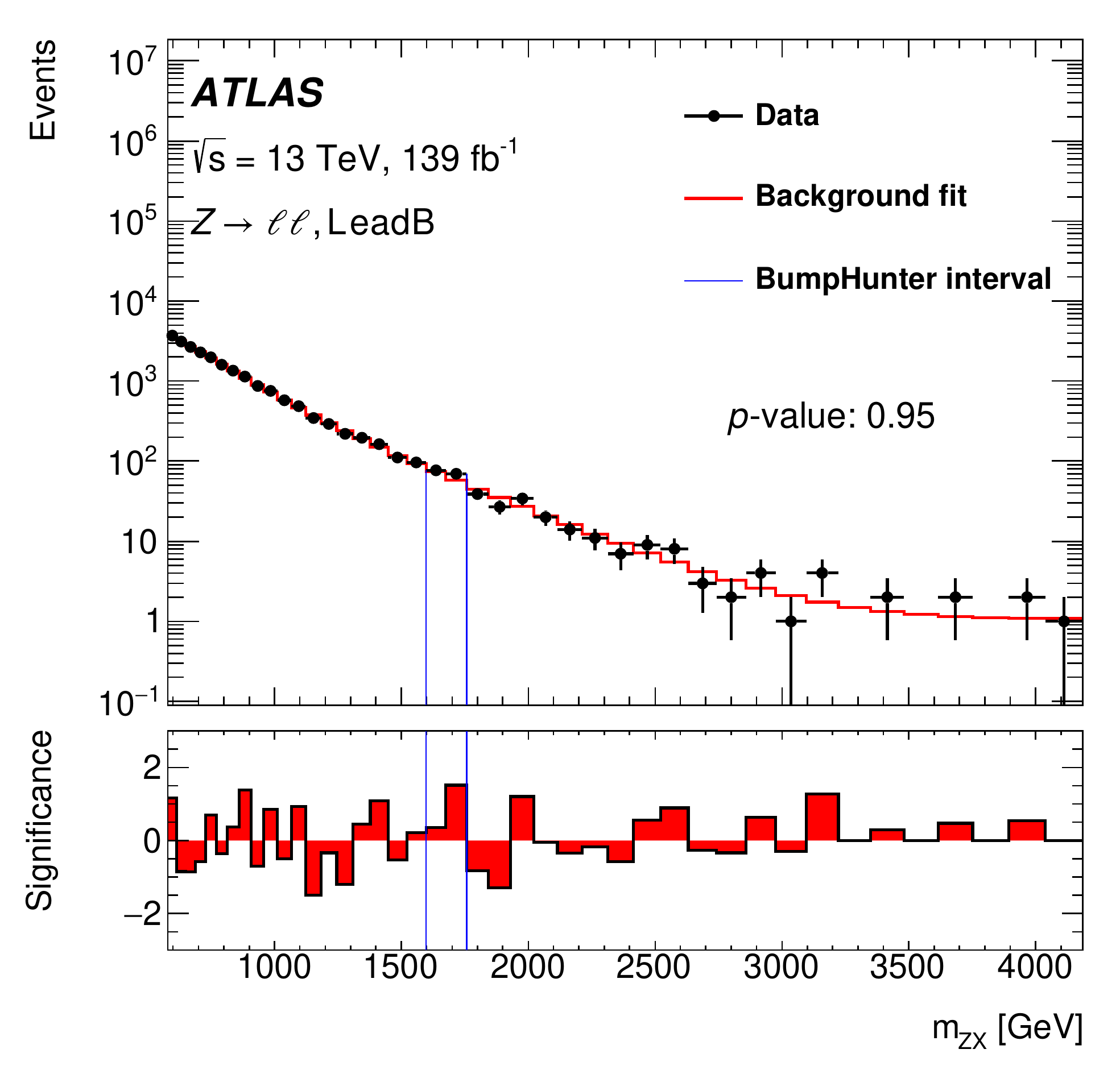} \label{fig:bh-mzx-b}}
\vspace{-.5\baselineskip}
\hfill
\subfloat[]{\includegraphics[scale=0.33]{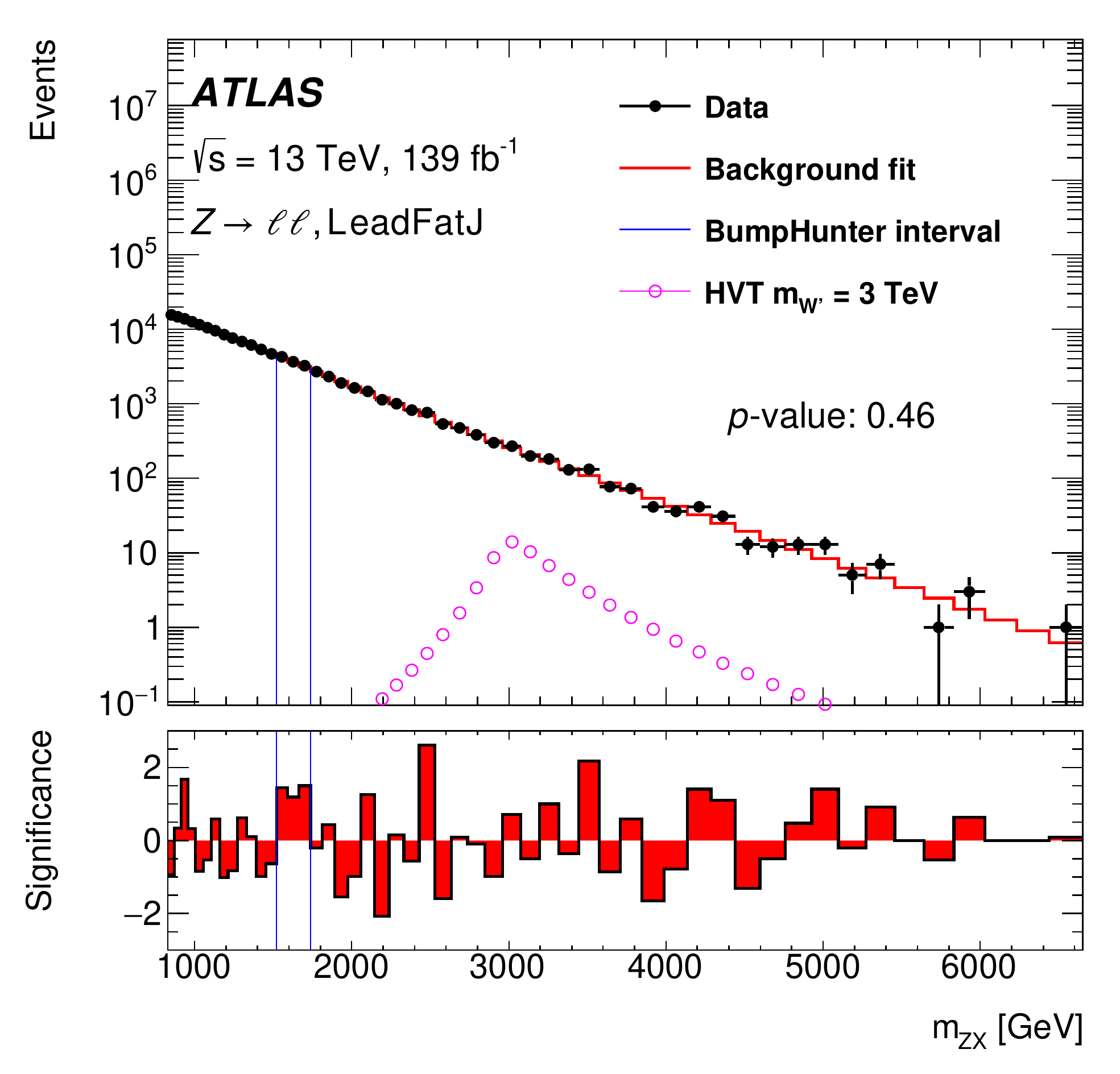} \label{fig:bh-mzx-c}}
\subfloat[]{\includegraphics[scale=0.33]{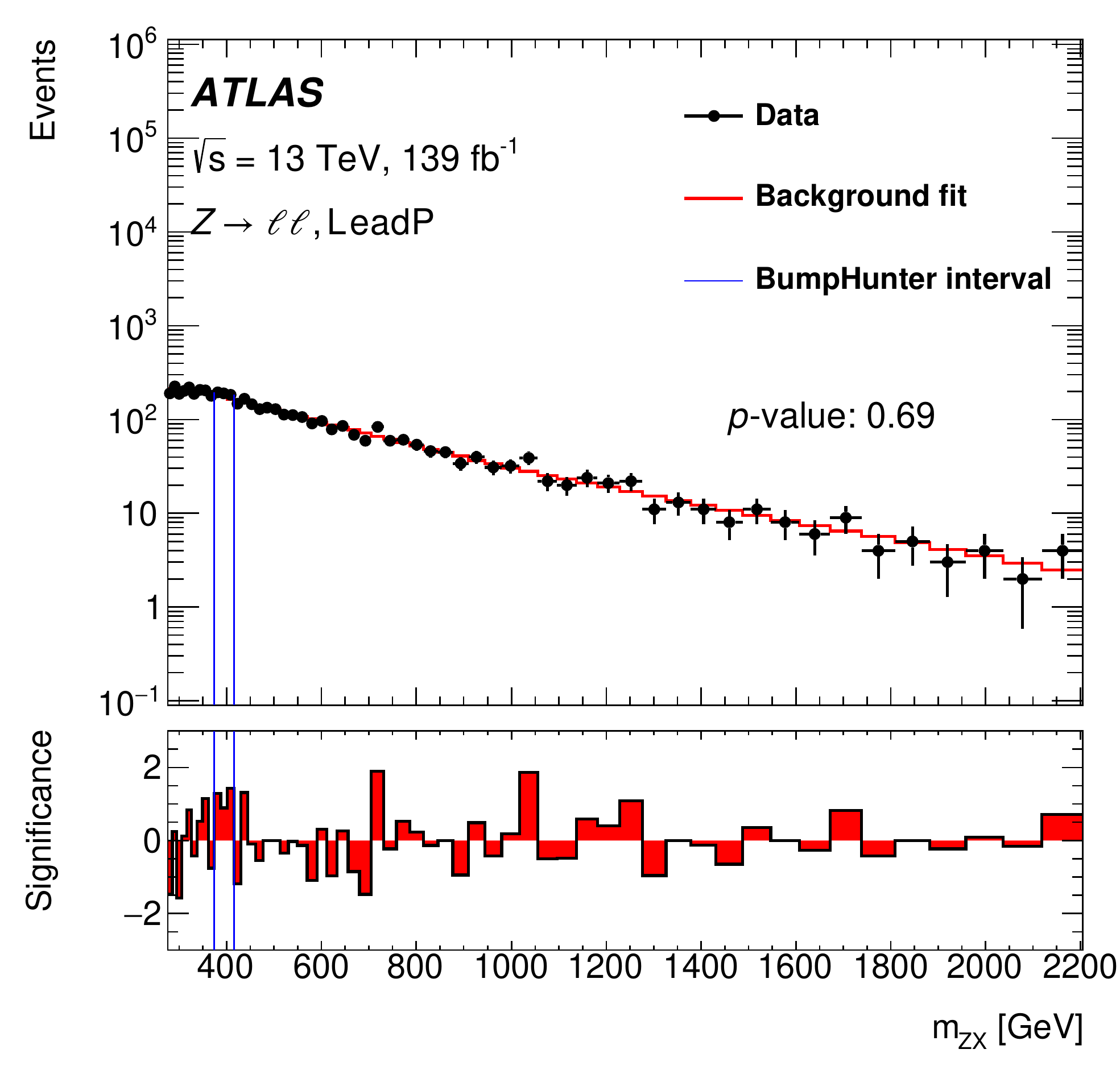} \label{fig:bh-mzx-d}}
\vspace{-.5\baselineskip}
\hfill
\subfloat[]{\includegraphics[scale=0.33]{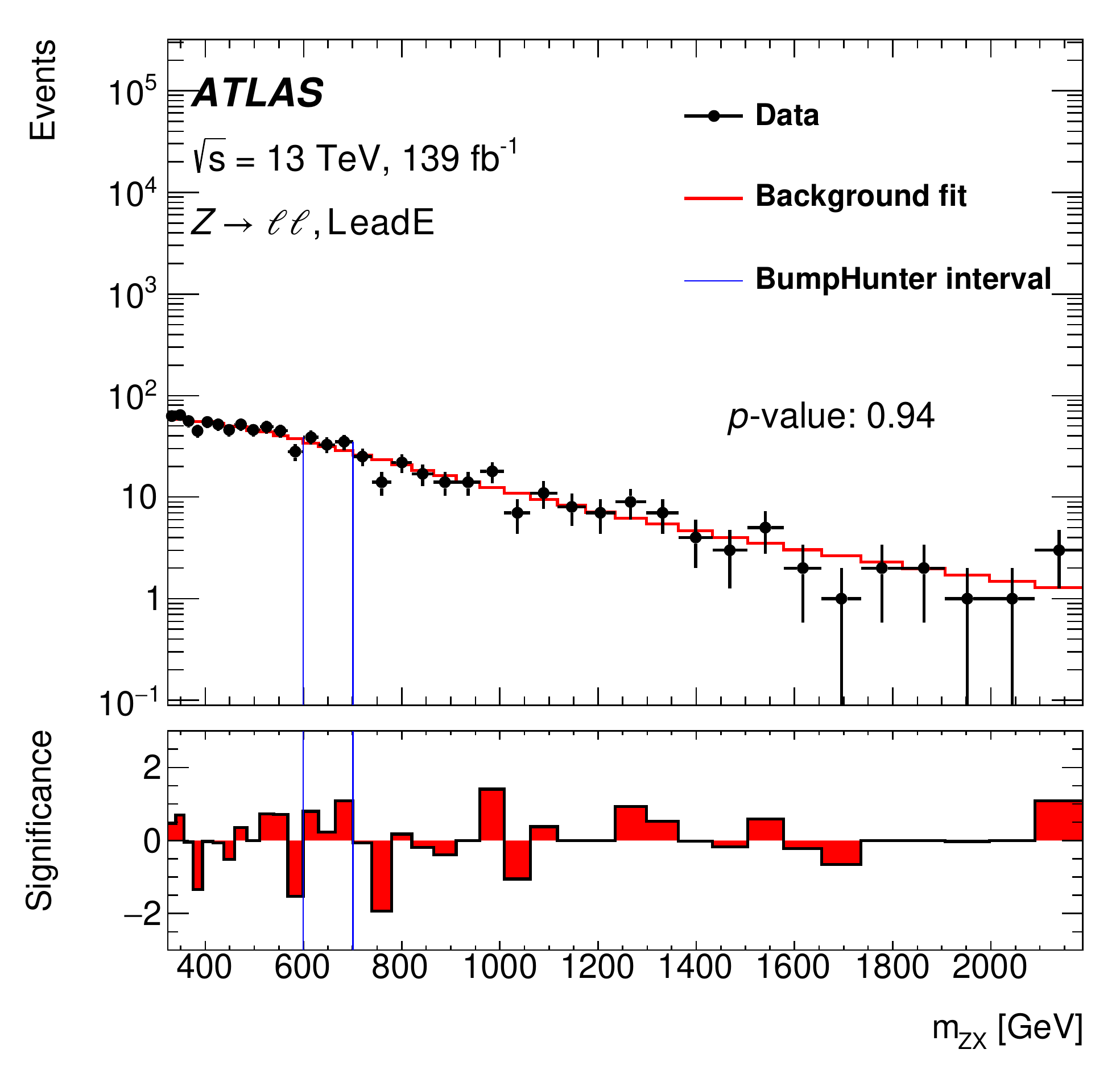} \label{fig:bh-mzx-e}}
\subfloat[]{\includegraphics[scale=0.33]{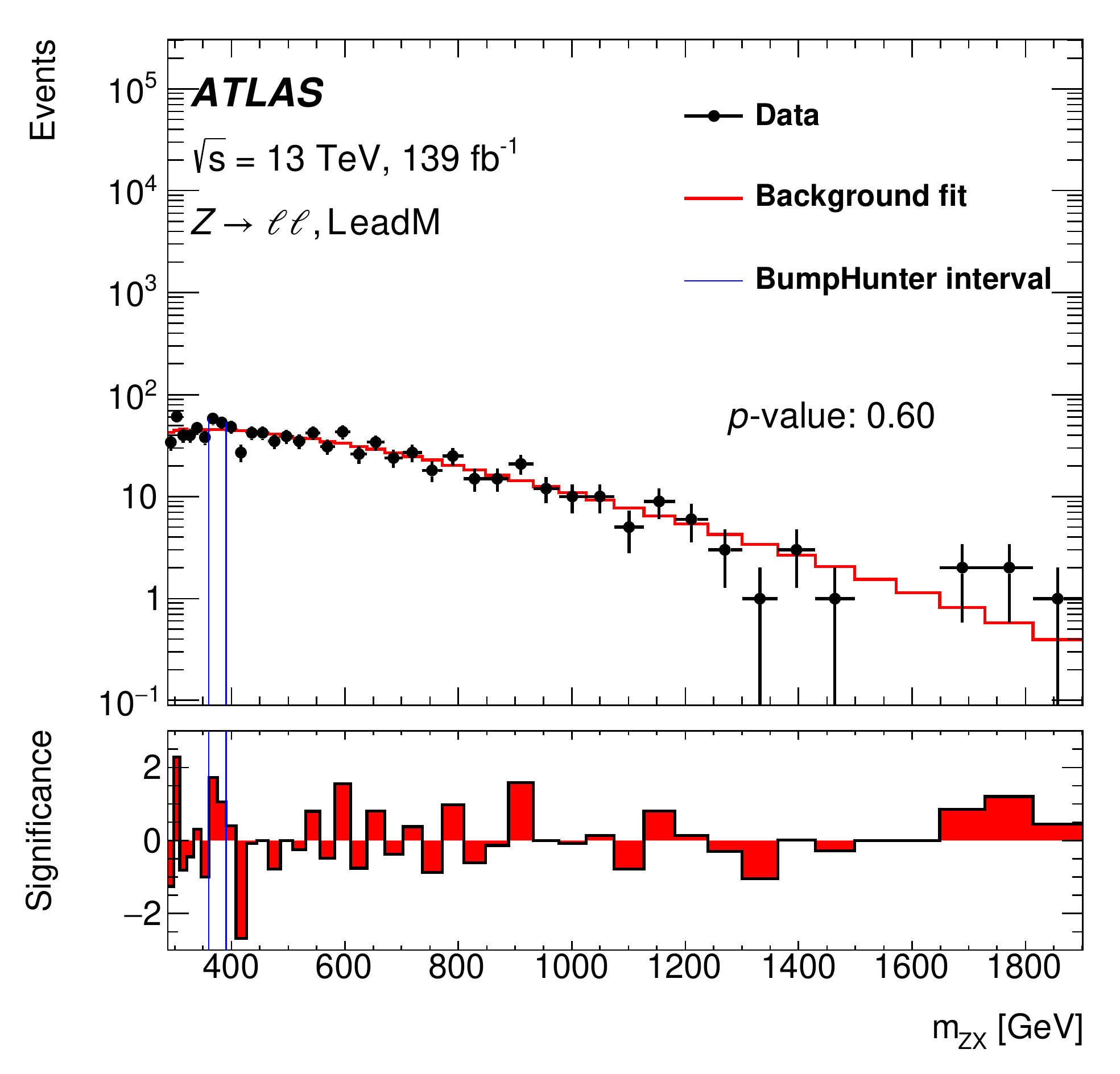} \label{fig:bh-mzx-f}}
\vspace{-.5\baselineskip}
\caption{The $m_{ZX}$ spectra in the six exclusive categories: \protect\subref{fig:bh-mzx-a} leading small-$R$ jet (Gaussian-shaped signals similar to those in the $m_X$ spectrum are also shown), \protect\subref{fig:bh-mzx-b} leading $b$-jet, \protect\subref{fig:bh-mzx-c} leading large-$R$ jet, where an HVT signal at 3~\TeV\ corresponding to a cross section of 1\,fb is also shown, \protect\subref{fig:bh-mzx-d} leading photon, \protect\subref{fig:bh-mzx-e} leading electron, and \protect\subref{fig:bh-mzx-f} leading muon. The vertical lines indicate the most discrepant interval identified by the BH test from the initial step and the $p$-values correspond to those shown in the fourth column in Table~\ref{tab:bhfit}.}
\label{fig:bh-mzx}
\end{figure}

\begin{figure}[htbp]
\centering
\subfloat[]{\includegraphics[scale=0.33]{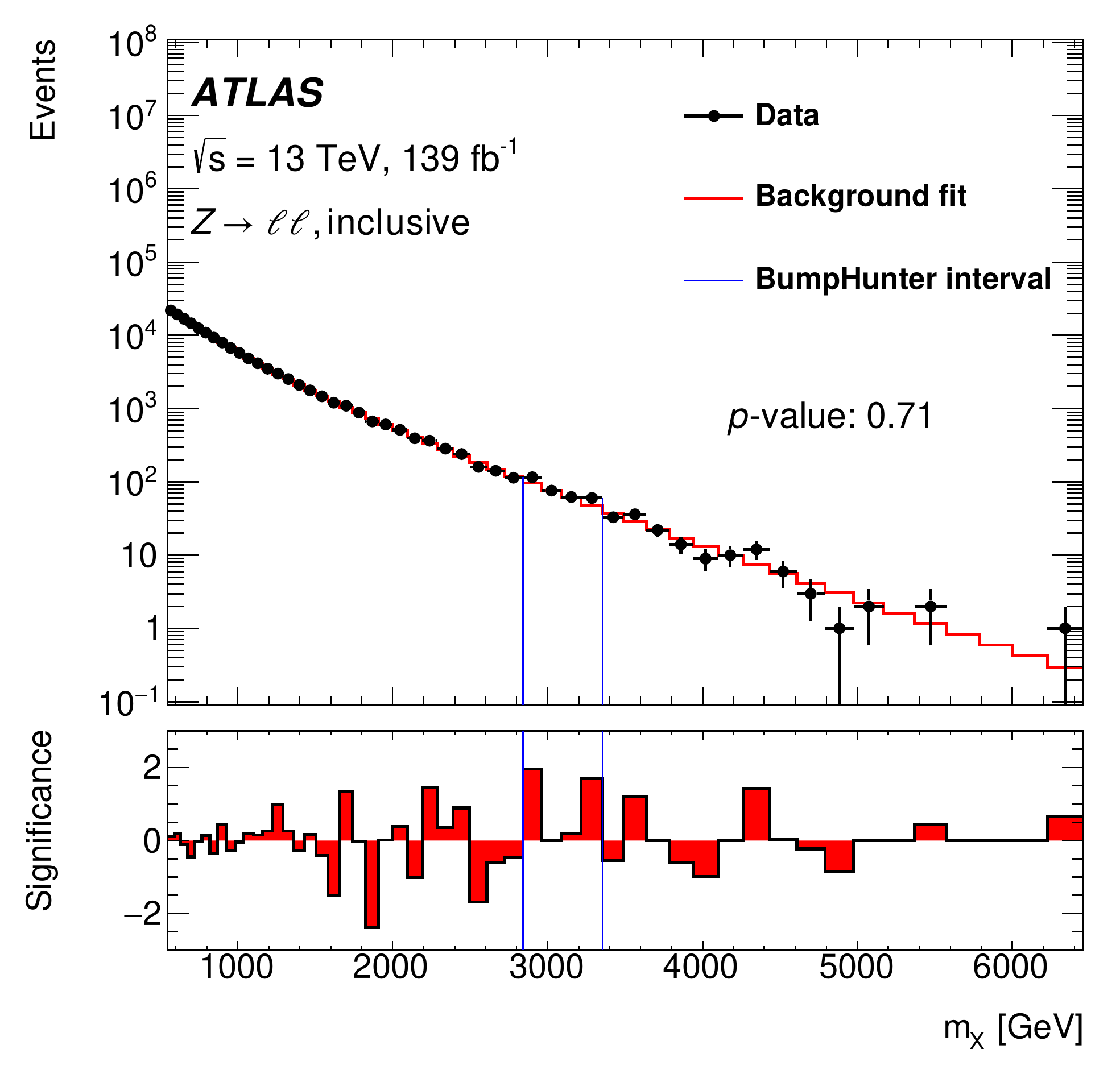} \label{fig:bh-comb-a}}
\subfloat[]{\includegraphics[scale=0.33]{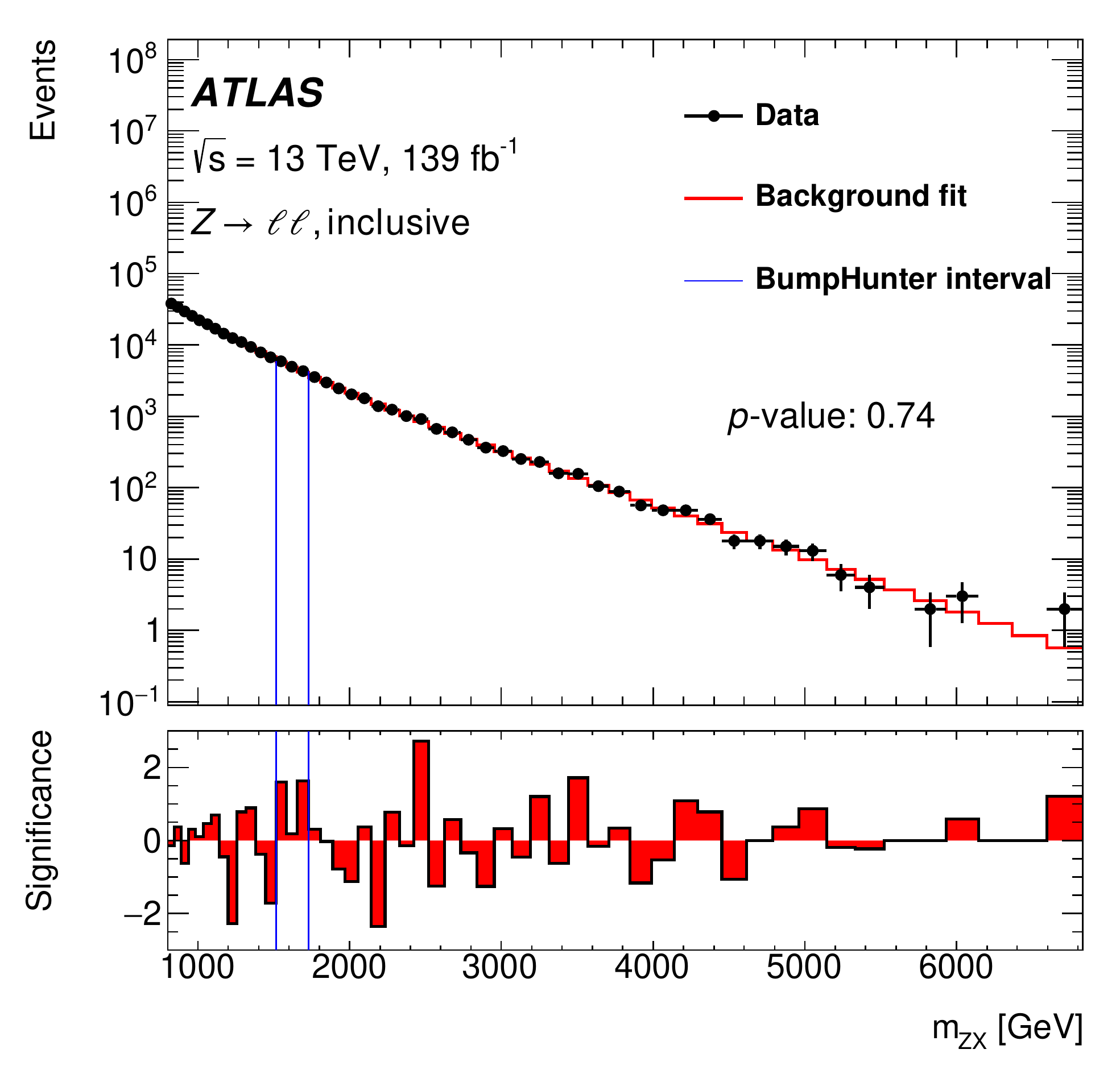} \label{fig:bh-comb-b}}
\vspace{-.5\baselineskip}
\caption{Mass spectra of \protect\subref{fig:bh-comb-a} $m_X$ and \protect\subref{fig:bh-comb-b} $m_{ZX}$ in the inclusive category. The vertical lines indicate the most discrepant interval identified by the BH test from the initial step and the $p$-values correspond to those shown in the fourth column in Table~\ref{tab:bhfit}.}
\label{fig:bh-comb}
\end{figure}

Using the fitted background estimate, the BH algorithm is applied to all the spectra within the signal-sensitive mass ranges.
The signal-sensitive mass ranges and the resulting BH $p$-values are given in Table~\ref{tab:bhfit}. The BH results using the backgrounds estimated by the fit in the full fitting range are also shown graphically in Figures~\ref{fig:bh-mx} and \ref{fig:bh-mzx} for the $m_X$ and $m_{ZX}$ spectra, respectively, in the six exclusive event categories. Figure~\ref{fig:bh-comb} shows the mass spectra of the inclusive event category.
Gaussian-shaped signals with three different widths at mass values around 1, 2 and 3\,\TeV\ in the $m_X$ and $m_{ZX}$ spectra are shown as examples in the leading small-$R$-jet category. The signal yields are normalised to correspond to a cross-section value of 1\,fb. An HVT signal at 3\,\TeV\ is also shown in the relevant $m_{ZX}$ mass spectrum of the leading large-$R$-jet category.
The largest excesses are at around 280\,\GeV\ in the $m_X$ spectrum of the leading muon category and at around 1.6\,\TeV\ in the $m_{ZX}$ spectrum of the leading large-$R$-jet category with a BH $p$-value of 0.48 and 0.46, respectively. With updated background estimates after excluding the initial BH interval, the corresponding new BH $p$-values are 0.08 and 0.10, respectively. They are still above the threshold value of 0.01, meaning that there is no significant excess in any of the mass spectra and event categories.
 
\clearpage

\section{Model-independent and model-dependent interpretations}
 
Since no significant excesses are observed in any of the mass spectra,
exclusion upper limits are derived at 95\% confidence level (CL) for model-independent
results based on Gaussian-shaped signals with relative width values of 3\%, 5\% and 10\%, as well as for HVT signal interpretations.
 
\subsection{Systematic uncertainty and statistical methodology}
 
The statistical uncertainty of the fit due to the limited size of the data sample is considered as a systematic uncertainty affecting the data-driven background determination.
 
For signals, the spurious-signal uncertainty corresponds to an envelope of absolute spurious-signal yields over the mass spectrum under consideration. The envelope is obtained from a smooth parameterisation passing through the local maxima of the spurious-signal yields. However, given the differences in the shape and normalisation of the mass spectra between data and simulation, the implementation uses the absolute uncertainty scaled to keep the relative uncertainty unchanged.
 
The spurious signal's systematic uncertainty and the statistical uncertainty of the background function fit are the dominant sources of uncertainty. Additional systematic uncertainties for the HVT signal modelling include experimental, theoretical and luminosity uncertainties. The large-$R$-jet-related uncertainties are the dominant source of experimental uncertainty.
The theoretical uncertainties include contributions due to missing higher-order corrections,
PDF uncertainties, and the uncertainty in the strong coupling constant \alphas.
The effect of missing higher-order corrections is estimated by
varying the renormalisation and factorisation scales by factors of 0.5 and 2. Similarly, the PDF and \alphas are varied to estimate the effect of their uncertainties.
The uncertainty in the combined 2015--2018 integrated luminosity is 1.7\%~\cite{ATLAS-CONF-2019-021}, obtained using the LUCID-2 detector~\cite{Avoni:2018iuv} for the primary luminosity measurement.
 
For each mass spectrum, a likelihood function is constructed as a product of Poisson distributions corresponding to different mass bins. Upper limits on the signal strength $\mu_s$ in terms of the cross-section times acceptance times branching ratio, $\sigma\times\mathcal{B}\times\mathcal{A}$, are extracted at 95\% CL using the CLs method~\cite{Read:2002hq} with a binned profile likelihood ratio as the test statistic.
The free parameters of the background functional forms described in Section~\ref{sec:search} are treated as nuisance parameters, which are optmised for each resonance mass value by the likelihood function without constraints during the fit~\cite{CMS-EXO-16-056}. The systematic uncertainties are also included as nuisance parameters, each with a Gaussian constraint except for the integrated luminosity, which has a log-normal constraint. Generally, each of the two dominant uncertainties mentioned above degrades the expected upper limits by around 10$-$20\% but can reach up to 50\% at low mass regions.
 
\clearpage
\subsection{Constraints on a generic Gaussian-shaped signal for a model-independent interpretation}
 
For a model-independent interpretation, Gaussian-shaped signals with relative width values of 3\%, 5\% and 10\% are studied. The cross section value of the signals is normalised to 1\,fb.
The exclusion upper limits are obtained by adjusting the normalisation of the signal strength $\mu_\text{s}$ of a template representing the Gaussian signal in question until a
CL$_\text{s}$ $p$-value of 0.05 is reached. The obtained $\mu_\text{s}$ value is thus the 95\% CL upper limit on the event yield of the Gaussian-shaped portion of a signal in an appropriately chosen BSM model. This $\mu_\text{s}$ value is then divided by the integrated luminosity ${\cal L}$ to obtain the upper limit on $\sigma\times\mathcal{B}\times\mathcal{A}$. For the Gaussian-shaped signals, the $\mathcal{B}\times\mathcal{A}$ value cannot be specified as those must be determined in the context of a specific BSM model.
 
\begin{figure}[htbp]
\centering
\subfloat[]{\includegraphics[scale=0.35]{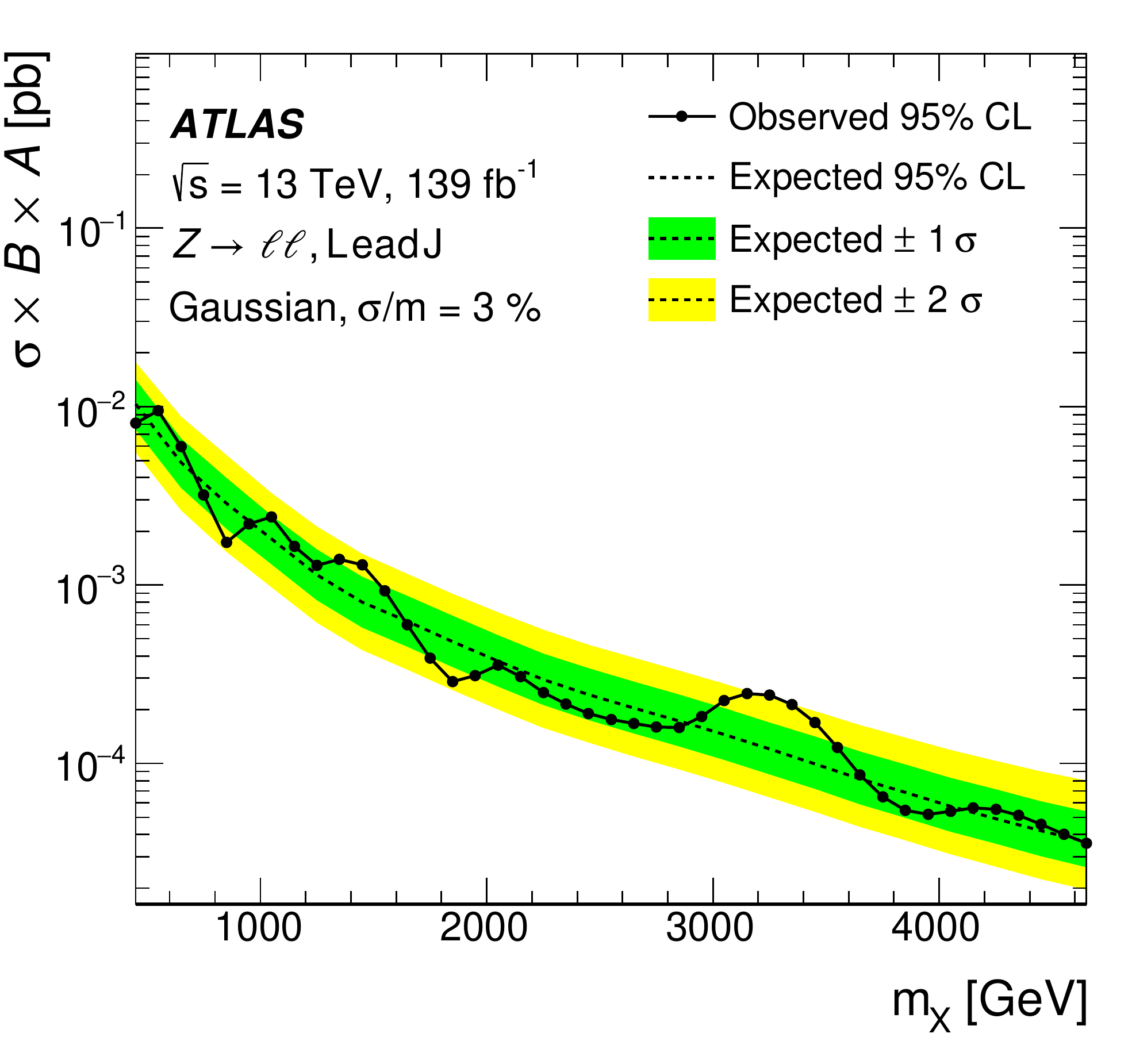} \label{fig:limit-mx-a}}
\subfloat[]{\includegraphics[scale=0.35]{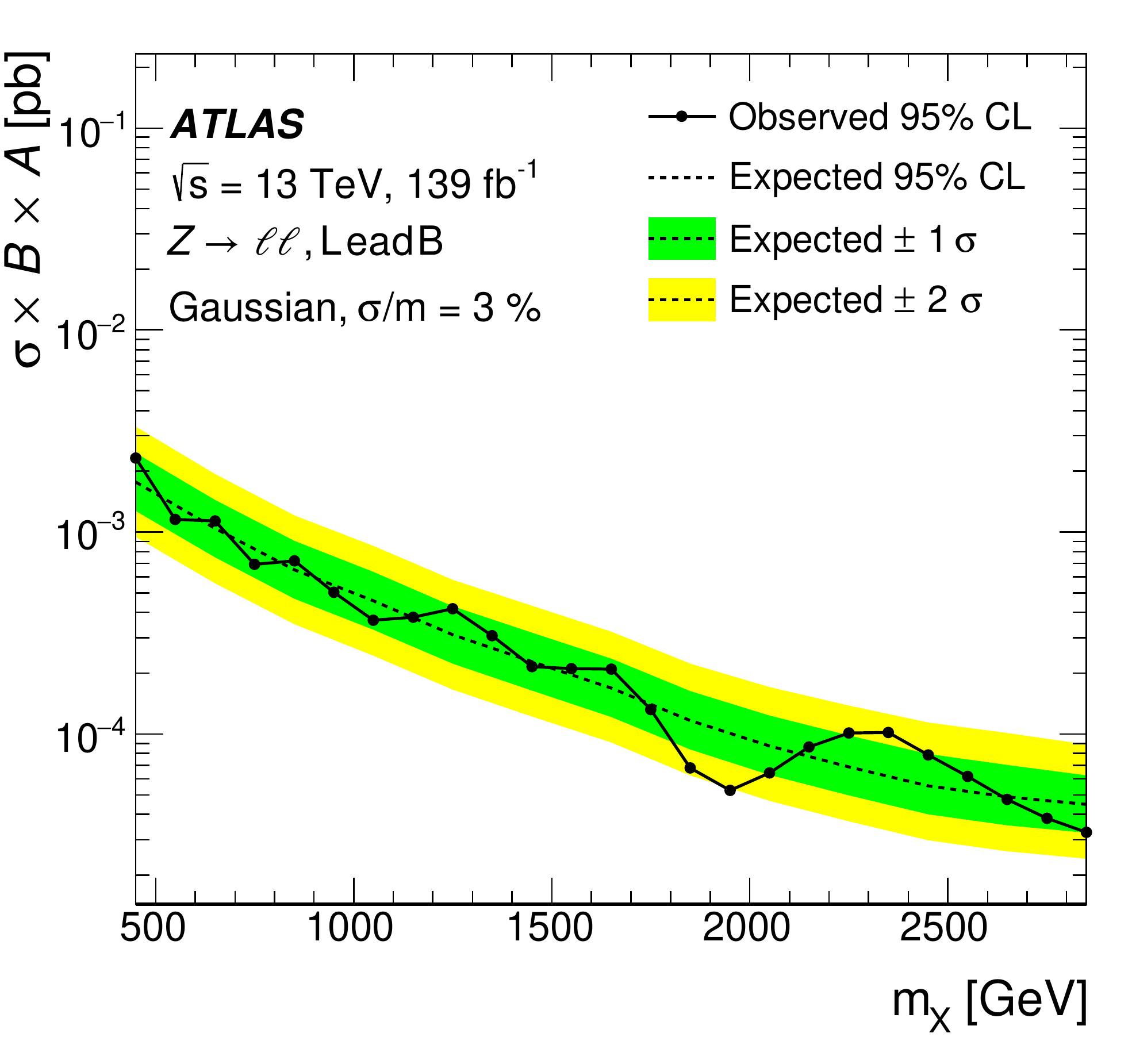}\label{fig:limit-mx-b}}
\vspace{-.5\baselineskip}
\subfloat[]{\includegraphics[scale=0.35]{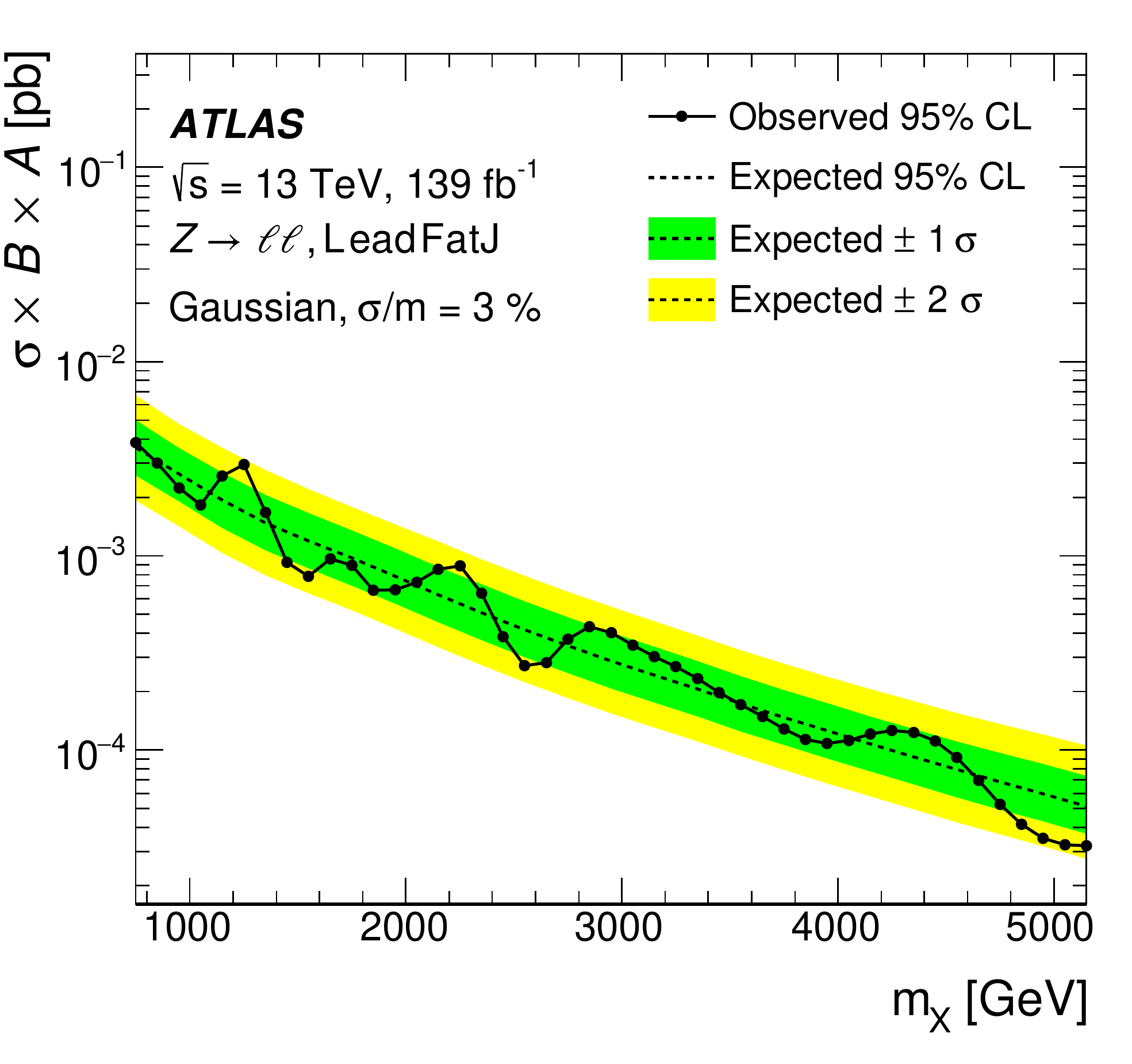}\label{fig:limit-mx-c}}
\subfloat[]{\includegraphics[scale=0.35]{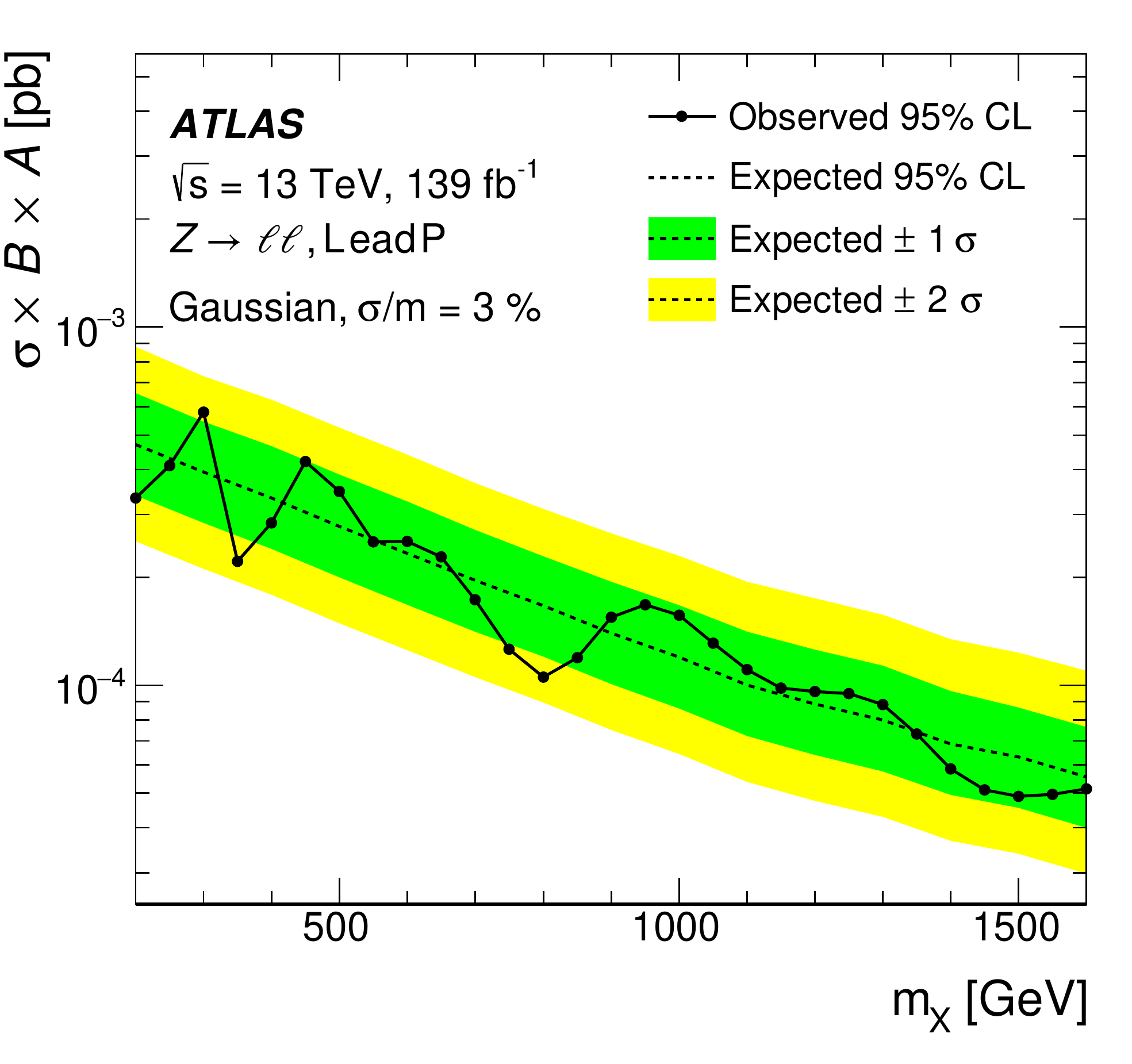}\label{fig:limit-mx-d}}
\vspace{-.5\baselineskip}
\subfloat[]{\includegraphics[scale=0.35]{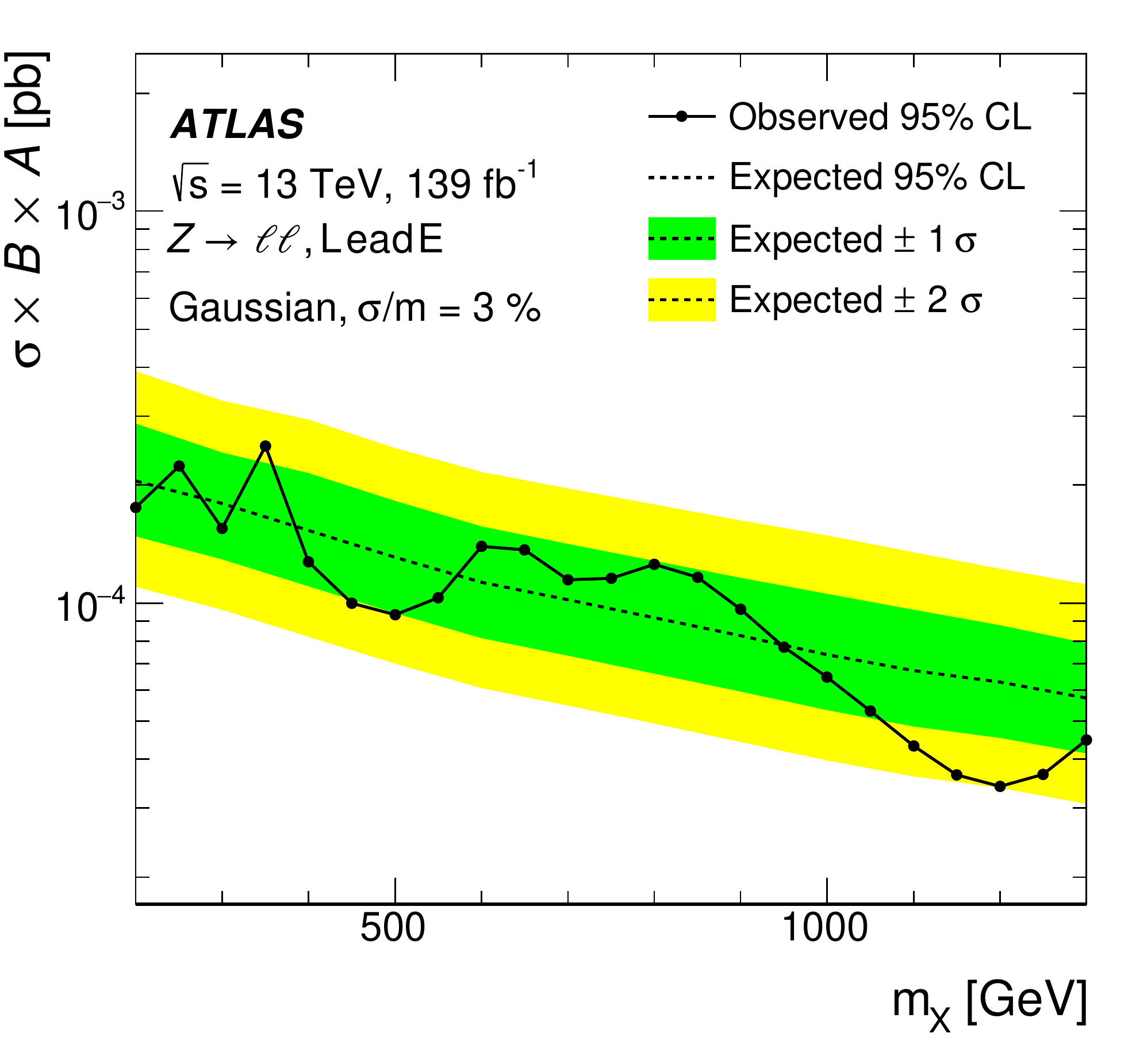}\label{fig:limit-mx-e}}
\subfloat[]{\includegraphics[scale=0.35]{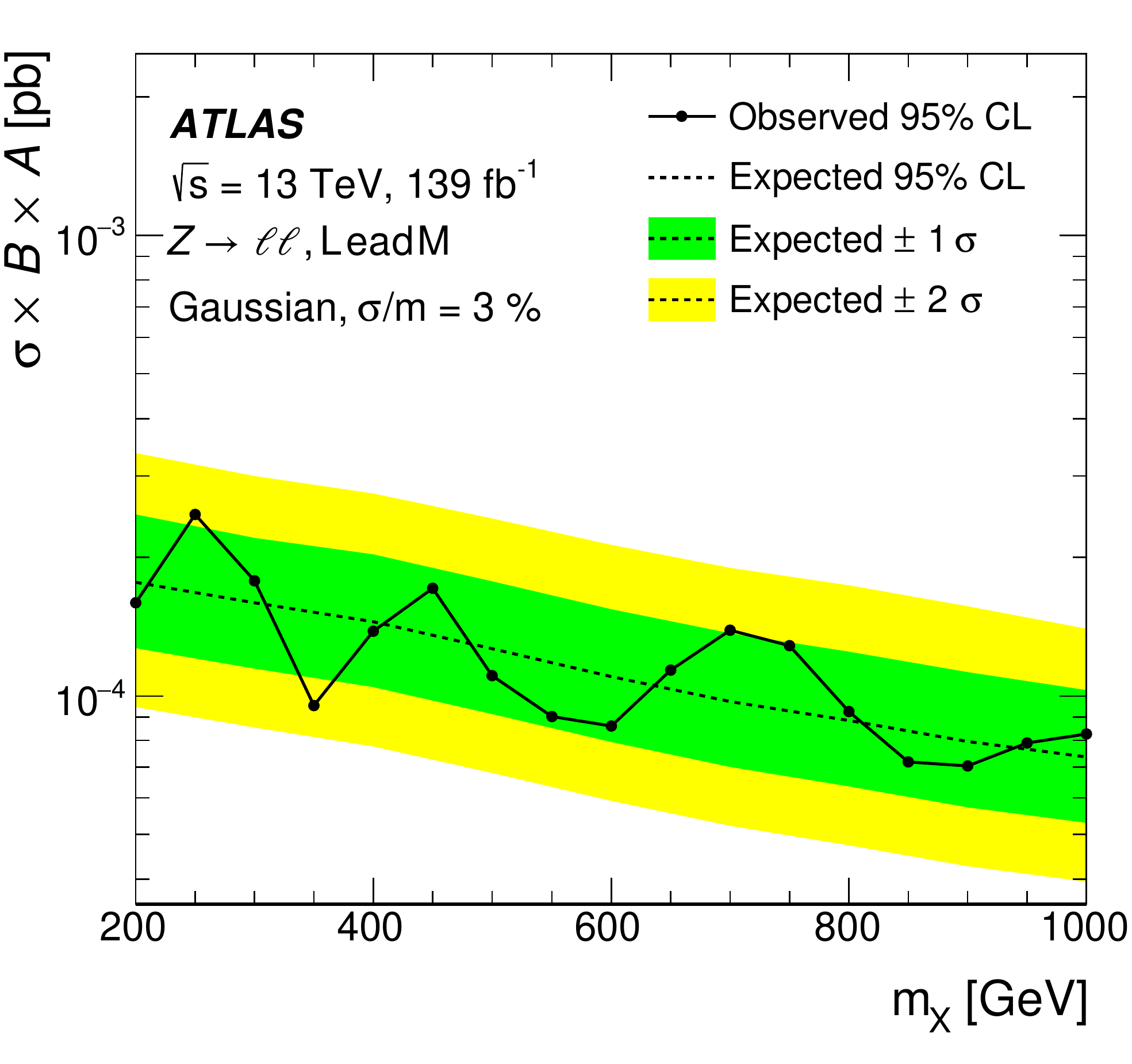}\label{fig:limit-mx-f}}
\vspace{-.5\baselineskip}
\caption{Upper limits at 95\% CL on the cross section times branching fraction times acceptance for a Gaussian-shaped signal with a relative width value of 3\% as a function of $m_X$ in the \protect\subref{fig:limit-mx-a} leading small-$R$-jet, \protect\subref{fig:limit-mx-b} leading $b$-jet, \protect\subref{fig:limit-mx-c} leading large-$R$-jet, \protect\subref{fig:limit-mx-d} leading photon, \protect\subref{fig:limit-mx-e} leading electron, and \protect\subref{fig:limit-mx-f} leading muon categories.}
\label{fig:limit-mx}
\end{figure}

\begin{figure}[htbp]
\centering
\subfloat[]{\includegraphics[scale=0.35]{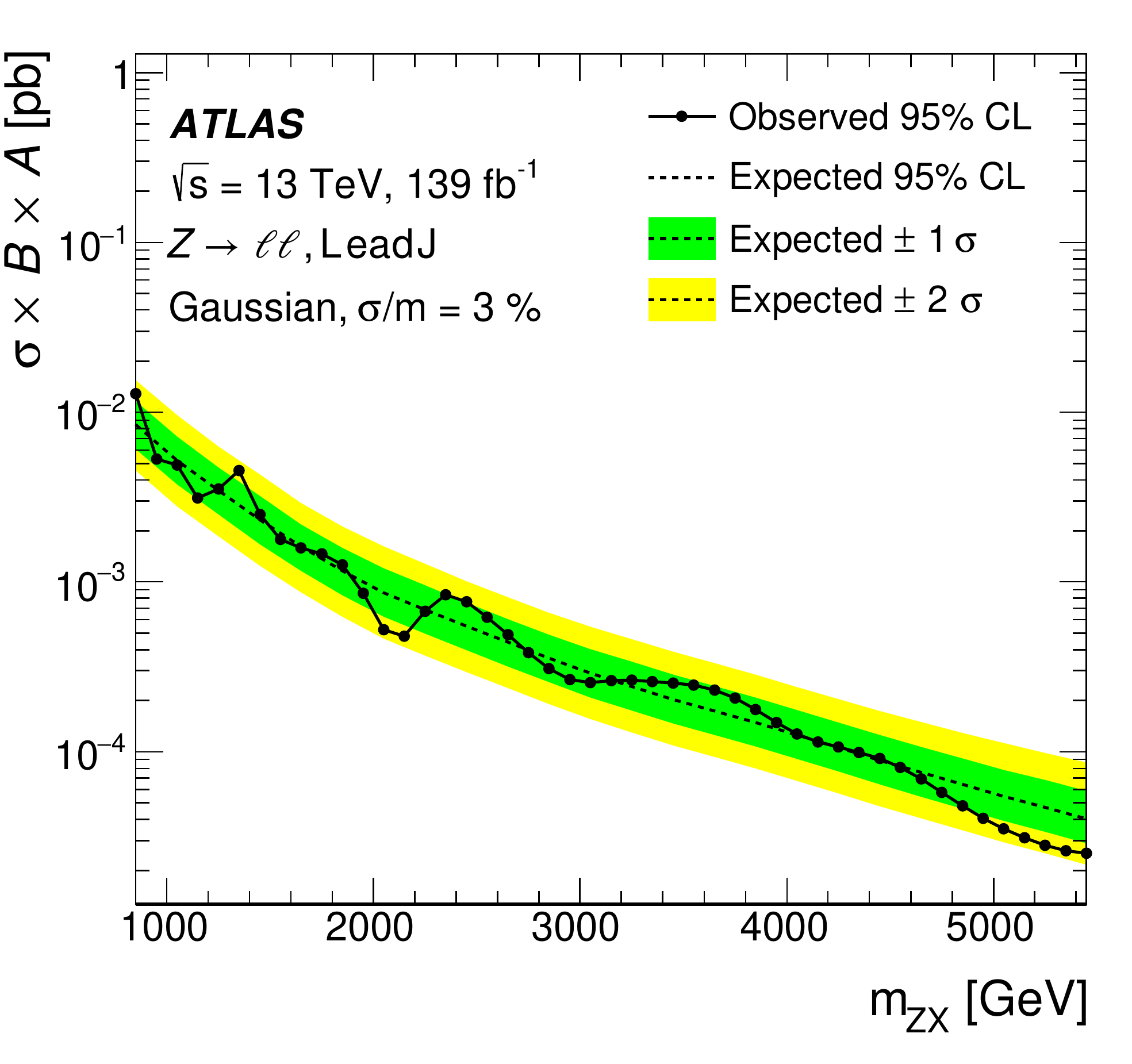} \label{fig:limit-mzx-a}}
\subfloat[]{\includegraphics[scale=0.35]{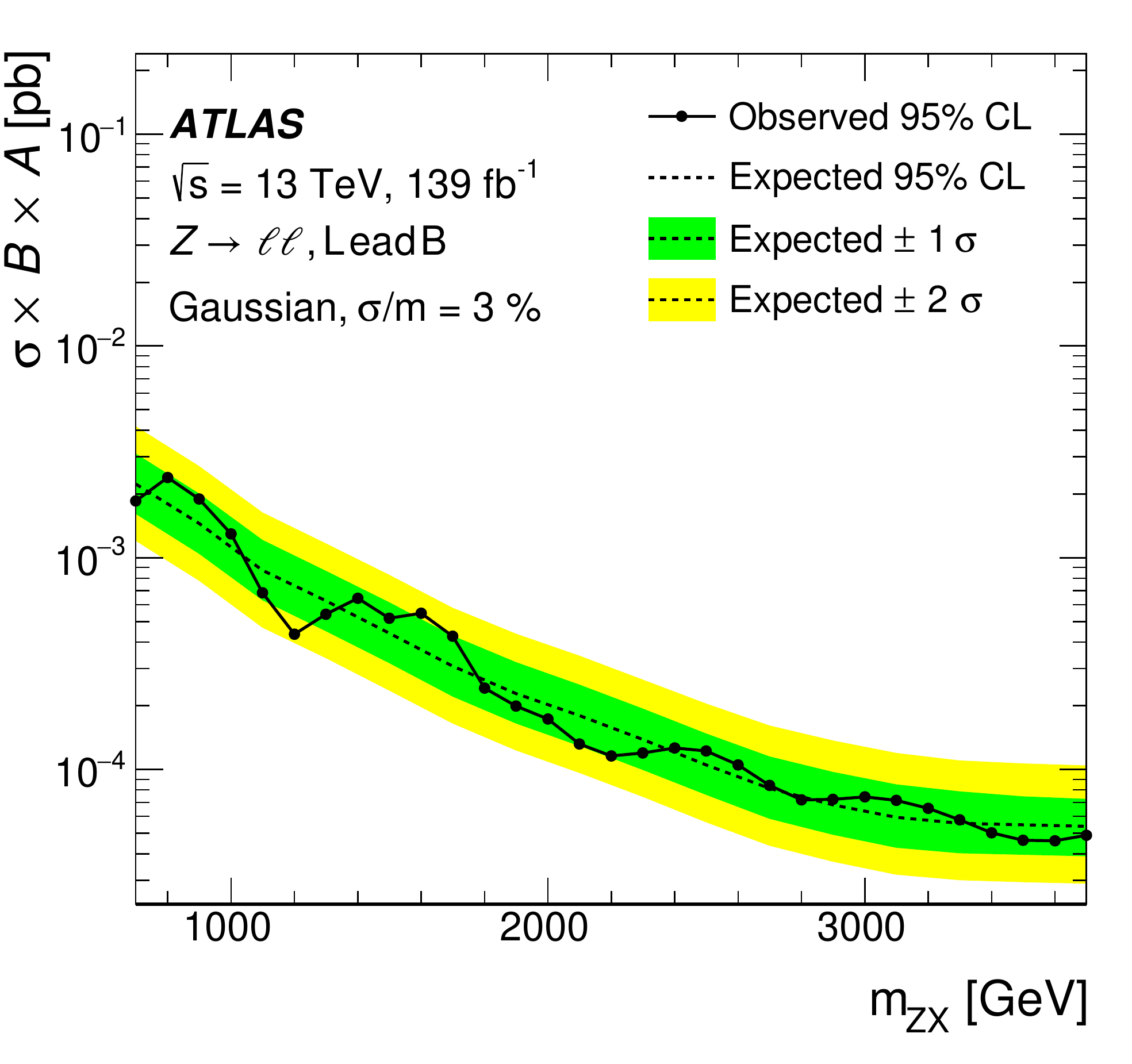}\label{fig:limit-mzx-b}}
\vspace{-.5\baselineskip}
\subfloat[]{\includegraphics[scale=0.35]{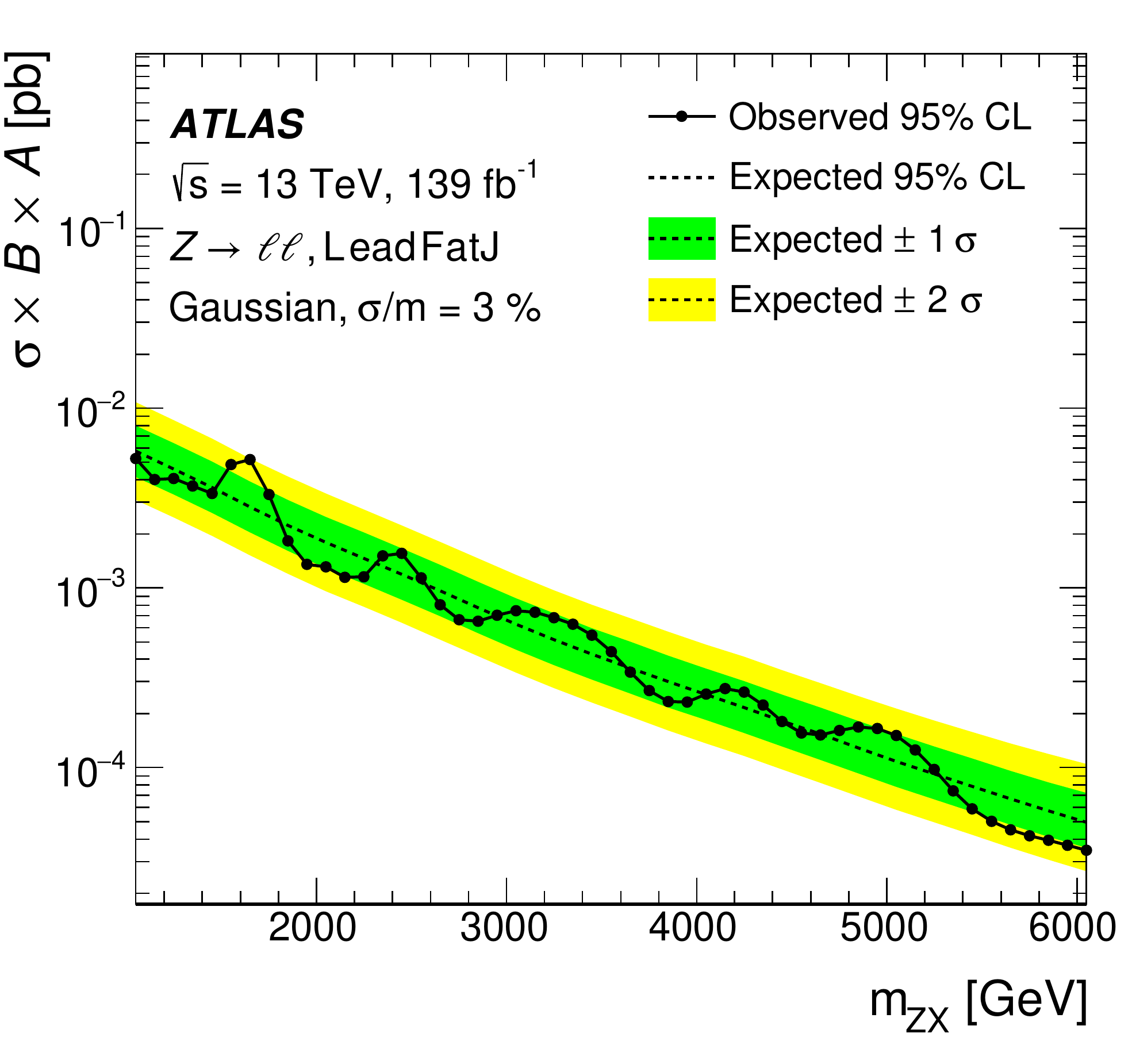}\label{fig:limit-mzx-c}}
\subfloat[]{\includegraphics[scale=0.35]{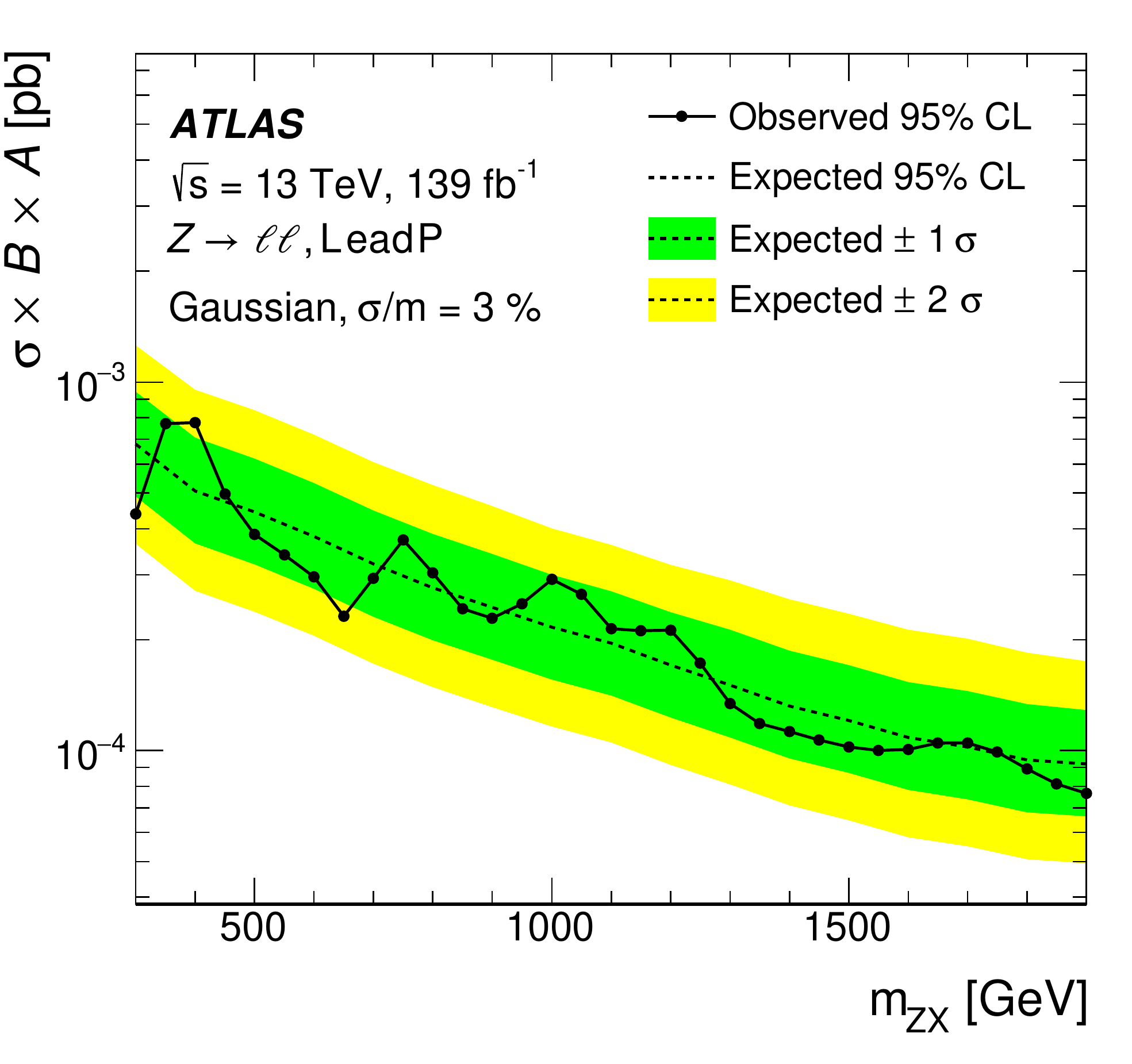}\label{fig:limit-mzx-d}}
\vspace{-.5\baselineskip}
\subfloat[]{\includegraphics[scale=0.35]{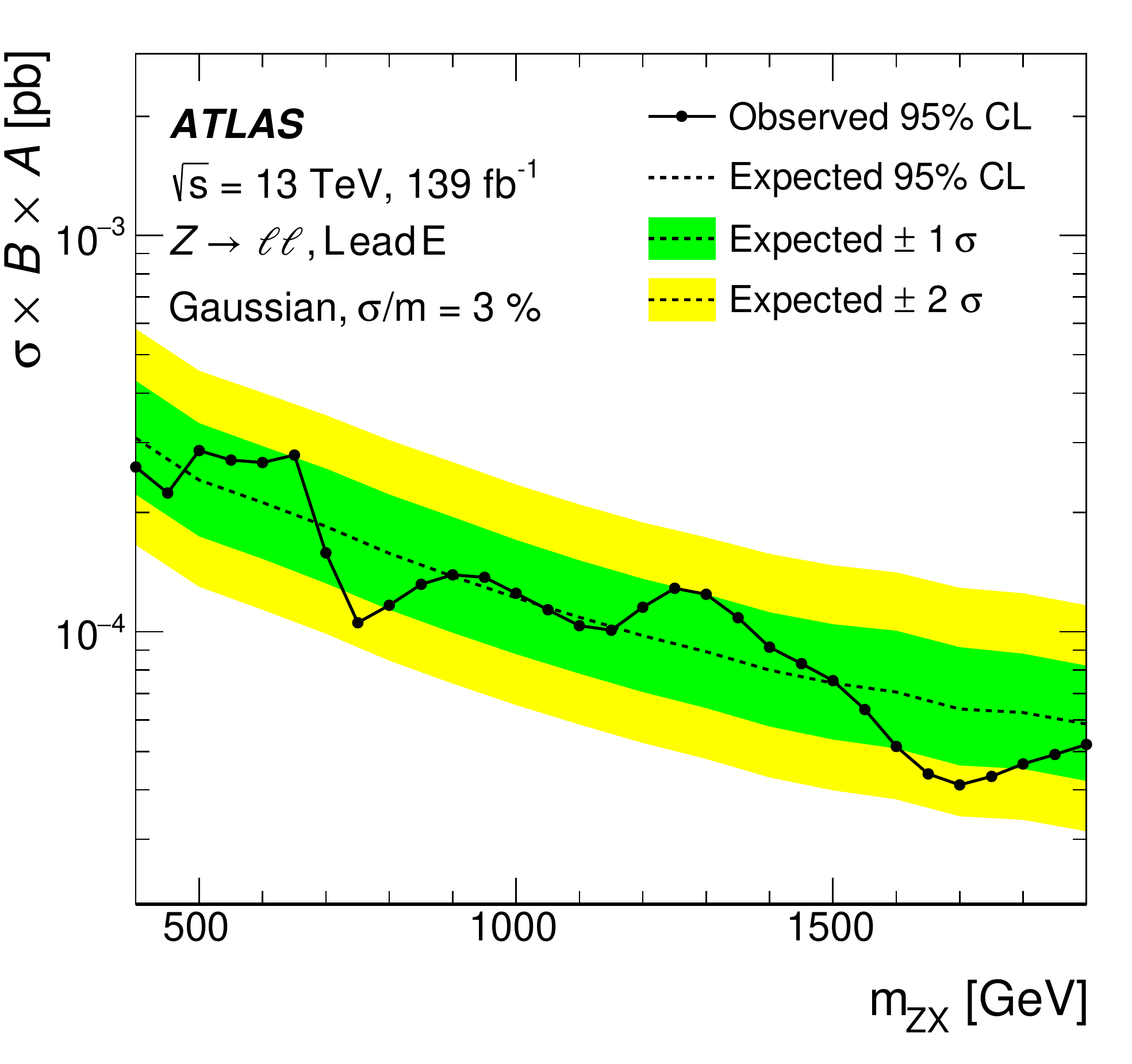}\label{fig:limit-mzx-e}}
\subfloat[]{\includegraphics[scale=0.35]{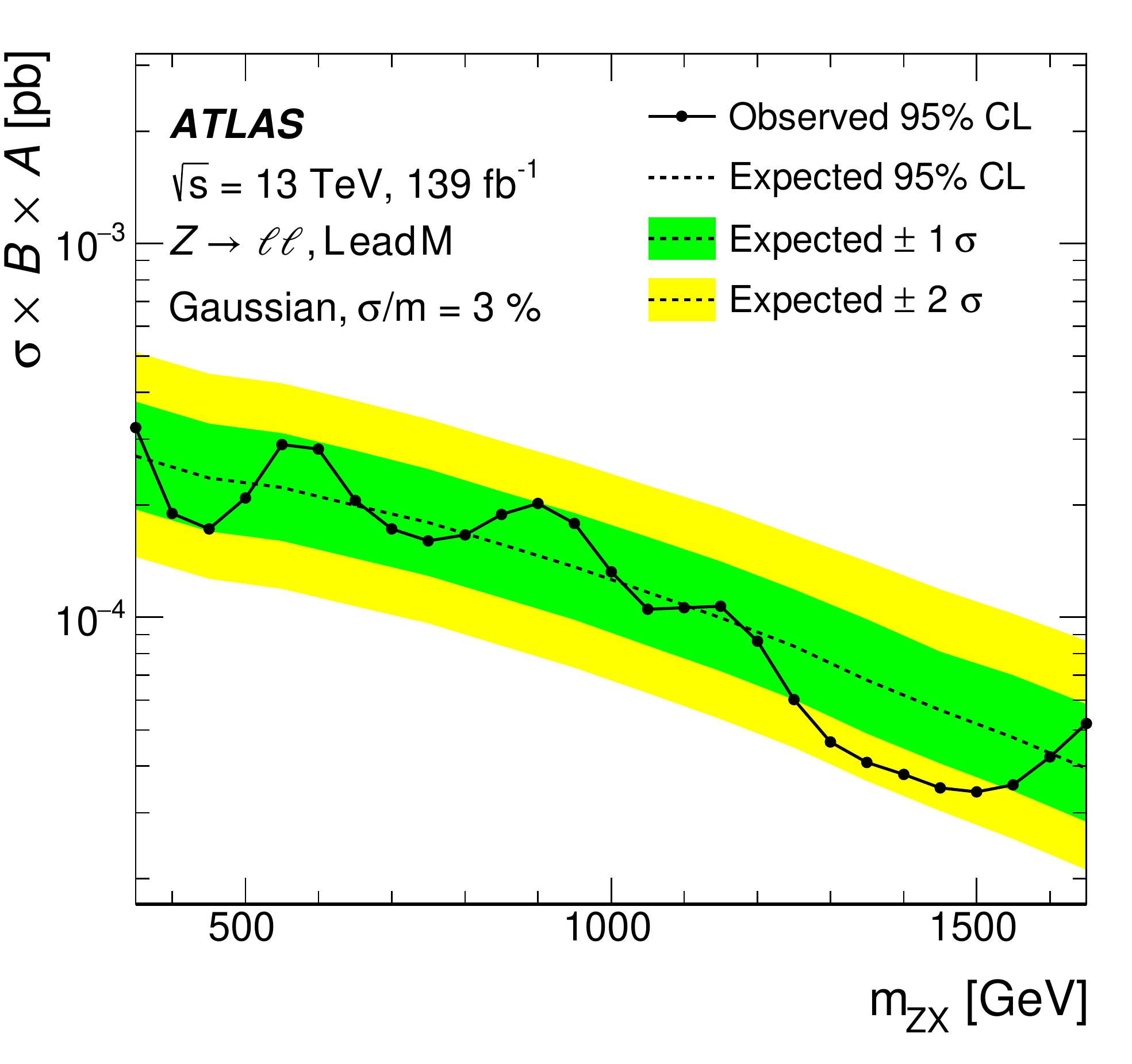}\label{fig:limit-mzx-f}}
\vspace{-.5\baselineskip}
\caption{Upper limits at 95\% CL on the cross section times branching fraction times acceptance for a Gaussian-shaped signal with a relative width value of 3\% as a function of $m_{ZX}$ in the \protect\subref{fig:limit-mzx-a} leading small-$R$-jet, \protect\subref{fig:limit-mzx-b} leading $b$-jet, \protect\subref{fig:limit-mzx-c} leading large-$R$-jet, \protect\subref{fig:limit-mzx-d} leading photon, \protect\subref{fig:limit-mzx-e} leading electron, and \protect\subref{fig:limit-mzx-f} leading muon categories.}
\label{fig:limit-mzx}
\end{figure}

\begin{figure}[htbp]
\centering
\subfloat[]{\includegraphics[scale=0.35]{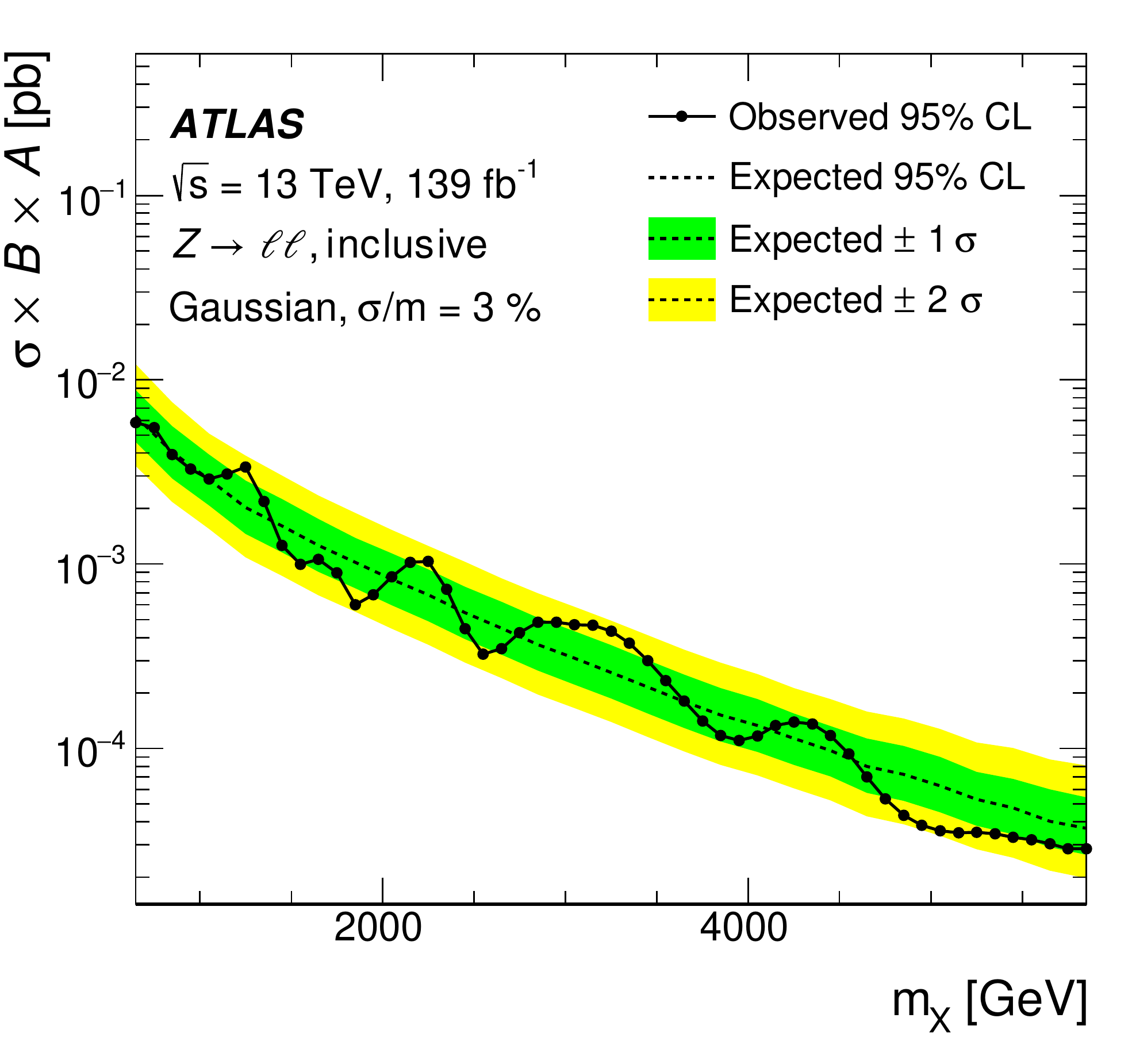} \label{fig:limit-comb-a}}
\subfloat[]{\includegraphics[scale=0.35]{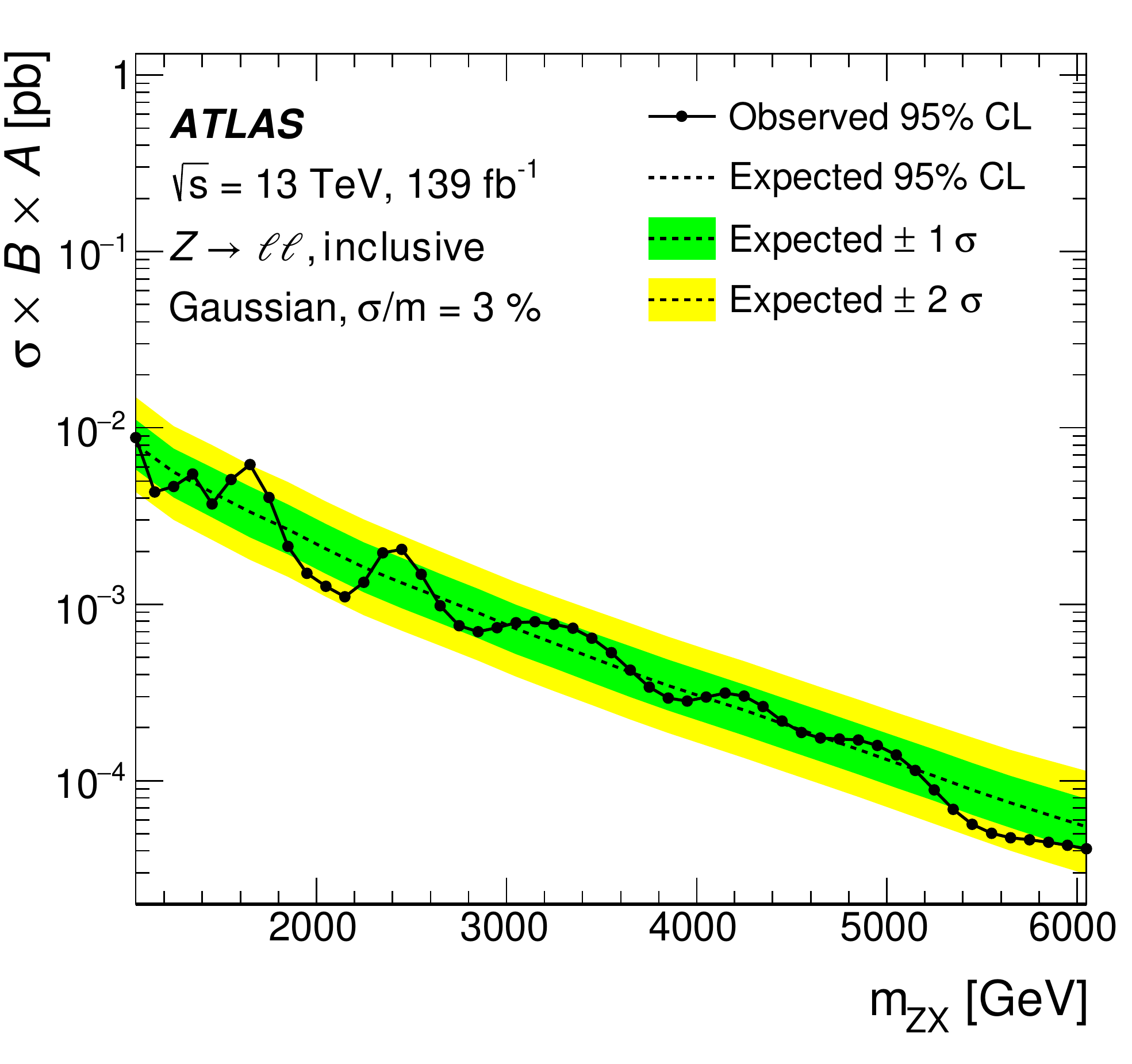} \label{fig:limit-comb-b}}
\vspace{-.5\baselineskip}
\caption{Upper limits at 95\% CL on the cross section times branching fraction times acceptance for a Gaussian-shaped signal with a relative width value of 3\% as functions of \protect\subref{fig:limit-comb-a} $m_X$ and \protect\subref{fig:limit-comb-b} $m_{ZX}$ in the inclusive category.}
\label{fig:limit-comb}
\end{figure}

\begin{figure}[htbp]
\centering
\subfloat[]{\includegraphics[scale=0.34]{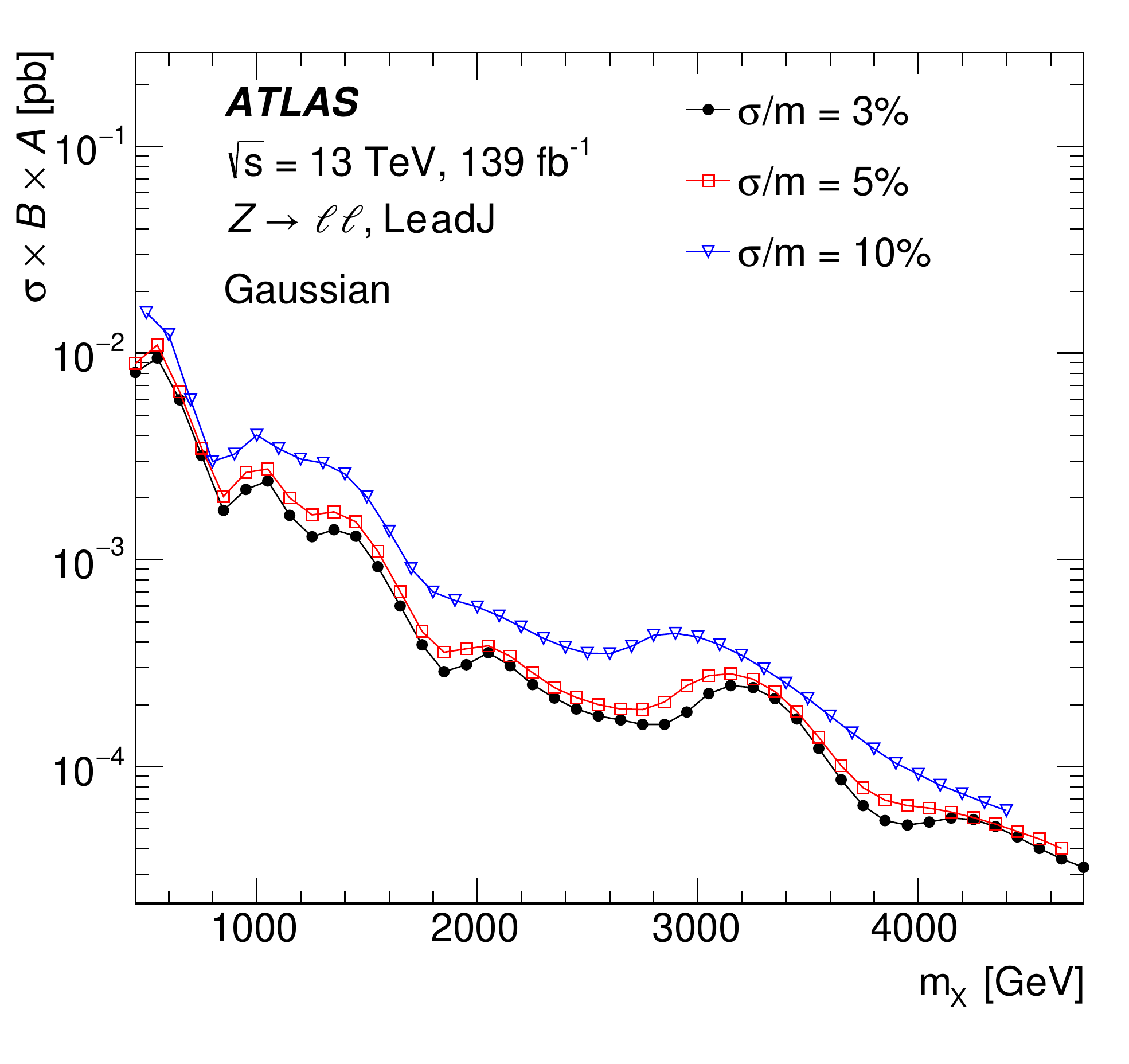} \label{fig:limit-comp-mx-a}}
\subfloat[]{\includegraphics[scale=0.34]{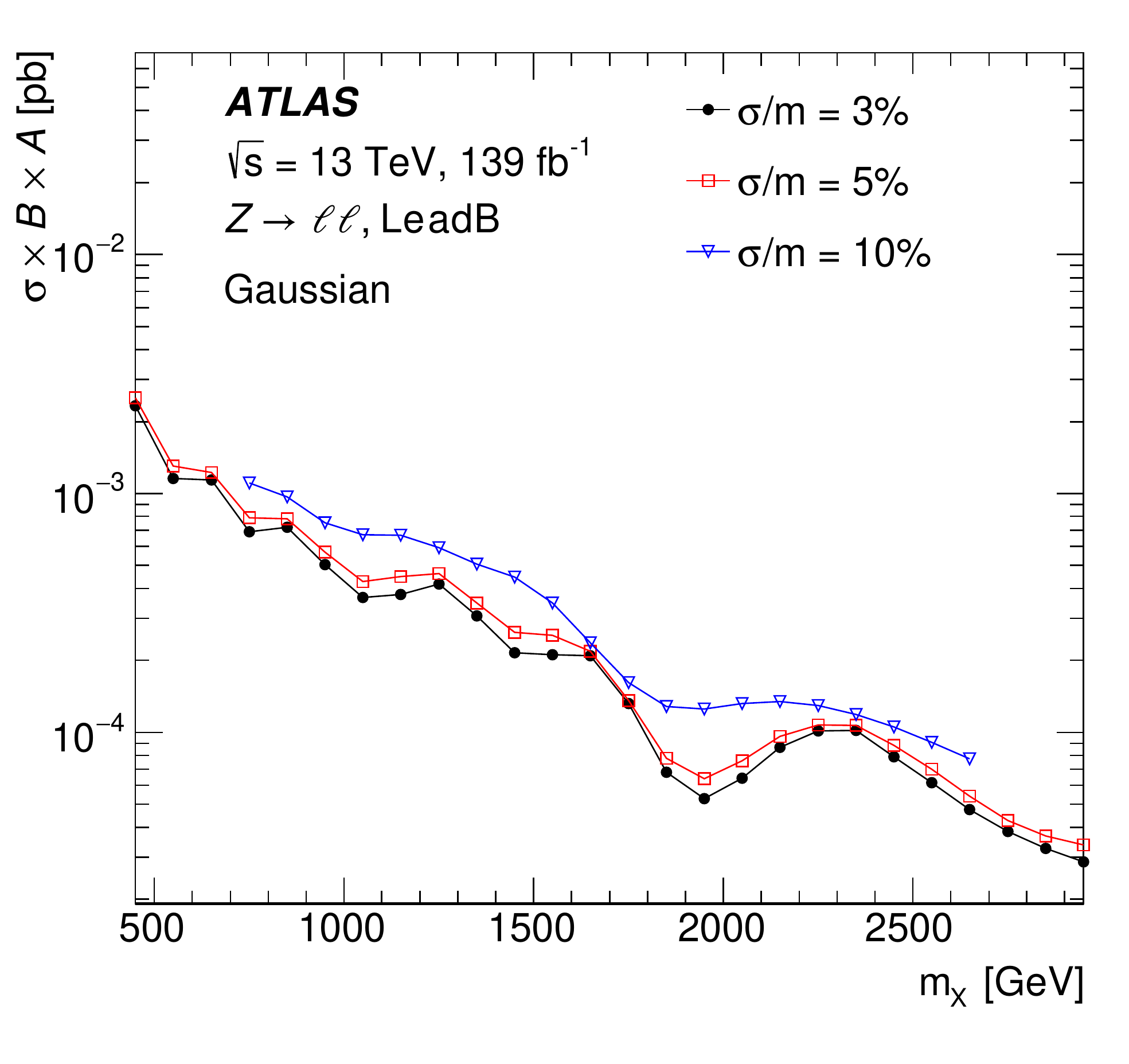} \label{fig:limit-comp-mx-b}}
\vspace{-.5\baselineskip}
\subfloat[]{\includegraphics[scale=0.34]{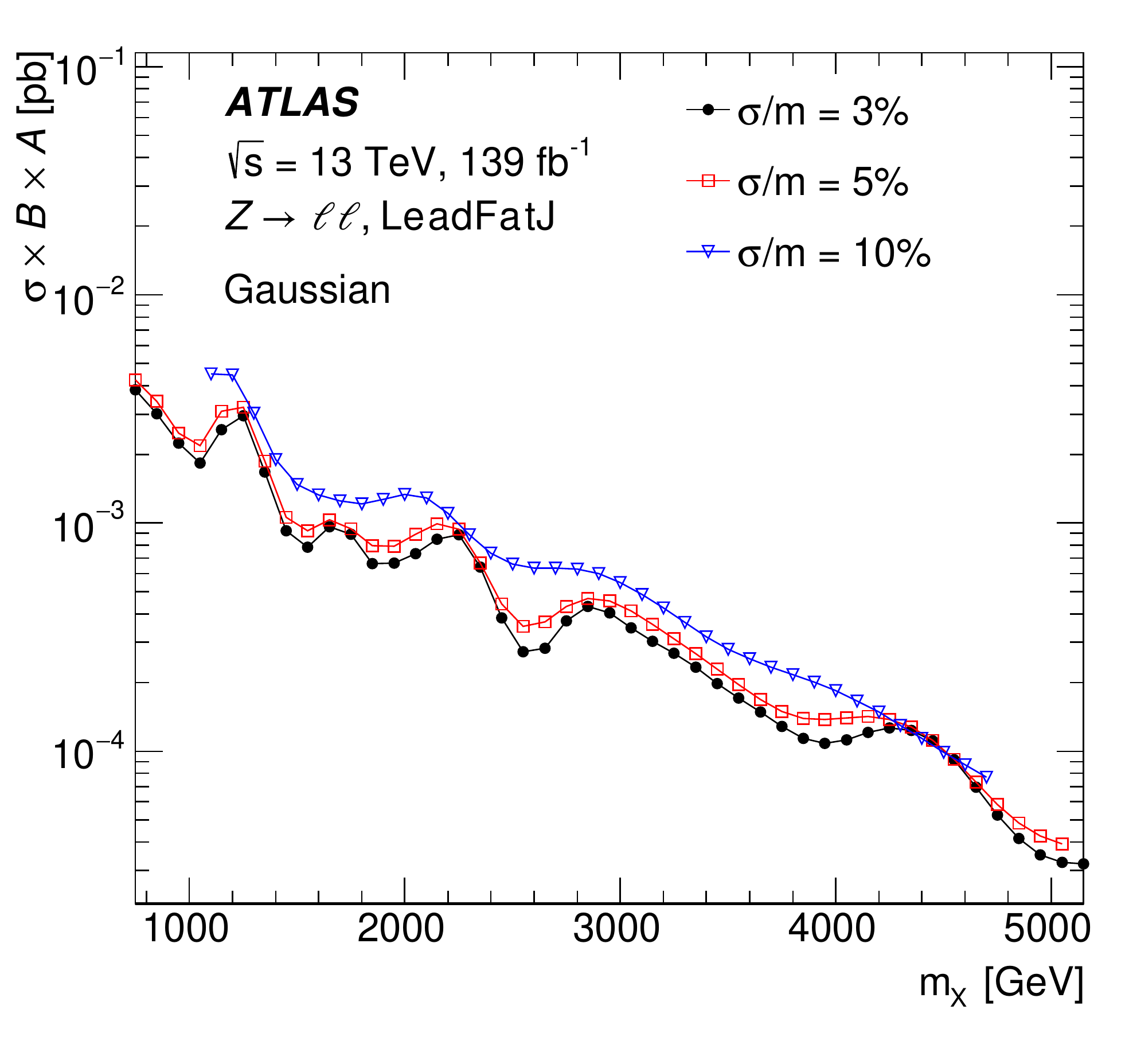} \label{fig:limit-comp-mx-c}}
\subfloat[]{\includegraphics[scale=0.34]{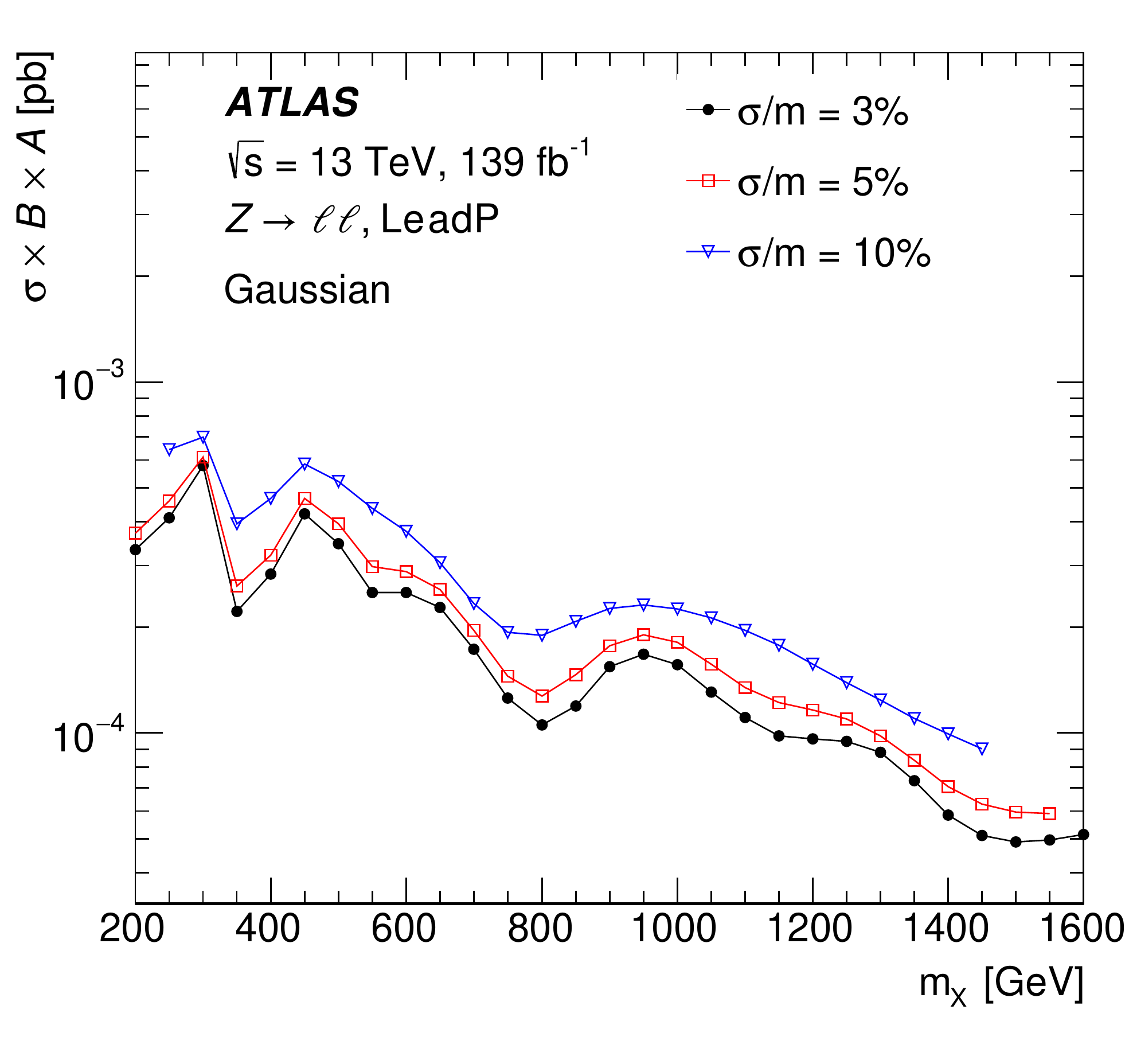} \label{fig:limit-comp-mx-d}}
\vspace{-.5\baselineskip}
\subfloat[]{\includegraphics[scale=0.34]{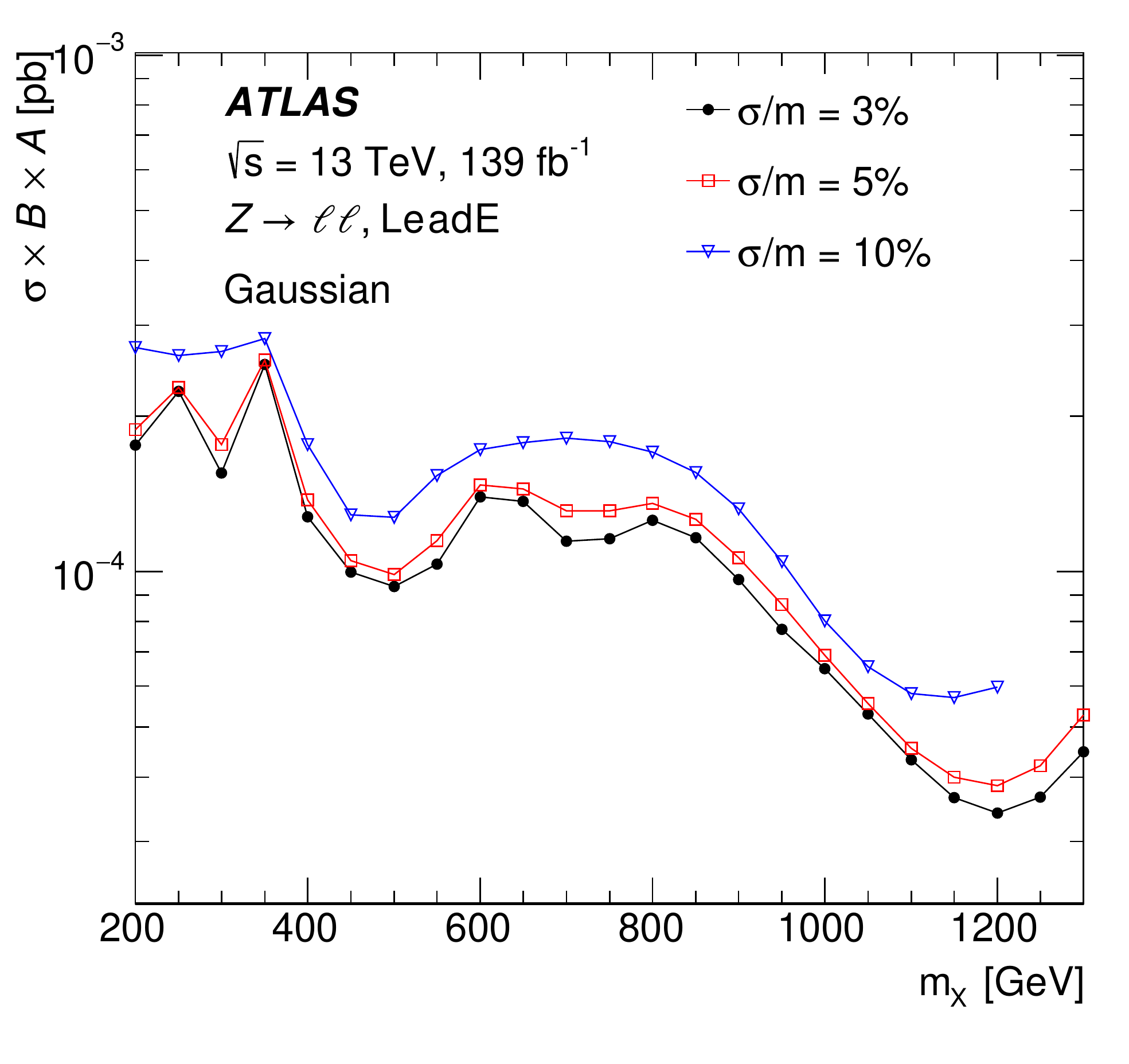} \label{fig:limit-comp-mx-e}}
\subfloat[]{\includegraphics[scale=0.34]{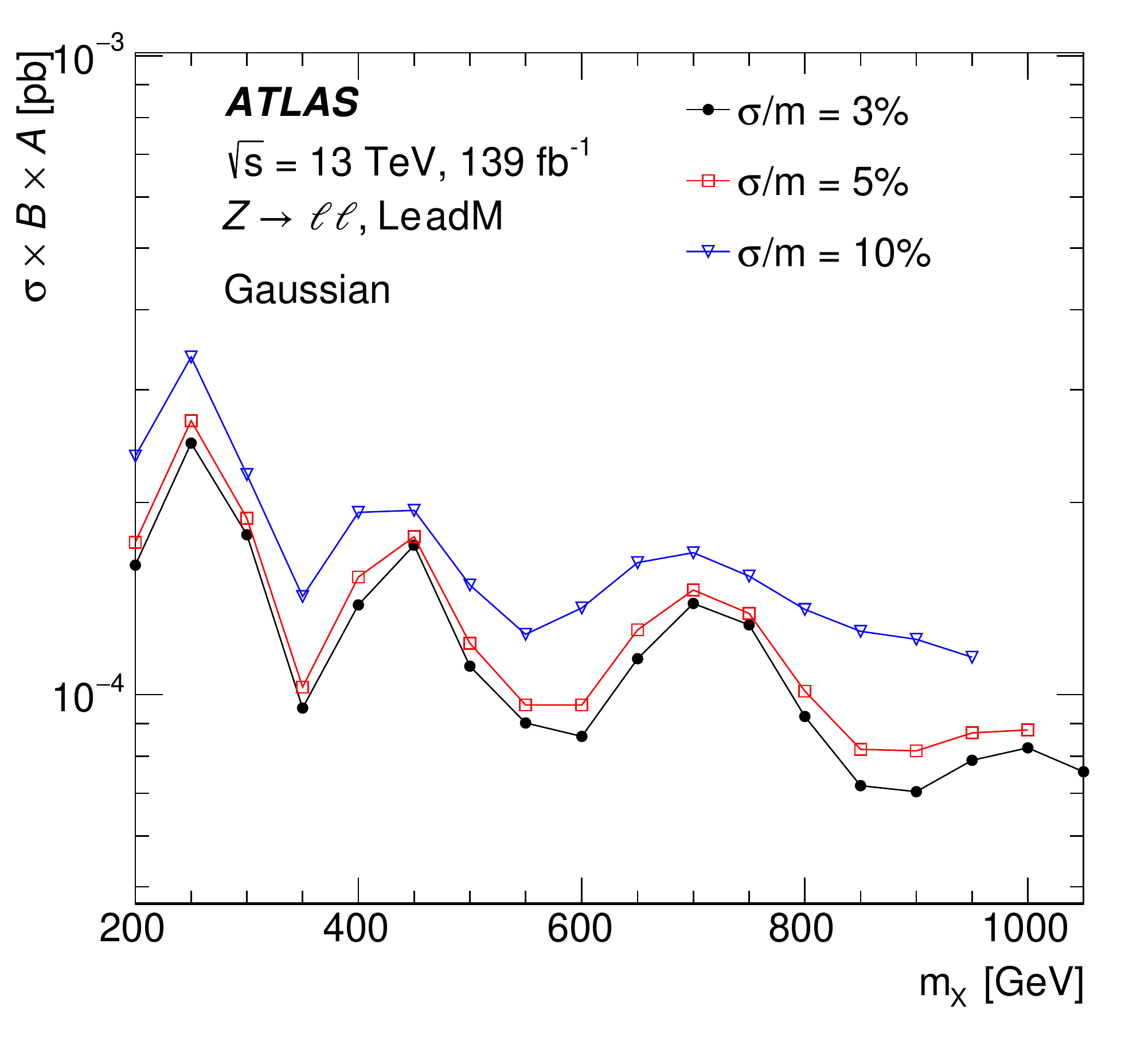} \label{fig:limit-comp-mx-f}}
\vspace{-.5\baselineskip}
\caption{Comparison of observed upper limits at 95\% CL on the cross section times branching fraction times acceptance for a Gaussian-shaped signals with relative width values of 3\%, 5\% and 10\% as a function of $m_X$ in the \protect\subref{fig:limit-comp-mx-a} leading small-$R$-jet, \protect\subref{fig:limit-comp-mx-b} leading $b$-jet, \protect\subref{fig:limit-comp-mx-c} leading large-$R$-jet, \protect\subref{fig:limit-comp-mx-d} leading photon, \protect\subref{fig:limit-comp-mx-e} leading electron, and \protect\subref{fig:limit-comp-mx-f} leading muon categories.}
\label{fig:limit-comp-mx}
\end{figure}

\begin{figure}[htbp]
\centering
\subfloat[]{\includegraphics[scale=0.34]{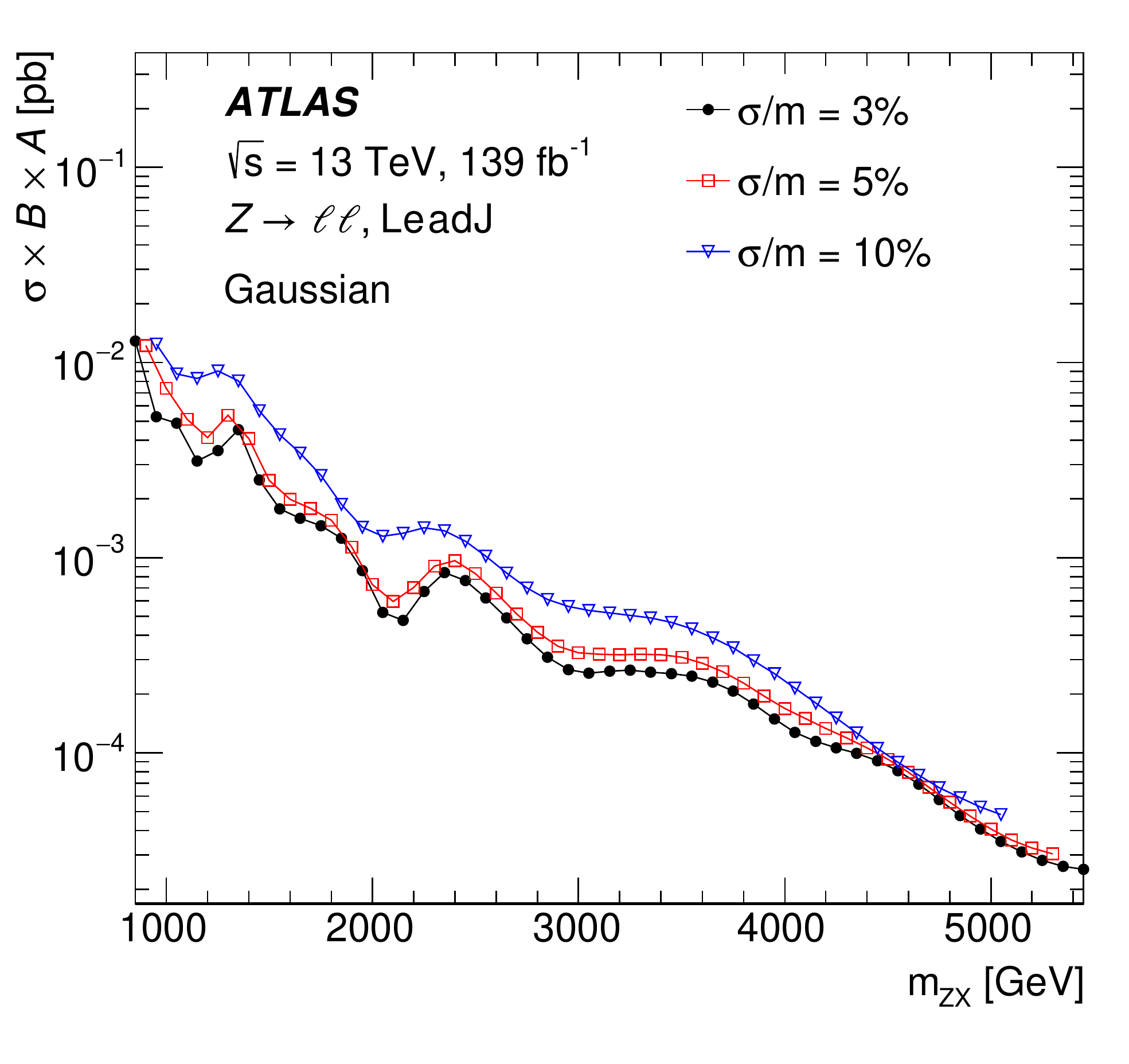} \label{fig:limit-comp-mzx-a}}
\subfloat[]{\includegraphics[scale=0.34]{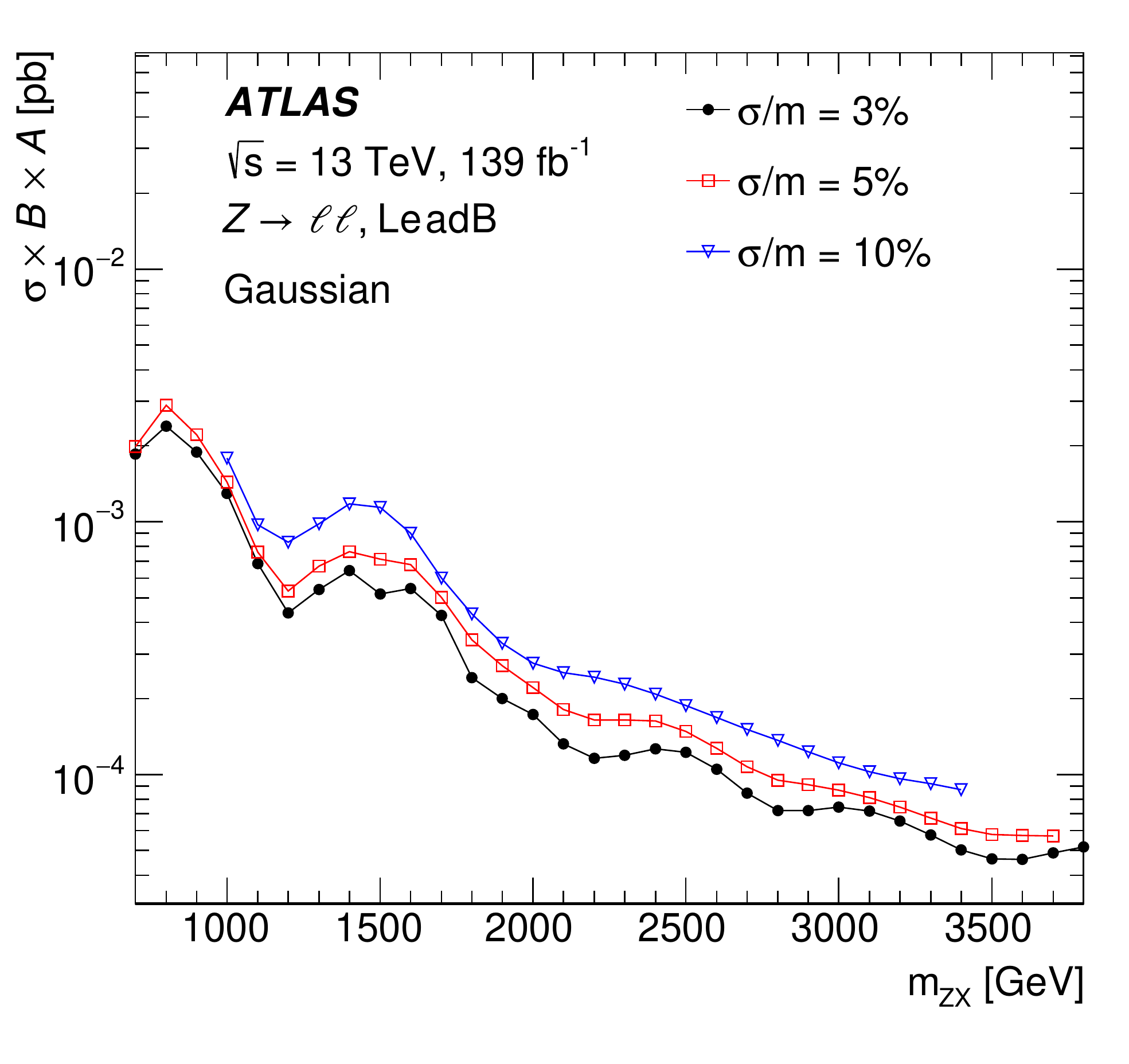} \label{fig:limit-comp-mzx-b}}
\vspace{-.5\baselineskip}
\subfloat[]{\includegraphics[scale=0.34]{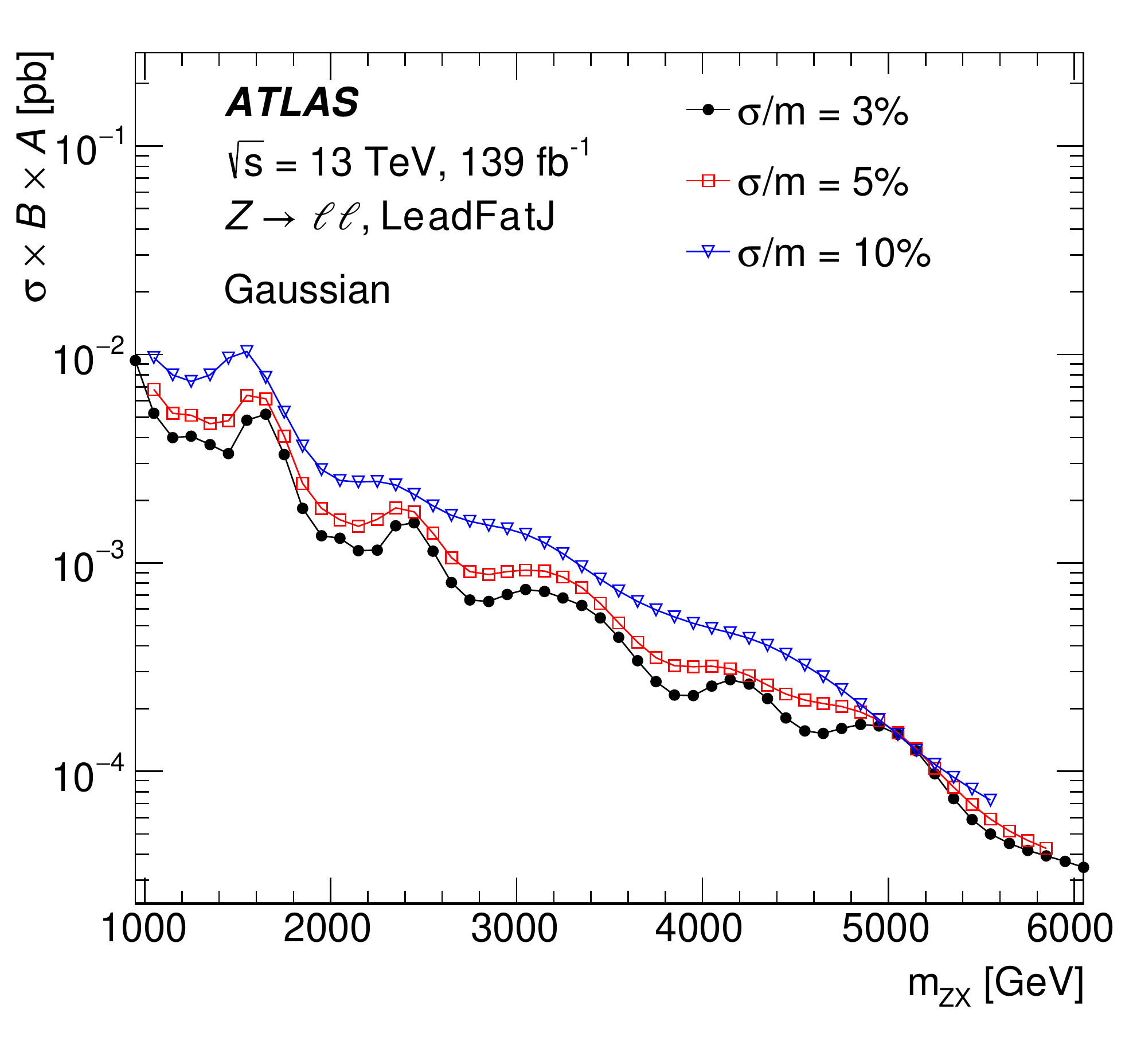} \label{fig:limit-comp-mzx-c}}
\subfloat[]{\includegraphics[scale=0.34]{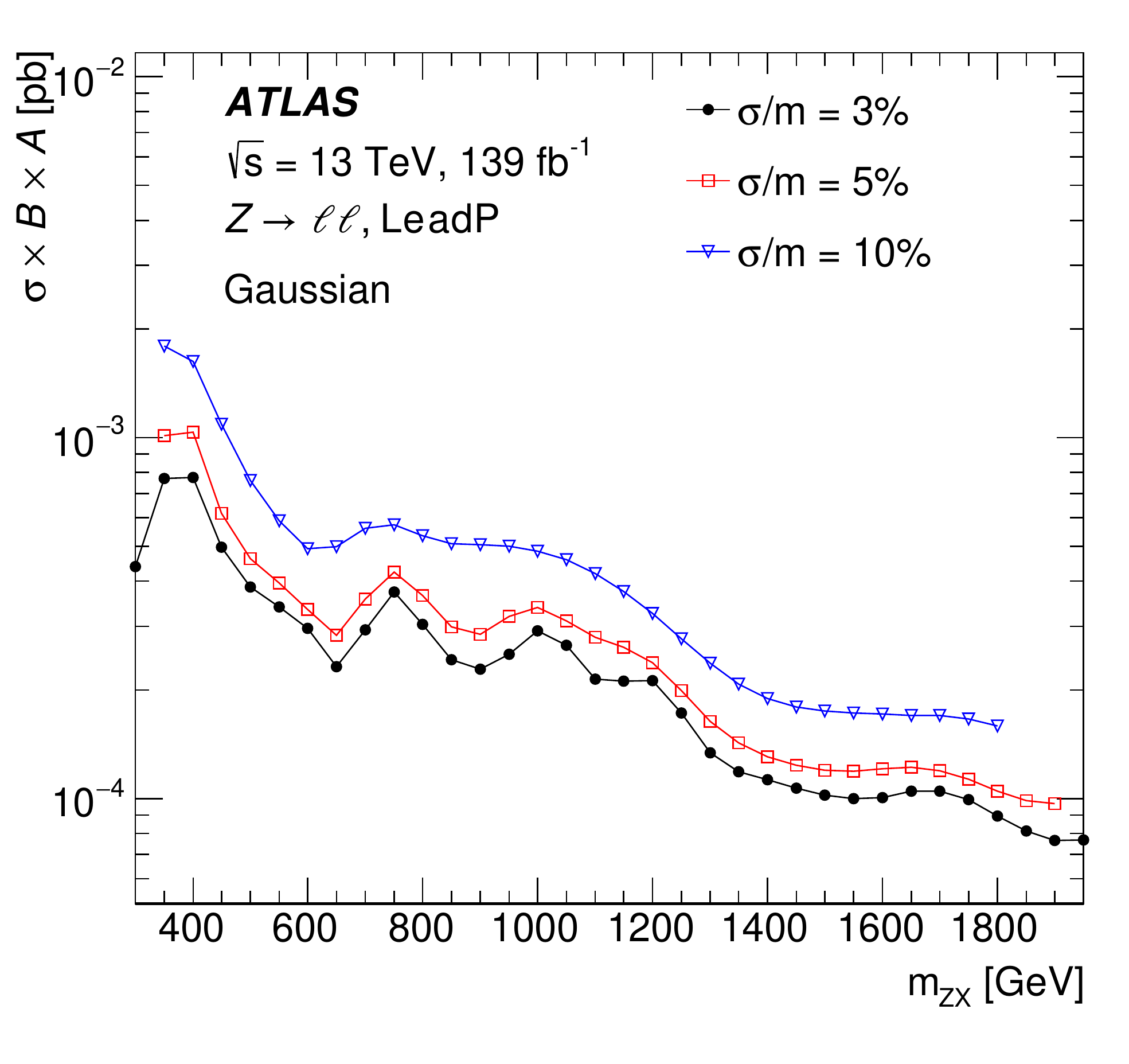} \label{fig:limit-comp-mzx-d}}
\vspace{-.5\baselineskip}
\subfloat[]{\includegraphics[scale=0.34]{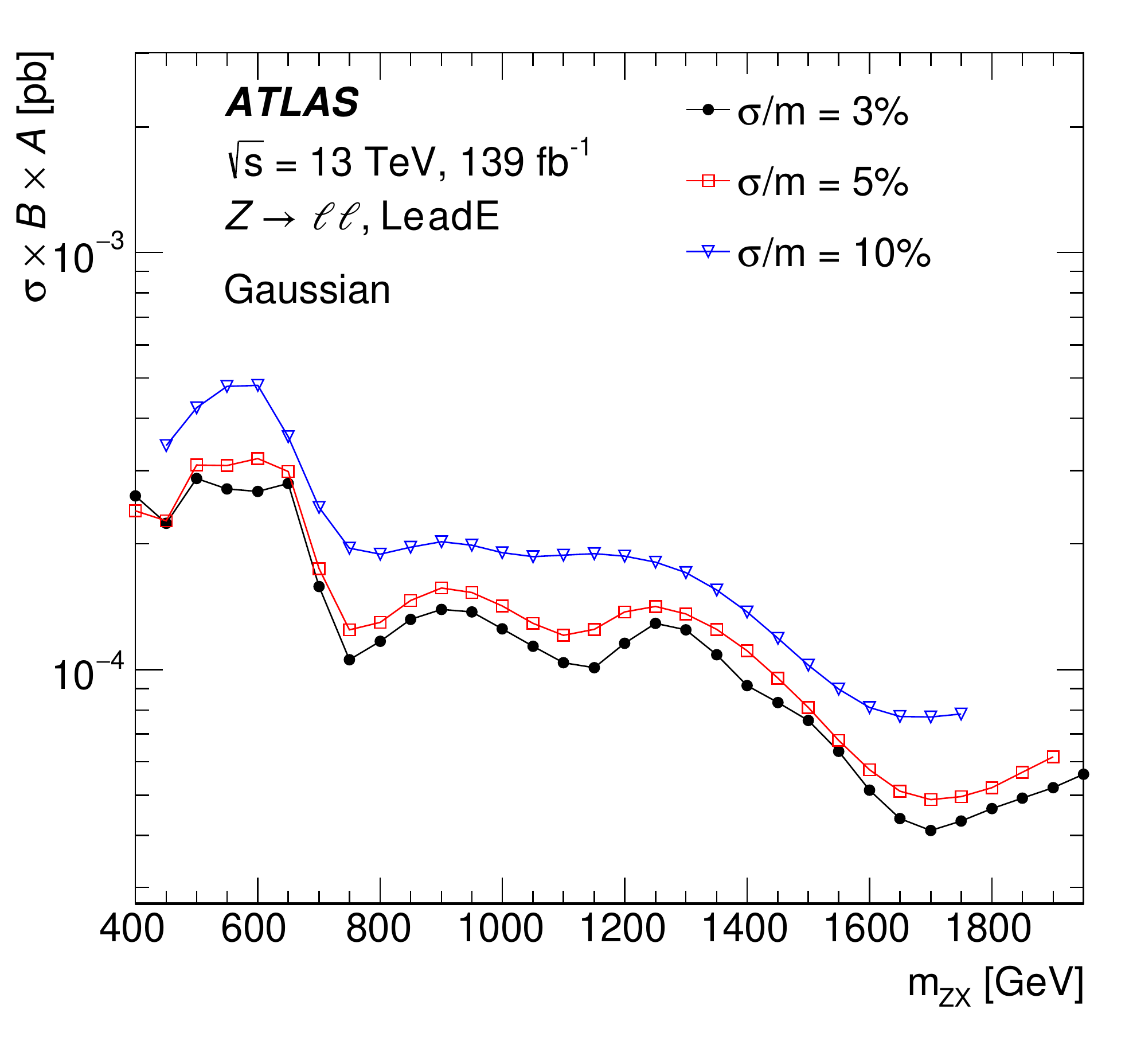} \label{fig:limit-comp-mzx-e}}
\subfloat[]{\includegraphics[scale=0.34]{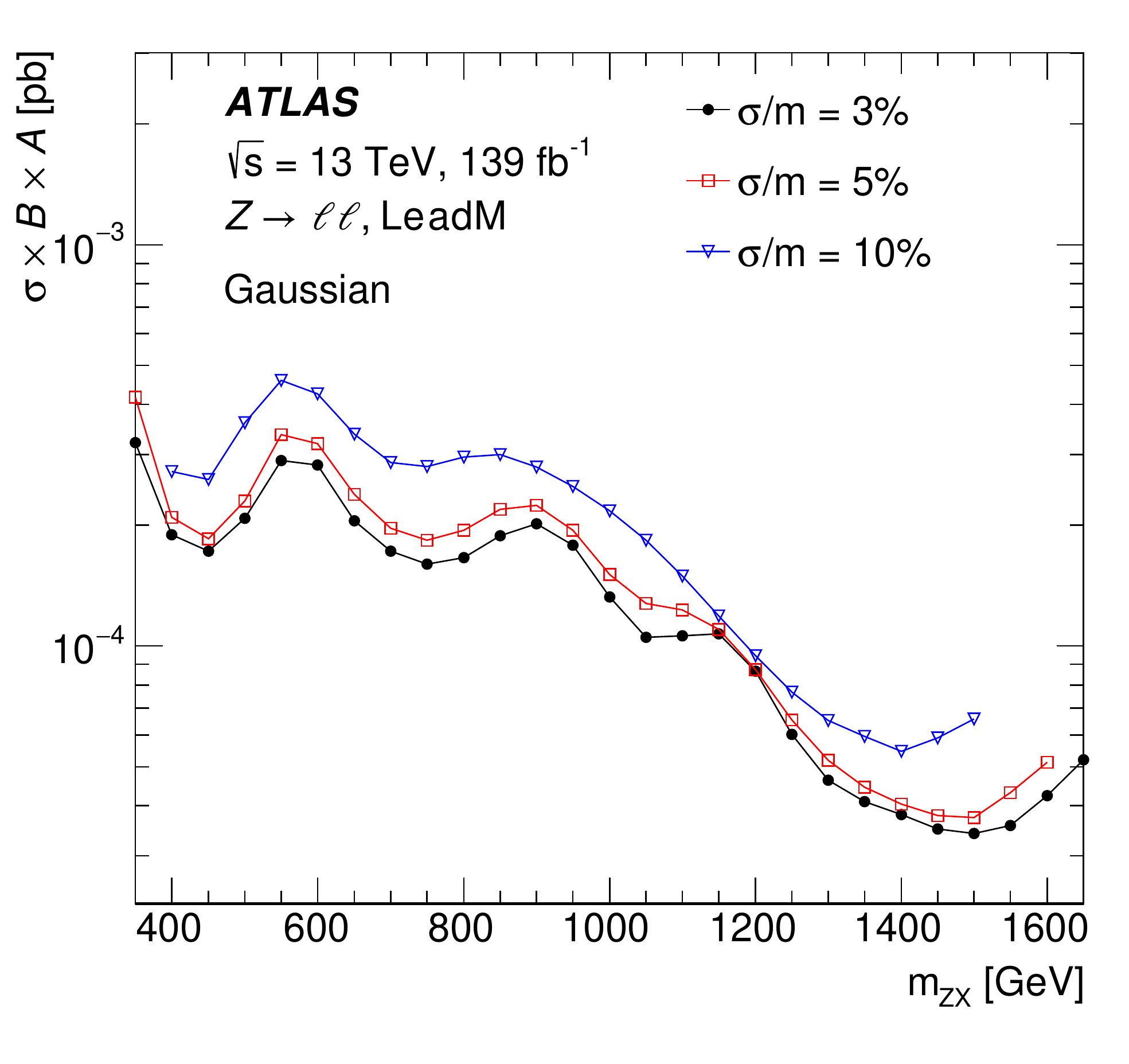} \label{fig:limit-comp-mzx-f}}
\vspace{-.5\baselineskip}
\caption{Comparison of observed upper limits at 95\% CL on the cross section times branching fraction times acceptance for a Gaussian-shaped signals with relative width values of 3\%, 5\% and 10\% as a function of $m_{ZX}$ in the \protect\subref{fig:limit-comp-mzx-a} leading small-$R$-jet, \protect\subref{fig:limit-comp-mzx-b} leading $b$-jet, \protect\subref{fig:limit-comp-mzx-c} leading large-$R$-jet, \protect\subref{fig:limit-comp-mzx-d} leading photon, \protect\subref{fig:limit-comp-mzx-e} leading electron, and \protect\subref{fig:limit-comp-mzx-f} leading muon categories.}
\end{figure}

\begin{figure}[htbp]
\centering
\subfloat[]{\includegraphics[scale=0.34]{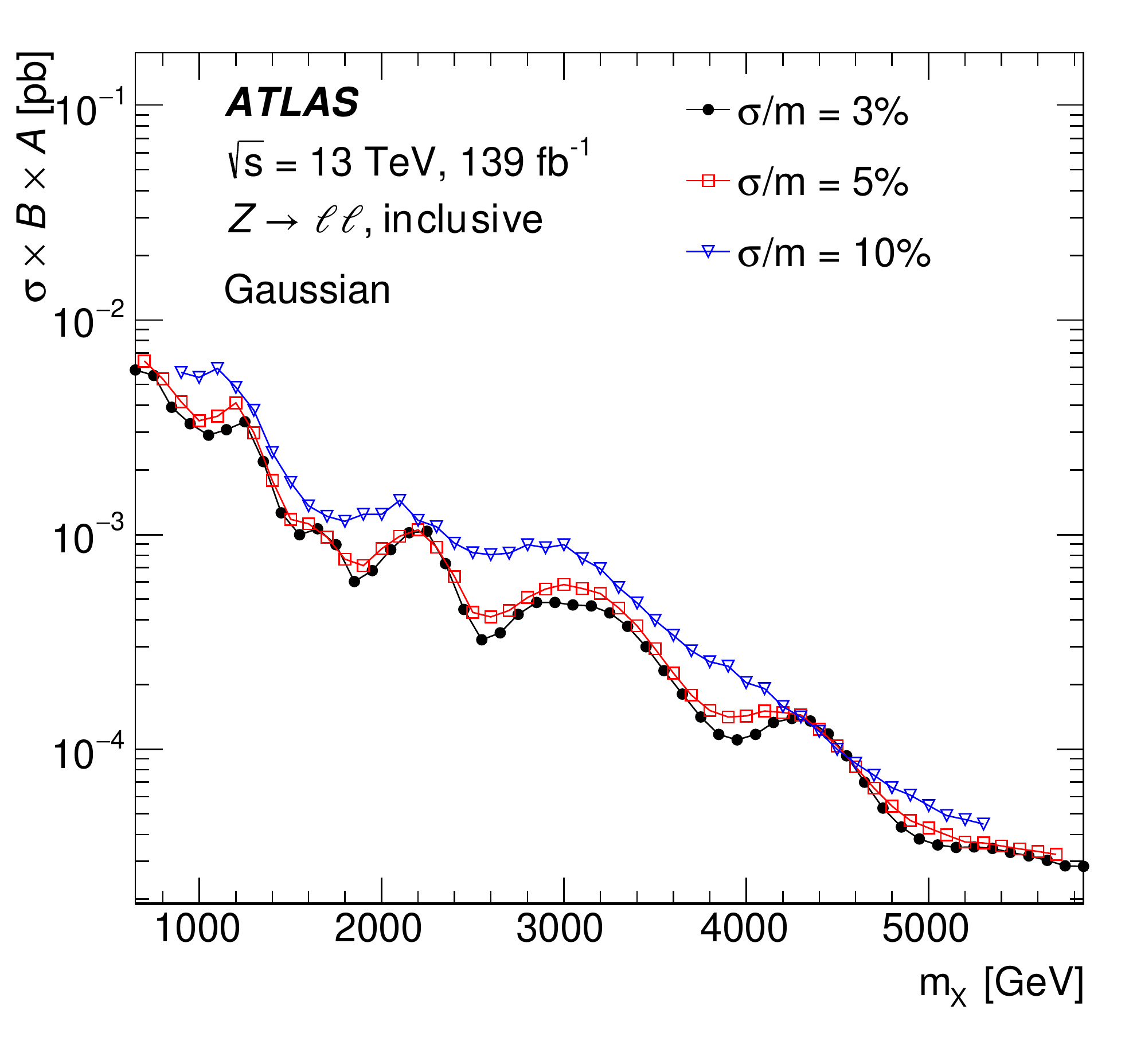} \label{fig:limit-comp-a}}
\subfloat[]{\includegraphics[scale=0.34]{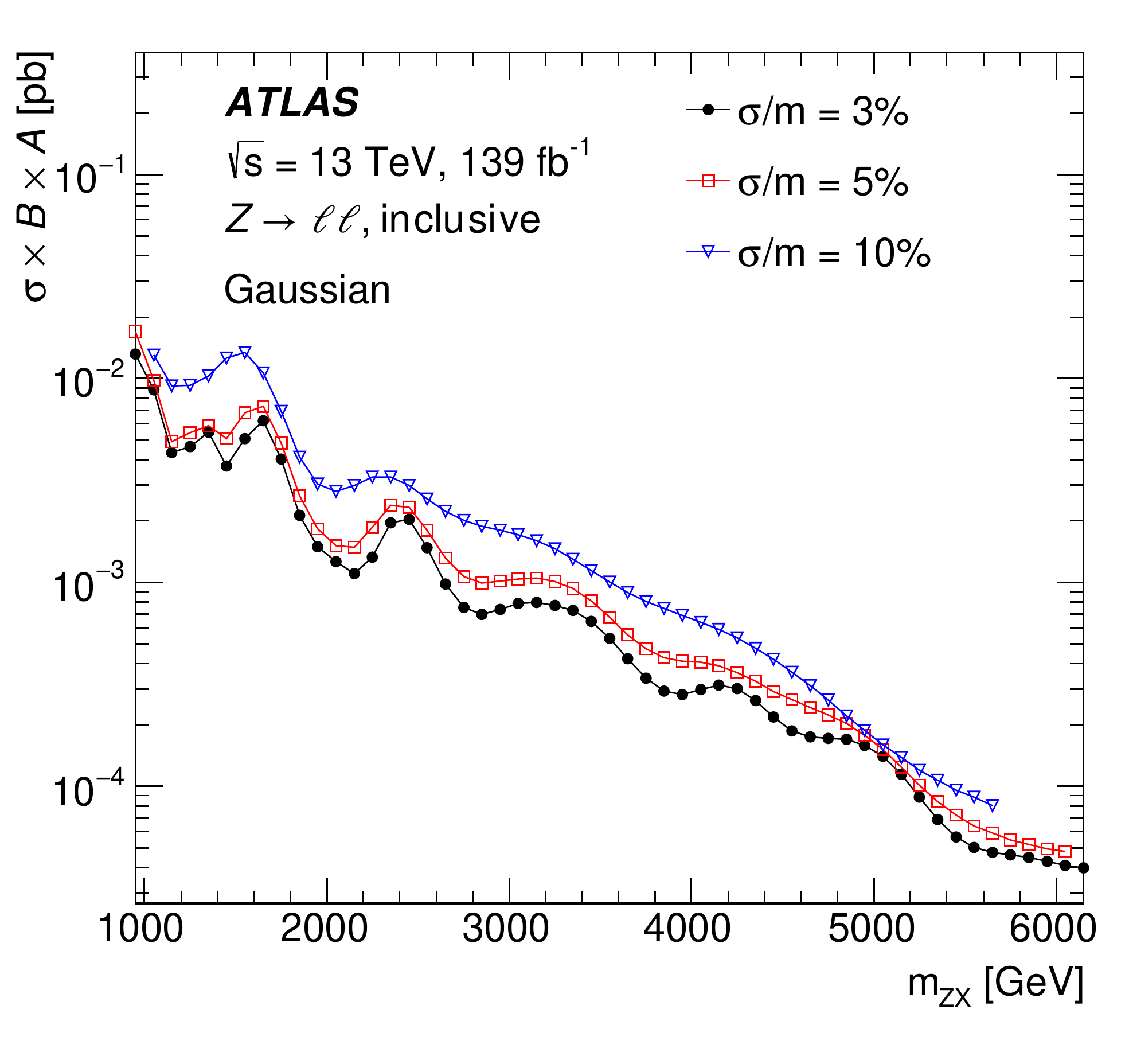} \label{fig:limit-comp-b}}
\vspace{-.5\baselineskip}
\caption{Comparison of observed upper limits at 95\% CL on the cross section times branching fraction times acceptance for a Gaussian-shaped signals with relative width values of 3\%, 5\% and 10\% as a function of \protect\subref{fig:limit-comp-a} $m_X$ and \protect\subref{fig:limit-comp-b} $m_{ZX}$ in the inclusive category.}
\label{fig:limit-comp}
\end{figure}

The resulting expected and observed limits at 95\% CL on the Gaussian-shaped signal with a relative width value of 3\% are shown as a function of $m_X$ in Figure~\ref{fig:limit-mx} and $m_{ZX}$ in Figure~\ref{fig:limit-mzx} for the six exclusive event categories and in Figure~\ref{fig:limit-comb} for the inclusive event category. Similar limits are also derived for width values of 5\% and 10\%. Comparisons between the observed limits for the three width values are presented in Figures~\ref{fig:limit-comp-mx}--\ref{fig:limit-comp} for each category. They show the general feature that the limits for the smallest  width are the most stringent and cover the largest mass range. Depending on the event categories, the upper limits on $\sigma\times\mathcal{B}\times\mathcal{A}$ for a Gaussian-shaped signal with a relative width of 3\% are 0.16~fb and 0.03~fb for $m_X$ values of 200~\GeV\ and 5.9~\TeV, respectively. The upper limits are 0.3~fb and 0.04~fb for $m_{ZX}$ values of 300~\GeV\ and 6~\TeV, respectively. The limits are calculated with the asymptotic approximation to the distribution of the test statistic~\cite{Cowan:2010js}. The results are compared with the alternative approach using pseudo-experiments and the difference is found to be well below 15\%, much smaller than the other dominant uncertainties. The limits are not corrected for efficiencies. Typical values of acceptance times efficiency can be found in~\cite{hepdata}.
 
These limits should be used when long low-mass off-shell tails from PDFs and non-perturbative effects on the narrow resonance signal shape can be safely truncated or neglected and, after applying the selection described in Sections~\ref{sec:recon} and \ref{sec:selection} (the main selection being $66<m_{\ell\ell}<116$\,\GeV\
and $\pTZ>100$\,\GeV), the reconstructed mass distribution approximates a Gaussian distribution.
All reconstructed objects passing the object selection including initial-state radiation (ISR) must be included in $m_X$ and $m_{ZX}$.
ISR broadens the reconstructed mass, reducing sensitivity as outlined for the HVT model in the next section.
A detailed description of how to use these limits is given in Ref.~\cite{EXOT-2013-11}.
 
\clearpage
\subsection{Constraint on a $W^\prime\to ZW$ signal in an HVT model}
 
\begin{figure}[htbp]
\centering
\includegraphics[scale=0.4]{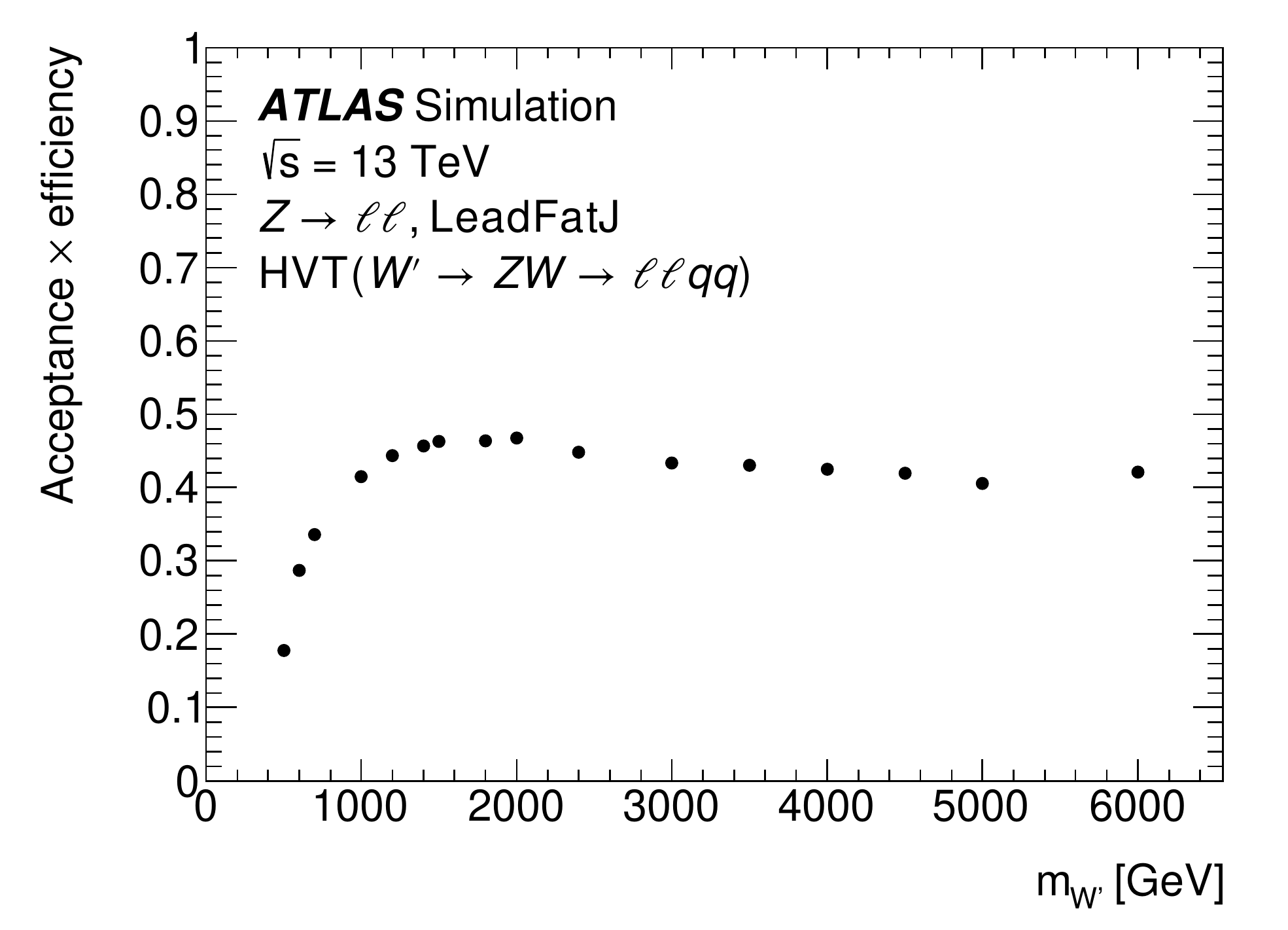}
\caption{Acceptance times efficiency for a $W^\prime\to ZW\to \ell\ell qq$ signal in an HVT model as a function of $m_{W^\prime}$ in the leading large-$R$-jet  category.}
\label{fig:eff_hvt}
\end{figure}

Although this generic analysis is not optimised for a particular model, it can also be used to constrain a specific model, for example the $W^\prime \to ZW\to \ell\ell qq$ signal in the HVT model. The leptonic final state of the $W$ boson decays is not used -- the presence of the neutrino causes the reconstructed mass to be significantly shifted from the generated value and does not improve the analysis sensitivity. The acceptance times efficiency values as a function of $m_{ZX}$ in the dominant leading large-$R$-jet category are shown in Figure~\ref{fig:eff_hvt}.
 
\begin{figure}[htbp]
\centering
\includegraphics[scale=0.4]{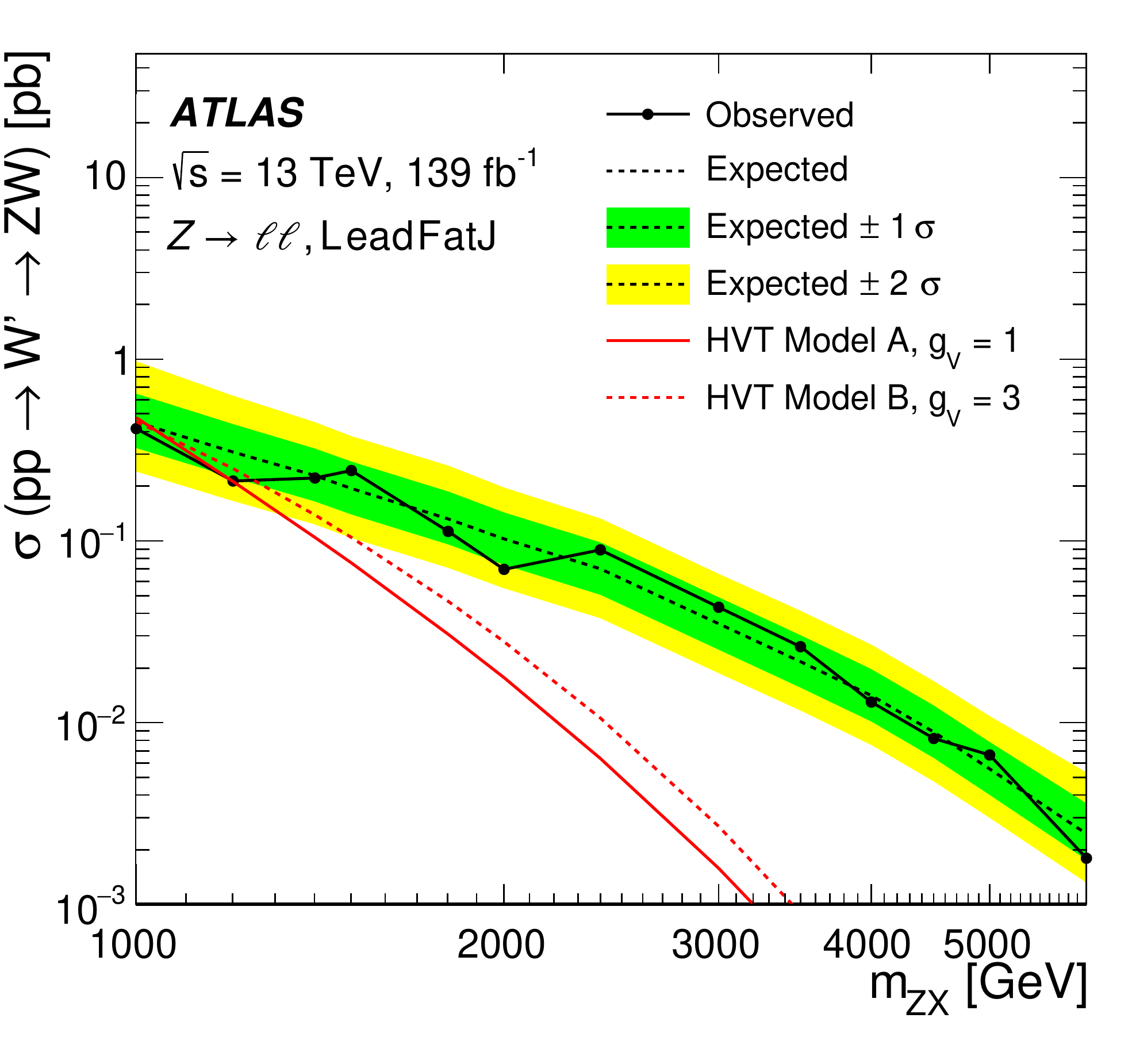}
\caption{Upper limits at 95\% CL on the cross section times the branching fraction ($W^\prime\to ZW$) for the HVT signal as a function of $m_{ZX}$ in the lead-$R$-jet  event category. The full (dashed) red curves correspond to the theoretical predictions of HVT model A (B).}
\label{fig:limit-hvt}
\end{figure}

The derived observed and expected upper limits at 95\% CL on the cross section times the branching fraction of the $W^\prime\to ZW \to \ell\ell qq$ decay for the HVT signal are shown in Figure~\ref{fig:limit-hvt}. The limits exclude an HVT signal lighter than 1.2\,\TeV\ for both models A and B. The two models differ in their coupling parameter $g_V$ to the SM gauge bosons, with a value of 1 and 3, respectively~\cite{1402.4431}. These limits are weaker than those in Ref.~\cite{HDBS-2018-10}, where boson tagging was applied to increase the signal sensitivity,  illustrating the trade-off between generality and sensitivity.
 
When constructing $m_{ZX}$, all reconstructed objects passing the object selection are included, even those arising from ISR. For the HVT signal, the expected cross section limits from reinterpreting the limits on a Gaussian-shaped signal are about 20\% stronger if ISR objects are excluded from the invariant mass calculation using generator-level information.
 
\FloatBarrier


\FloatBarrier

\section{Conclusion}
 
This article reports a novel generic search for resonances produced in association with, or decaying to, a $Z$ boson with $\pTZ > 100$\,\GeV\ and a subsequent decay into an $e^+e^-$ or $\mu^+\mu^-$ pair, using 139\,fb$^{-1}$ of proton--proton collisions recorded at $\sqrt{s}=13$\,TeV with the ATLAS detector at the Large Hadron Collider between 2015 and 2018. The invariant mass spectra of the system recoiling against the $Z$ boson and of the full final state including the $Z$ boson in six independent event categories, as well as the inclusive category, are analysed simultaneously. No significant excess is observed above the data-driven estimates of the smoothly falling distributions of the Standard Model background. Constraints on Gaussian-shaped signals with relative width values between 3\% and 10\% are derived. Depending on the event category, the search covers the mass range down to 200\,\GeV\ and up to 6\,\TeV, and the most stringent upper limits on the visible cross sections for a Gaussian-shaped signal with a relative width value of 3\% are about 0.16\,fb at the low mass boundary and 0.03\,fb at high mass boundary. In a model-dependent interpretation of a $W^\prime\to ZW$ signal in a heavy vector triplet model, signal mass values below 1.2\,\TeV\ are excluded using the leading large-$R$-jet event category.


\section*{Acknowledgements}


We thank CERN for the very successful operation of the LHC, as well as the
support staff from our institutions without whom ATLAS could not be
operated efficiently.
 
We acknowledge the support of
ANPCyT, Argentina;
YerPhI, Armenia;
ARC, Australia;
BMWFW and FWF, Austria;
ANAS, Azerbaijan;
CNPq and FAPESP, Brazil;
NSERC, NRC and CFI, Canada;
CERN;
ANID, Chile;
CAS, MOST and NSFC, China;
Minciencias, Colombia;
MEYS CR, Czech Republic;
DNRF and DNSRC, Denmark;
IN2P3-CNRS and CEA-DRF/IRFU, France;
SRNSFG, Georgia;
BMBF, HGF and MPG, Germany;
GSRI, Greece;
RGC and Hong Kong SAR, China;
ISF and Benoziyo Center, Israel;
INFN, Italy;
MEXT and JSPS, Japan;
CNRST, Morocco;
NWO, Netherlands;
RCN, Norway;
MEiN, Poland;
FCT, Portugal;
MNE/IFA, Romania;
MESTD, Serbia;
MSSR, Slovakia;
ARRS and MIZ\v{S}, Slovenia;
DSI/NRF, South Africa;
MICINN, Spain;
SRC and Wallenberg Foundation, Sweden;
SERI, SNSF and Cantons of Bern and Geneva, Switzerland;
MOST, Taiwan;
TENMAK, T\"urkiye;
STFC, United Kingdom;
DOE and NSF, United States of America.
In addition, individual groups and members have received support from
BCKDF, CANARIE, Compute Canada and CRC, Canada;
PRIMUS 21/SCI/017 and UNCE SCI/013, Czech Republic;
COST, ERC, ERDF, Horizon 2020 and Marie Sk{\l}odowska-Curie Actions, European Union;
Investissements d'Avenir Labex, Investissements d'Avenir Idex and ANR, France;
DFG and AvH Foundation, Germany;
Herakleitos, Thales and Aristeia programmes co-financed by EU-ESF and the Greek NSRF, Greece;
BSF-NSF and MINERVA, Israel;
Norwegian Financial Mechanism 2014-2021, Norway;
NCN and NAWA, Poland;
La Caixa Banking Foundation, CERCA Programme Generalitat de Catalunya and PROMETEO and GenT Programmes Generalitat Valenciana, Spain;
G\"{o}ran Gustafssons Stiftelse, Sweden;
The Royal Society and Leverhulme Trust, United Kingdom.
 
The crucial computing support from all WLCG partners is acknowledged gratefully, in particular from CERN, the ATLAS Tier-1 facilities at TRIUMF (Canada), NDGF (Denmark, Norway, Sweden), CC-IN2P3 (France), KIT/GridKA (Germany), INFN-CNAF (Italy), NL-T1 (Netherlands), PIC (Spain), ASGC (Taiwan), RAL (UK) and BNL (USA), the Tier-2 facilities worldwide and large non-WLCG resource providers. Major contributors of computing resources are listed in Ref.~\cite{ATL-SOFT-PUB-2021-003}.


\clearpage
\appendix

\clearpage
\section{Appendix}
\subsection{Merged-$e^+e^-$ identification}\label{app:merged-ee}
 
The electrons from a boosted $Z$ boson corresponding to the merged-$e^+e^-$ topology appear as a reconstructed small-$R$ jet in the EM  calorimeter, which differs, however, from a hadronic jet. Therefore, the key idea of the BDT analysis is to separate the merged-$e^+e^-$ system from hadronic jets.
The following variables are selected as inputs for the BDT analysis (the
requirements, where indicated, are applied prior to BDT training):
\begin{itemize}
\item $p^\mathrm{jet}_\mathrm{T}$: the transverse momentum of a jet (requiring $p^\mathrm{jet}_\mathrm{T}>450$\,\GeV),
\item $\eta^\mathrm{jet}$: the pseudorapidity of a jet (requiring $|\eta^\mathrm{jet}|<2.47$),
\item $m_\mathrm{jet}$: the invariant mass of a jet (requiring  the expected jet mass for a $Z\to e^+e^-$ decay to be around the $Z$ boson mass: $60 <m_\mathrm{jet}<120$\,\GeV),
\item $f^\mathrm{EM}_\mathrm{jet}$: the jet energy fraction deposited in the EM calorimeter, satisfying a $\pT$-dependent selection,
\item max$(E_\mathrm{layer}/E_\mathrm{jet})$: the maximum jet energy fraction deposited in a calorimeter layer,
\item $N^\mathrm{track}_\mathrm{jet}$: the number of tracks, with $p_\mathrm{T}>500$\,\MeV, associated to a jet using the ghost-association~\cite{Cacciari:2007fd} (requiring $1\leq N^\mathrm{track}_\mathrm{jet}\leq 7$),
\item $\overline{\Delta R}\cdot p_\mathrm{T}/m_\mathrm{jet}$: where $\overline{\Delta R}$ is defined as $\sum (\Delta R(\mathrm{jet, constituent})\cdot p_\mathrm{T}(\mathrm{constituent}))/\sum p_\mathrm{T}(\mathrm{constituent})$,
\item $n^\mathrm{constituent}_\mathrm{jet}$: the number of constituents of a jet,
\item $n_\mathrm{GSF\ tracks}$: the number of Gaussian Sum Filter (GSF)~\cite{ATLAS-CONF-2012-047}) tracks inside the jet cone $\Delta R\mathrm{(jet, track)} < 0.4$, which satisfy $p_\mathrm{T}>50$\,\GeV,
\item $\Sigma \mathrm{Q}_\mathrm{GSF\ tracks}$: the total charge of the GSF tracks satisfying the selections,
\item $n_{e/\gamma\textrm{~clusters}}$: the number of $e/\gamma$ clusters inside the jet cone which satisfy $p_\mathrm{T}>25$\,\GeV.
\end{itemize}
In this study, signal events are based on $Z(\to e^+e^-)$\,+\,jets MC samples requiring the merged-$e^+e^-$ candidate to match the true $Z$ boson within $\Delta R < 0.1$. The SM background events correspond to combined MC samples of $\gamma$\,+\,jets, $W(\to e\nu)$\,+\,jets and $W(\to e\nu)$\,+\,$\gamma$. Background processes containing only hadronic jets, e.g. the dijet background, are efficiently suppressed to a negligible level by the selection listed above, in particular by a $p_\mathrm{T}$-dependent $f^\mathrm{EM}_\mathrm{jet}$ requirement. The raw number of selected signal and background jets is 102\,593 and 8572, respectively. Events are required to fire either jet triggers or electron triggers. Events with two oppositely charged same-flavour leptons are vetoed since those events are analysed using the standard likelihood-based electron identification.
The jet energy fraction in the EM calorimeter must satisfy a $p_\mathrm{T}$-dependent selection, and at
least one but no more than seven tracks must be found in the jet cone. The candidate must match with at least one GSF track, and satisfy $N_\mathrm{PFO}-N_\mathrm{GSF\ tracks} < 2$
to suppress $\gamma$\,+\,jets background, where $N_\mathrm{PFO}$ is the number of particle-flow tracks with $p_\mathrm{T}>1$\,\GeV. At least one $e/\gamma$ cluster~\cite{EGAM-2018-01} must also be matched
with the jet. The cluster-based mass, $m_\mathrm{cls}$, different from $m_\mathrm{jet}$, is defined using the matched $e/\gamma$ cluster information. It is constructed with the leading-$p_\text{T}$ $e/\gamma$ cluster in the jet, to show the two-body decay property of the $Z\to e^+e^-$
process. The cluster-based mass is defined as $m_\mathrm{cls} = \frac{1}{2}\overline{\Delta R}p_\mathrm{T}^2/\sqrt{z(1-z)}$, where $z$ is the ratio of leading-cluster energy to jet energy. The jet must satisfy
$m_\mathrm{cls}>60$\,\GeV.
 
\begin{figure}[htb]
\centering
\subfloat[]{\includegraphics[width=0.49\columnwidth]{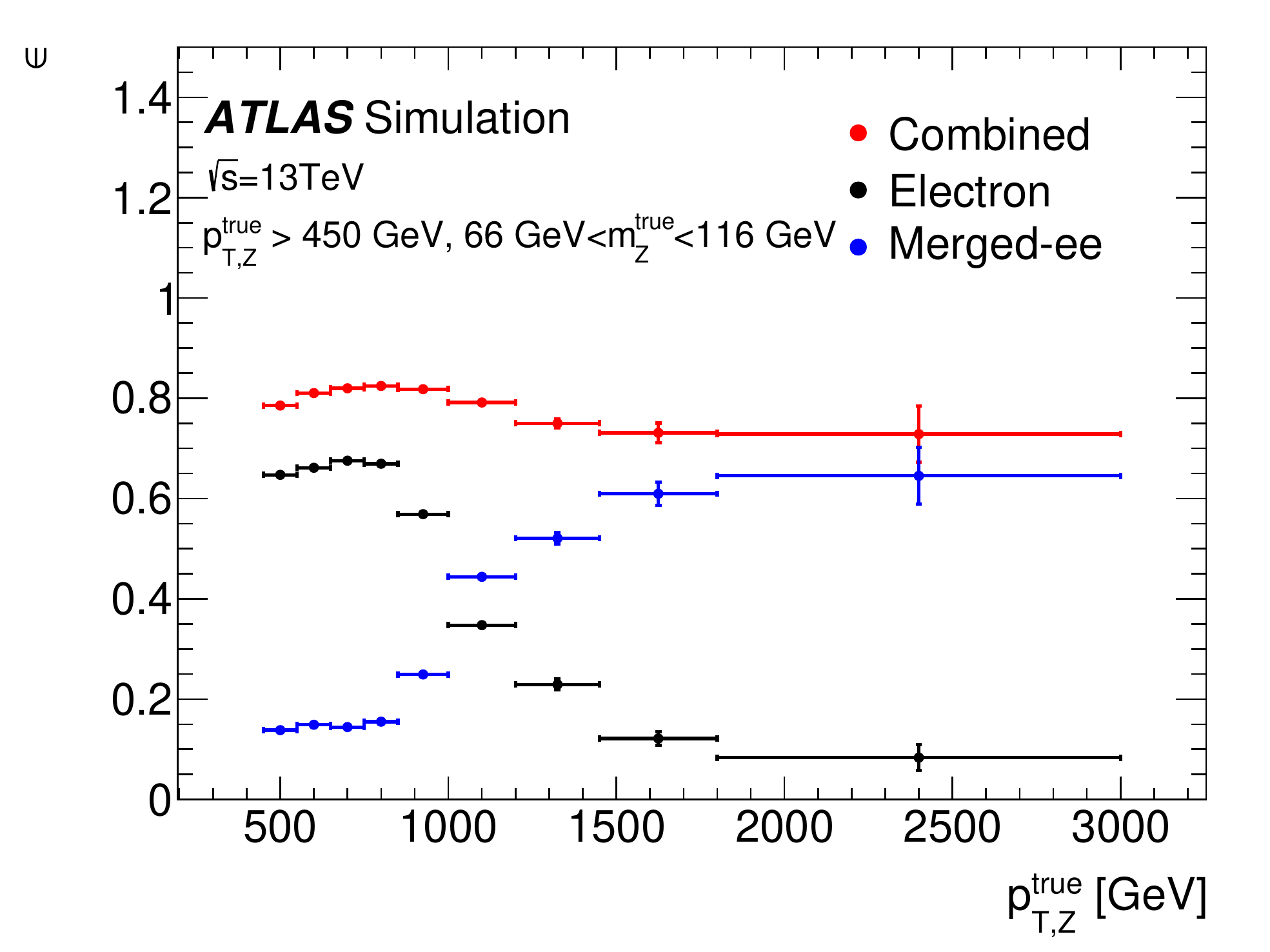} \label{fig:eidBDT-a}}
\subfloat[]{\includegraphics[width=0.49\columnwidth]{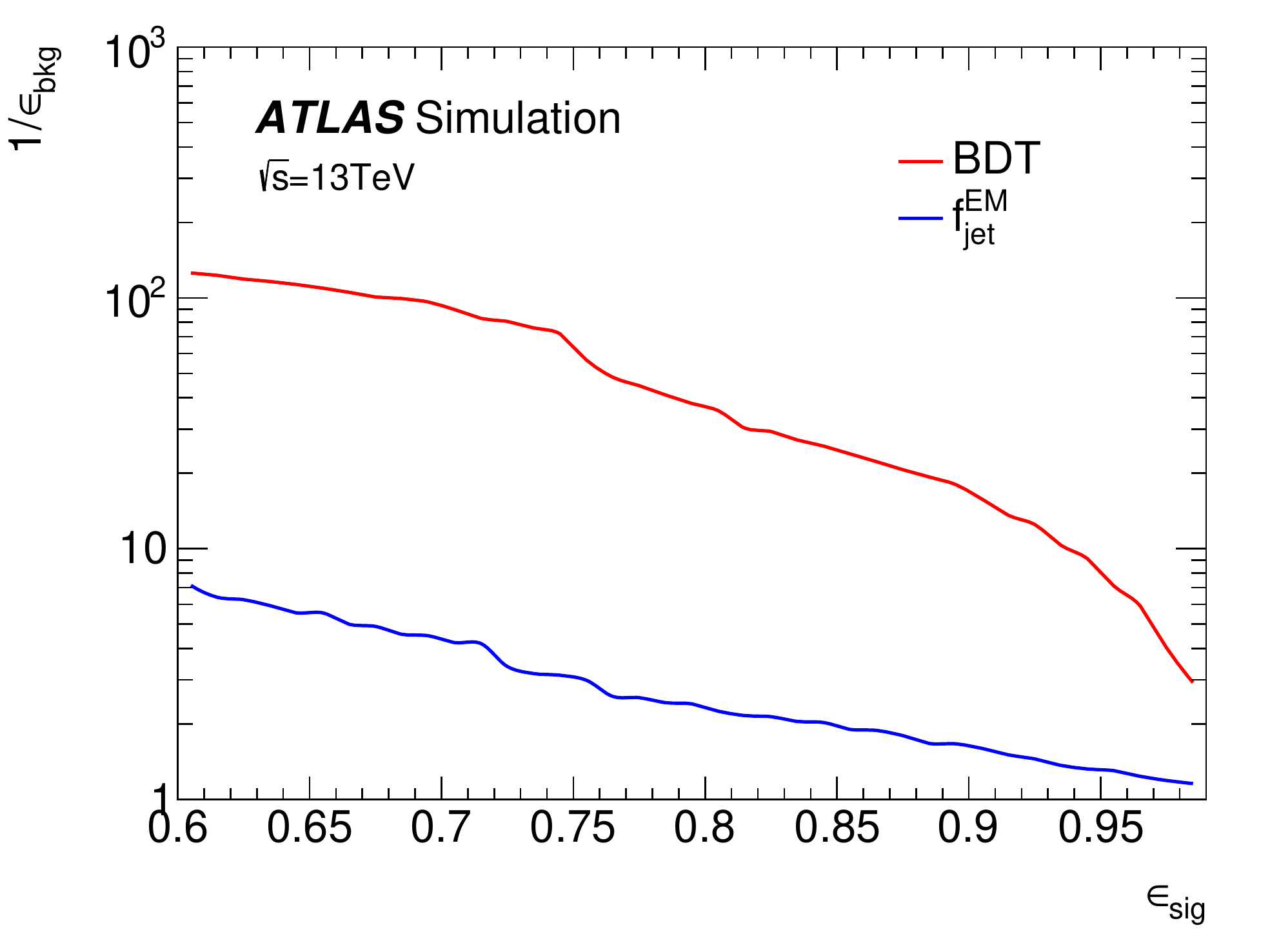} \label{fig:eidBDT-b}}
\vspace{-.5\baselineskip}
\caption{\protect\subref{fig:eidBDT-a} Comparison of the identification efficiencies using standard and merged-$e^+e^-$ reconstruction as a function of true $\pTZ$. \protect\subref{fig:eidBDT-b} Background rejection factor as a function of signal efficiency.  The red curve shows the BDT performance whereas the blue curve corresponds to that of a cut-based analysis relying only on the jet energy fraction deposited in the EM calorimeter.} \label{fig:eidBDT}
\end{figure}

Using a $p_\mathrm{T}$-dependent BDT-score selection, the merged-$e^+e^-$ identification efficiency as a function of $p^Z_\mathrm{T}$ is determined using the $Z(\to e^+e^-)$\,+\,jets MC samples and compared with that of the standard electron identification in Figure~\ref{fig:eidBDT} (left). The decrease in the efficiency of the standard electron identification at high  $\pTZ$ is recovered with the BDT-based merged-$e^+e^-$ identification, reaching a combined efficiency of 80\%. The background rejection factor for three types of processes surviving the analysis preselection is shown in Figure~\ref{fig:eidBDT} (right), illustrating the better performance of the BDT method in comparison with a cut-based method using $f_\mathrm{jet}^\mathrm{EM}$ at the same signal efficiency.
 
\subsection{Signal-sensitive mass range}\label{app:signal-range}
 
To obtain a mass range in a given event category in which the search is able to detect a signal, a set of 100 pseudo-experiments are generated using the fitted background functional form with Poisson fluctuations. Gaussian-shaped signals with intrinsic relative width values varying between 3\% and 10\% are injected, after taking into account detector resolution effects, on top of each pseudo-experiment. The injected signals correspond to a signal strength of $5\sigma$ and are spaced every 250\,\GeV\ in $m_{X}$ and $m_{ZX}$. Each of the 100 resulting mass spectra is treated as data and is studied following the search strategy described in Section~\ref{sec:strategy} in two steps.
 
\begin{figure}[htbp]
\centering
\subfloat[]{\includegraphics[scale=0.4]{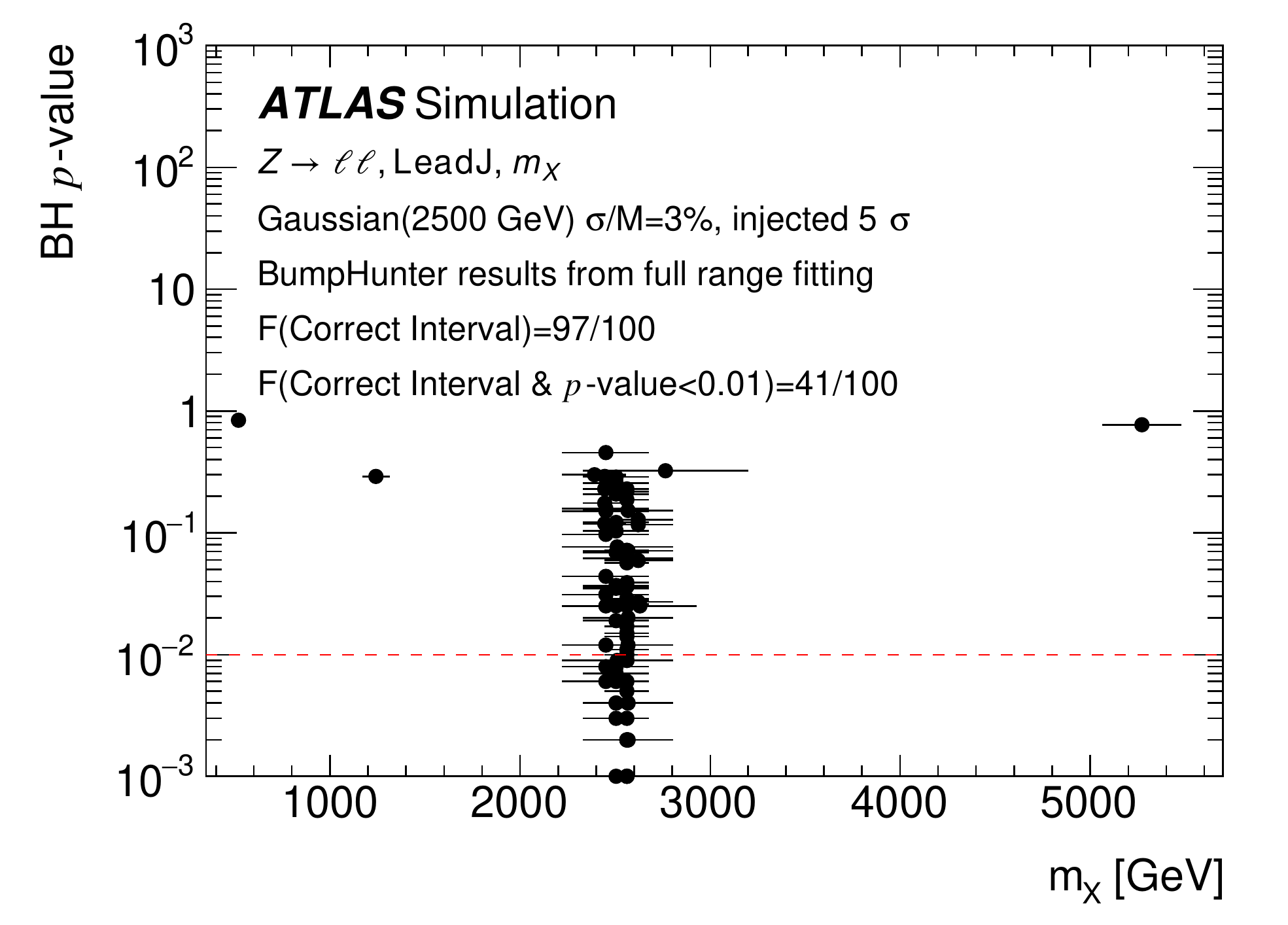} \label{appfig:injection-a}}
\subfloat[]{\includegraphics[scale=0.4]{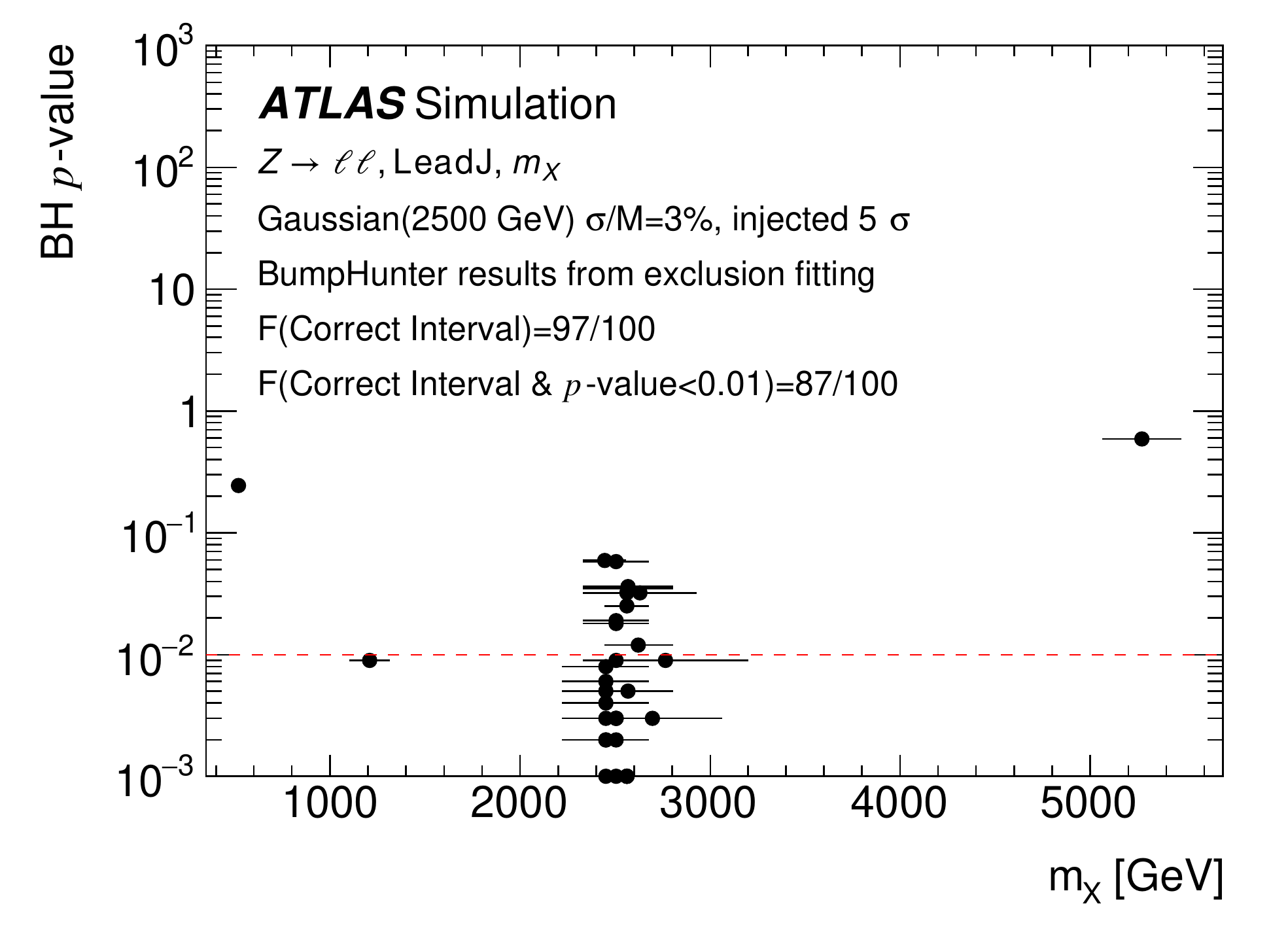} \label{appfig:injection-b}}
\vspace{-.5\baselineskip}
\caption{BH $p$-values of the 100 pseudo-experiments as a function of $m_X$ in the leading small-$R$-jet category for a $5\sigma$ injected Gaussian-shaped signal with a mass of 2.5\,\TeV\ and a relative width value of 3\%. The fractions ($F$) of the pseudo-experiments that have the correctly identified interval and BH $p$-values below the threshold of 0.01 (red dashed horizontal line) are indicated. \protect\subref{appfig:injection-a} The background is derived by the background-only fit in the full fitting range. \protect\subref{appfig:injection-b} The background is derived by the background-only fit in the range excluding the BH interval.}
\label{appfig:injection}
\end{figure}

As an example, the sensitive range of the $m_X$ spectrum in the leading small-$R$-jet category was found as follows. A Gaussian-shaped signal with a width value of 3\% is injected centred at 2.5\,\TeV\ with the number of signal events corresponding to a signal significance equal to $5\sigma$. In 97 of the 100 pseudo-experiments, the BH algorithm correctly identifies the location of the injected signal. However, only 41 pseudo-experiments yield a BH $p$-value below the threshold of 0.01 since the signal is partially absorbed in the background-only fit which is performed in the full fitting range. But if the BH interval which is derived using the background-only fit in the full fitting range is excluded from the fit range, the number of pseudo-experiments with a BH $p$-value below 0.01 increases to 87. For signal width values of 5\% and 10\%, this number is 76 and 57 respectively, showing the lower sensitivity for signals with larger widths. The results for a signal width value of 3\% are shown in Figure~\ref{appfig:injection}.
 
\begin{figure}[htbp]
\centering
\subfloat[]{\includegraphics[scale=0.4]{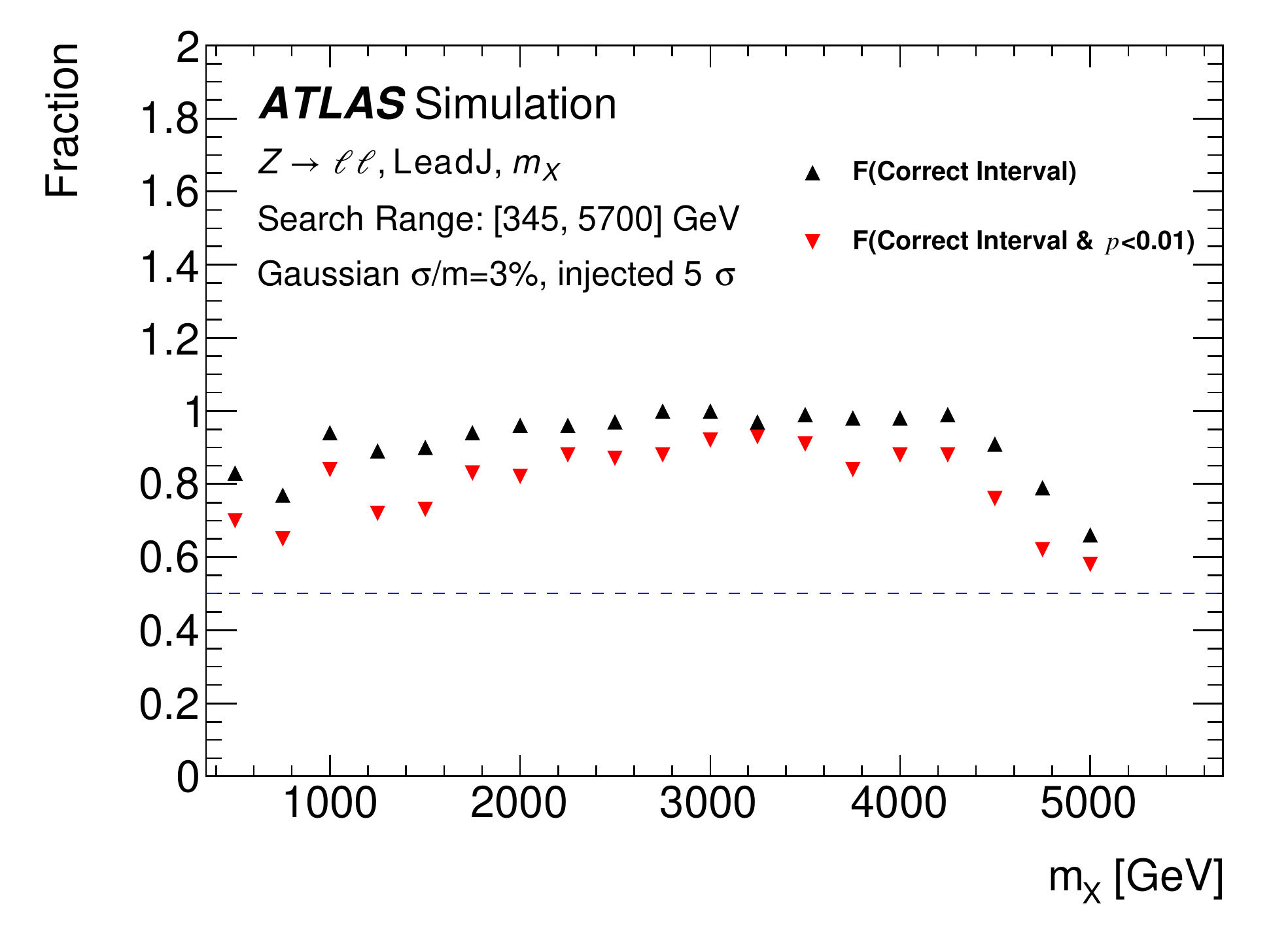} \label{appfig:scan-5sig-a}}
\subfloat[]{\includegraphics[scale=0.4]{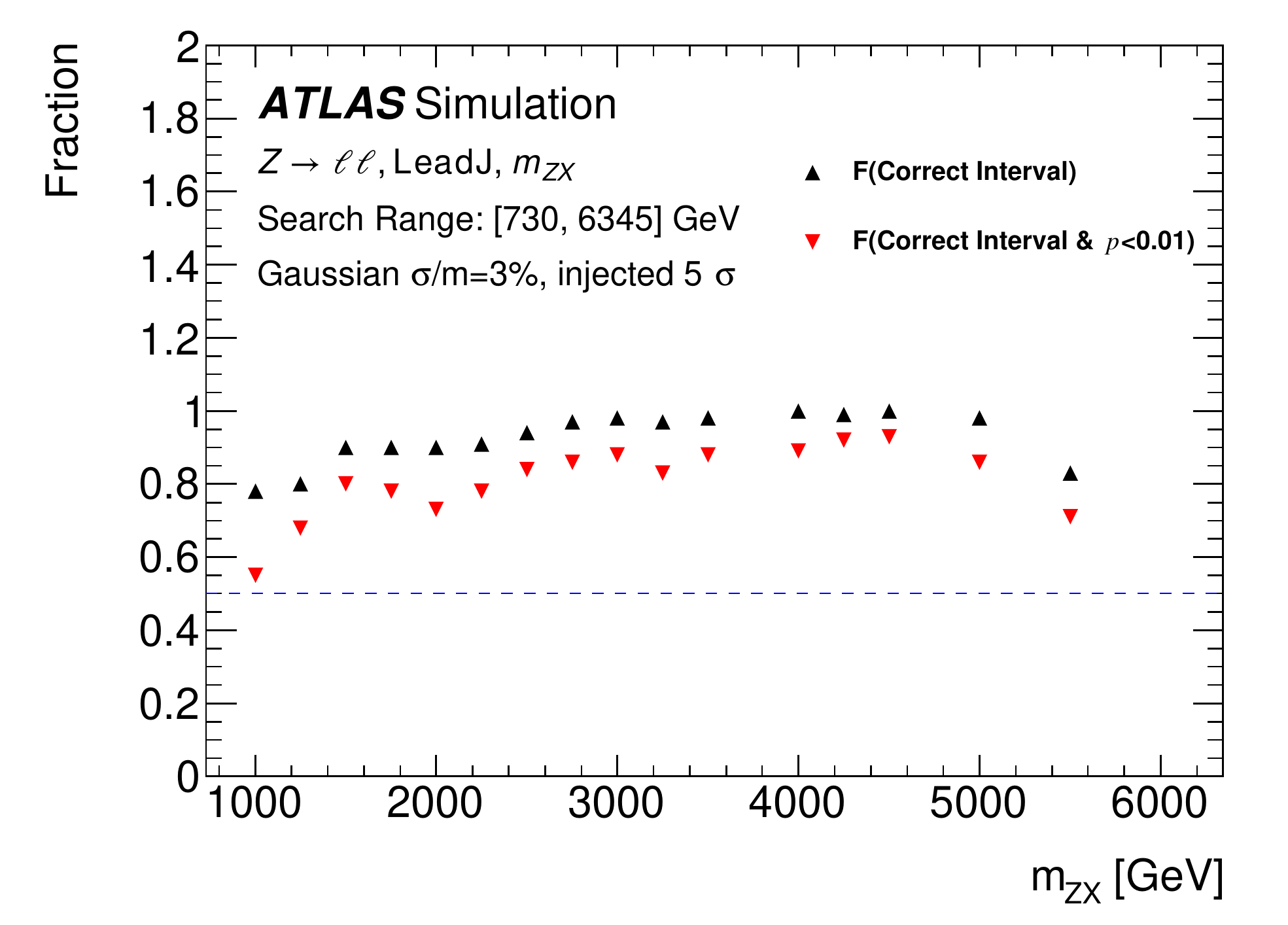} \label{appfig:scan-5sig-b}}
\vspace{-.5\baselineskip}
\caption{Fractions ($F$) of pseudo-experiments in which the detected BH interval agrees with the injected mass point and the BH $p$-value is below 0.01 as a function of \protect\subref{appfig:scan-5sig-a} $m_X$ and \protect\subref{appfig:scan-5sig-b} $m_{ZX}$ in the leading small-$R$-jet category for a $5\sigma$ injected Gaussian-shaped signal with a relative width of 3\%. The blue dashed horizontal line at 0.5 is used to define the signal-sensitive mass range.}
\label{appfig:scan-5sig}
\end{figure}

The study is also performed at other mass points for both the $m_X$ and $m_{ZX}$ spectra in all event categories. The fractions from the right panel of Figure~\ref{appfig:injection} for the leading small-$R$-jet category are presented in Figure~\ref{appfig:scan-5sig}. It shows that the sensitivity of the BH algorithm is best in the intermediate mass range and decreases at low and high mass values. The signal-sensitive mass range in this analysis is defined by requiring at least 50\% of the pseudo-experiments to have a correctly identified BH interval and a BH $p$-value below 0.01.
 
\subsection{Limit-sensitive mass range}\label{app:limit-range}
 
\begin{figure}[htbp]
\centering
\subfloat[]{\includegraphics[width=0.45\columnwidth]{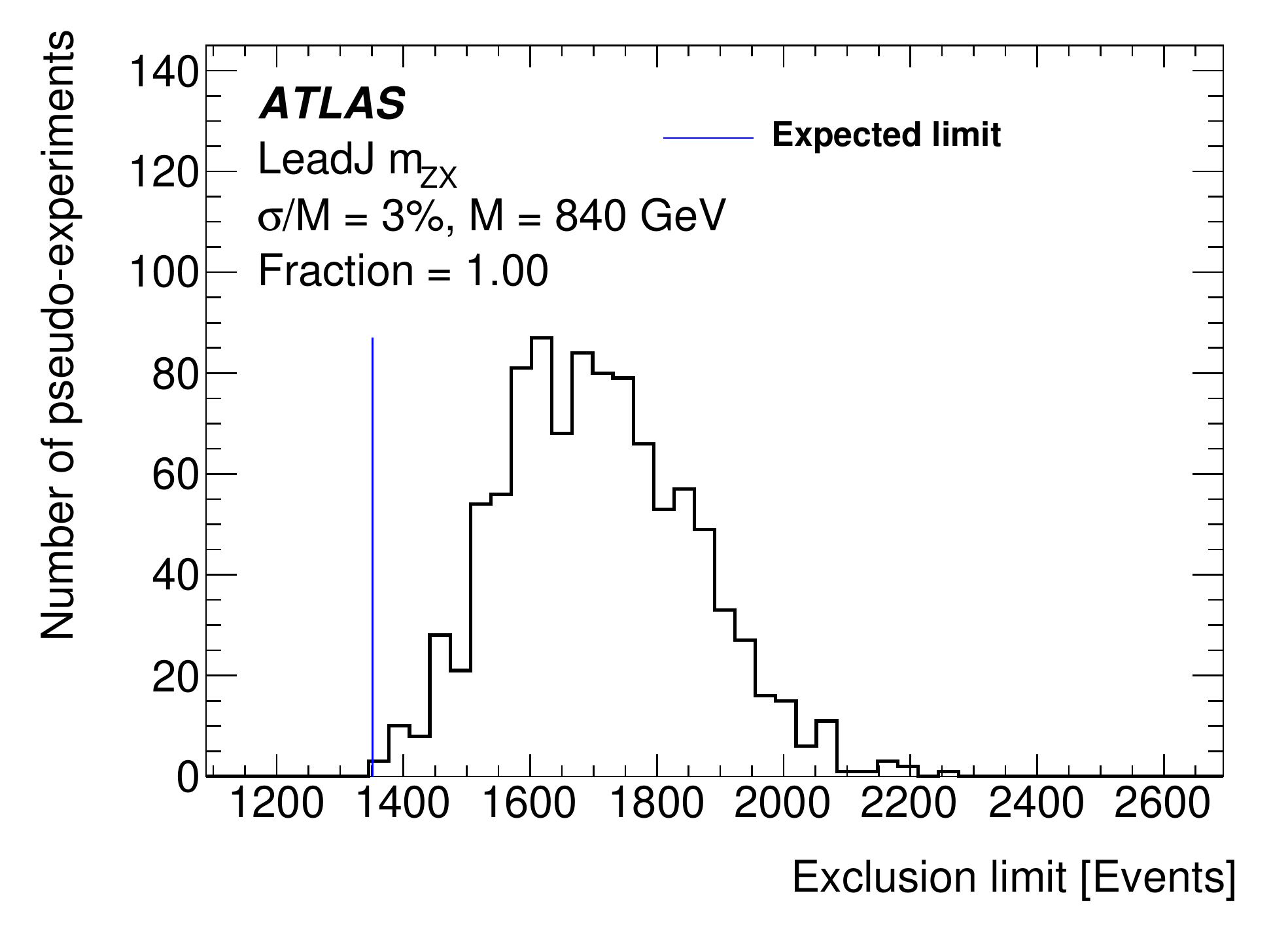} \label{appfig:limit-a}}
\subfloat[]{\includegraphics[width=0.45\columnwidth]{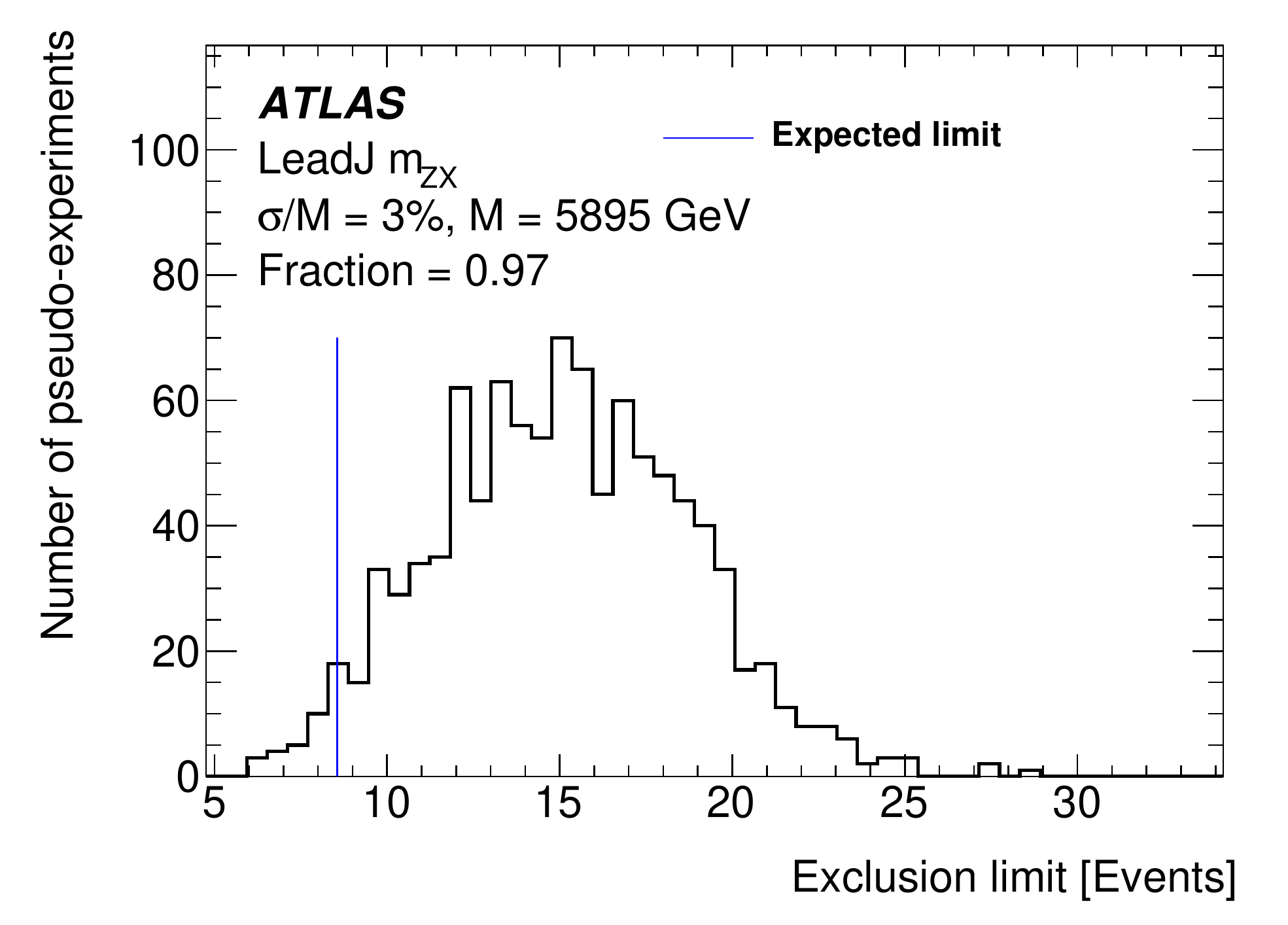} \label{appfig:limit-b}}
\vspace{-.5\baselineskip}
\hfill
\subfloat[]{\includegraphics[width=0.45\columnwidth]{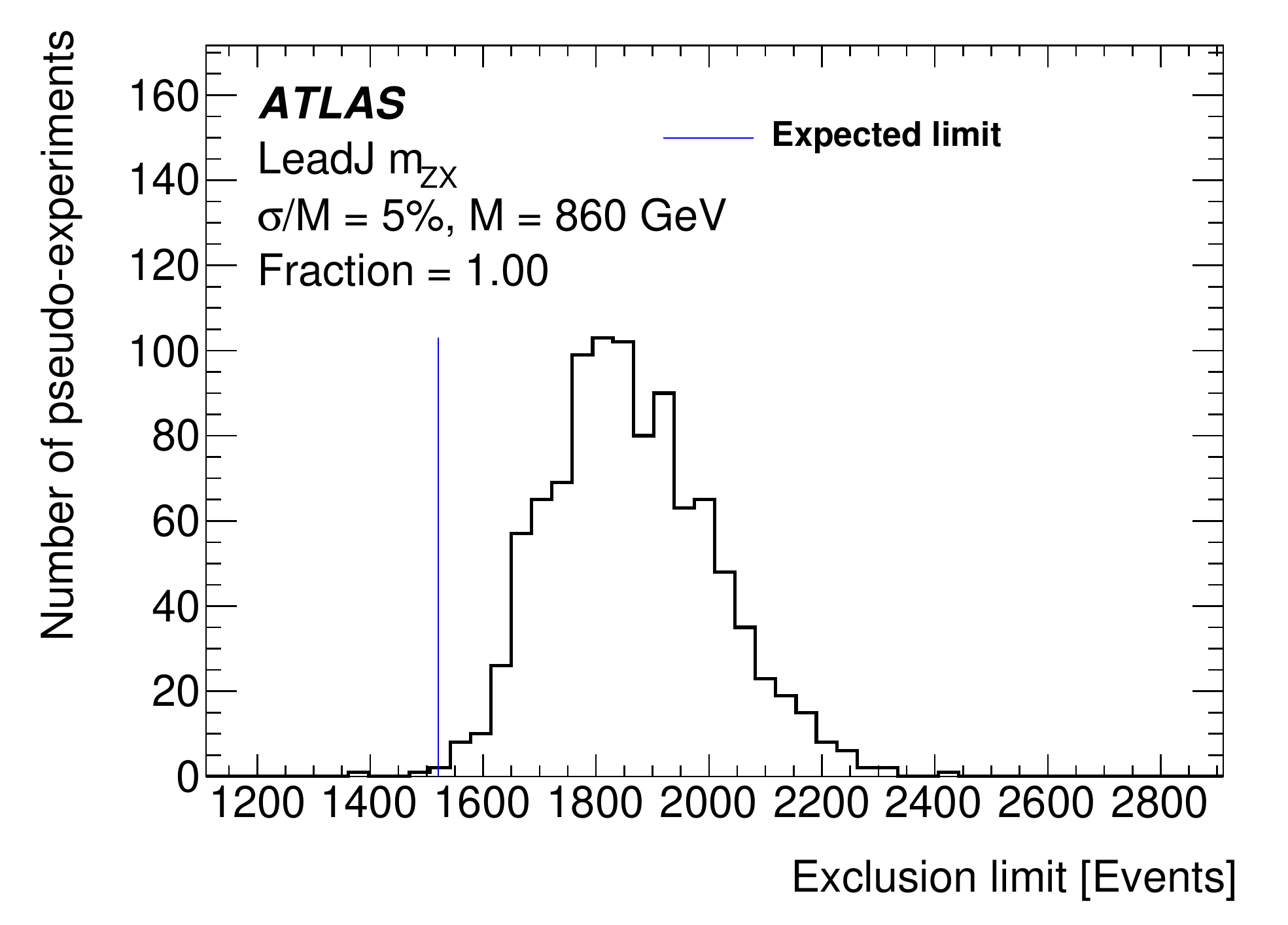} \label{appfig:limit-c}}
\subfloat[]{\includegraphics[width=0.45\columnwidth]{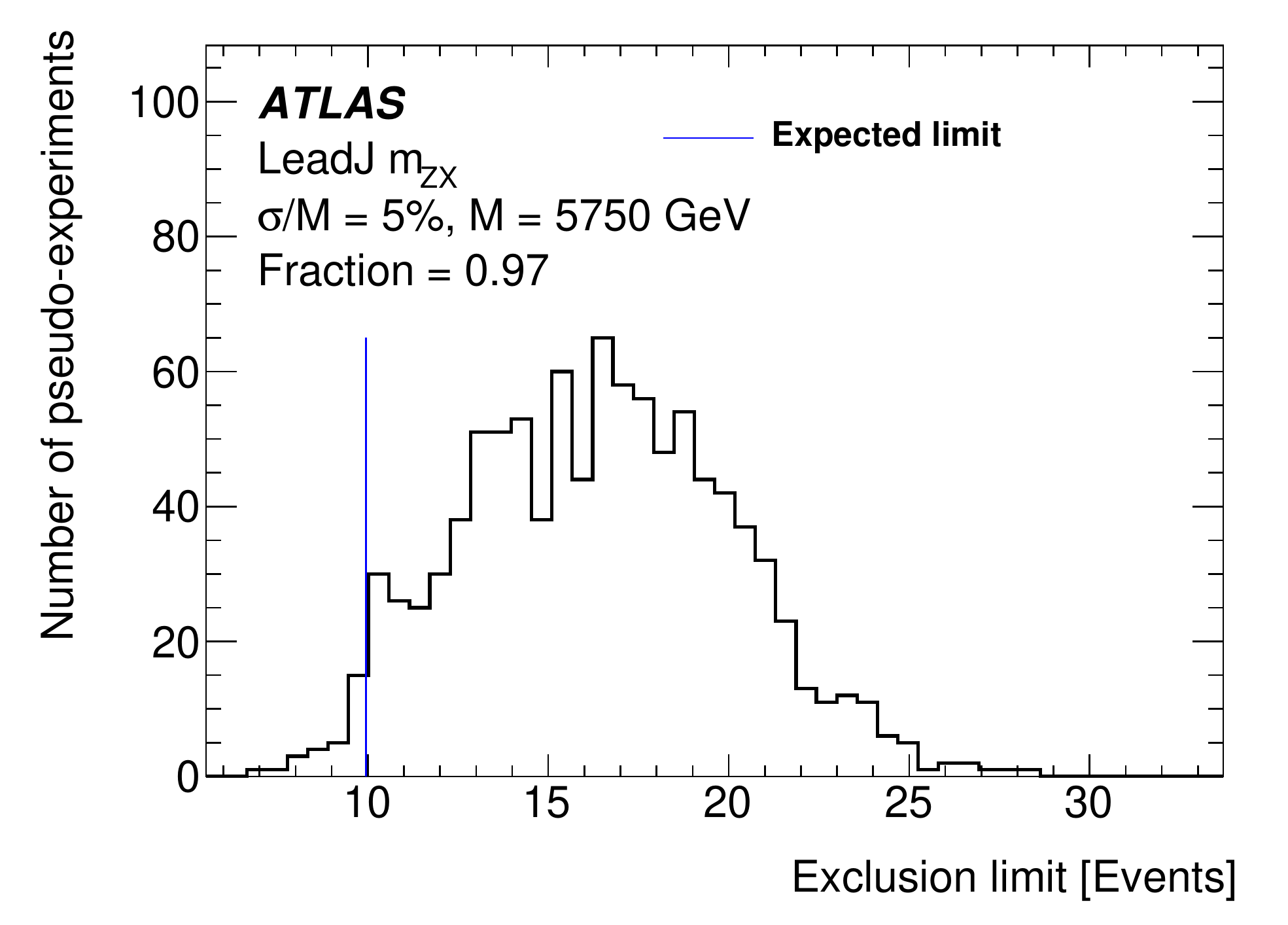}\label{appfig:limit-d}}
\vspace{-.5\baselineskip}
\hfill
\subfloat[]{\includegraphics[width=0.45\columnwidth]{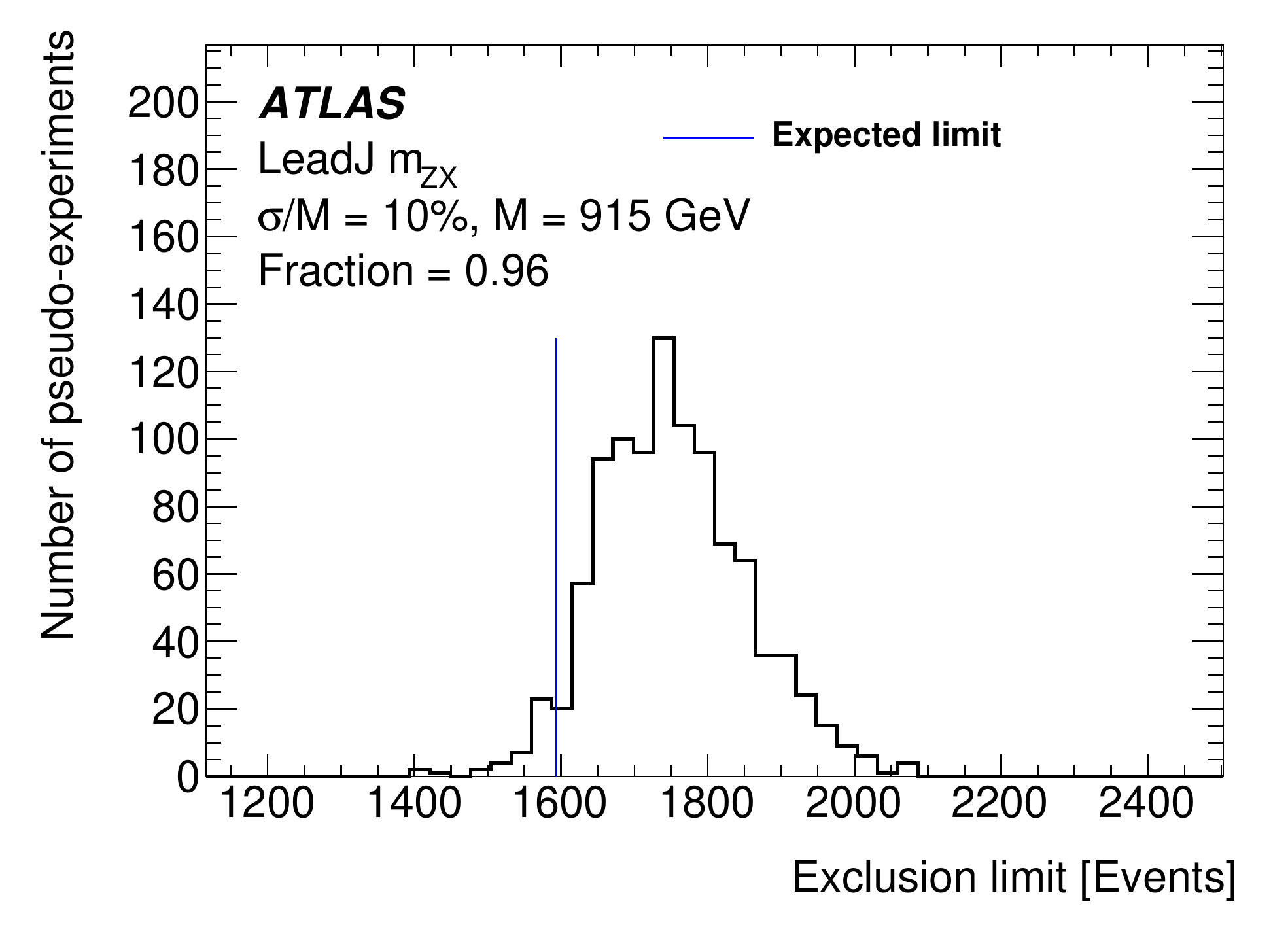} \label{appfig:limit-e}}
\subfloat[]{\includegraphics[width=0.45\columnwidth]{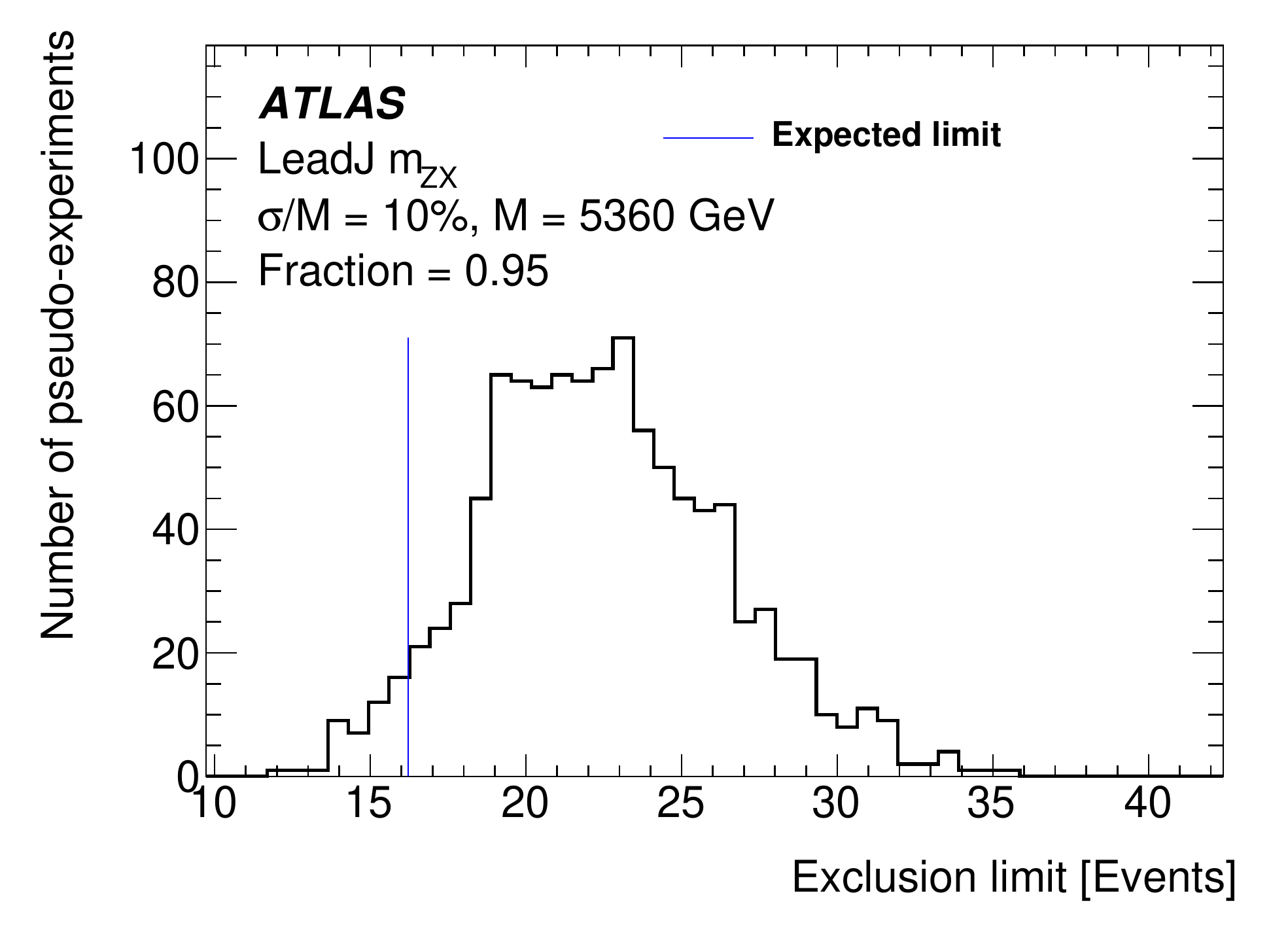} \label{appfig:limit-f}}
\vspace{-.5\baselineskip}
\caption{Distribution of exclusion upper limits on signal event yields at 95\% CL from 1000 pseudo-experiments for Gaussian-shaped signals with relative width values of
\protect\subref{appfig:limit-a}--\protect\subref{appfig:limit-b} 3\%, \protect\subref{appfig:limit-c}--\protect\subref{appfig:limit-d} 5\% and \protect\subref{appfig:limit-e}--\protect\subref{appfig:limit-f} 10\% at the \protect\subref{appfig:limit-a},\protect\subref{appfig:limit-c},\protect\subref{appfig:limit-e} lower and \protect\subref{appfig:limit-b},\protect\subref{appfig:limit-d},\protect\subref{appfig:limit-f} upper boundaries of the limit-sensitive mass range for the $m_{ZX}$ spectrum of the leading small-$R$-jet category. The vertical line corresponds to the expected nominal exclusion limit at 95\% CL.}
\label{appfig:limit}
\end{figure}

For a given event category, a background mass spectrum based on MC simulations can be used to derive an expected exclusion upper limit at 95\% CL for a Gaussian-shaped signal of a certain given width in a model-independent way. This is referred to as nominal exclusion limit in the following.
A set of 1000 pseudo-experiments are generated from the MC estimates with both the background and a signal with a signal strength set to the nominal exclusion limit value. For each pseudo-experiment, a signal+background fit is performed and an exclusion upper limit at 95\% CL is extracted. The distribution of the cross-section exclusion upper limits from the 1000 pseudo-experiments is compared with the nominal limit derived directly from the MC mass spectrum. The fraction of pseudo-experiments that have a higher upper limit is calculated. If it is larger than 95\%, the background modelling is sensible for the mass point under study and the corresponding exclusion upper limit is conservative. Given the limited size of each pseudo-experiment, a fraction larger than 90\% is considered acceptable. One example is shown in Figure~\ref{appfig:limit} for mass points at the lower and upper boundaries of the limit-sensitive mass range for the $m_{ZX}$ mass spectrum in the leading small-$R$-jet category for Gaussian-shaped signals with relative width values of 3\%, 5\% and 10\%. The results for all the mass spectra and event categories are summarised in Table~\ref{apptab:limit-range}. As expected, the limit-sensitive mass range is largest for signals with small widths. Most of these mass ranges can be applied directly to data when setting limits; an adjustment is only needed when the limit-sensitive mass range extends beyond the corresponding fit range in data.
 
\begin{table}[htbp]
\begin{center}
\caption{A list of mass spectra, event categories, relative width values of Gaussian-shaped signals and  limit-sensitive mass ranges and fractions, corresponding to the mass values in the previous column, of pseudo-experiments having exclusion upper limits higher than the nominal exclusion limit at 95\% CL.}
\label{apptab:limit-range}
\begin{tabular}{clccc}
\hline
Spectrum & Category &  Width [\%] & Limit range [\GeV]  & Fractions  \\\hline
$m_X$ & LeadJ &  3 & [430, 5195] & 0.986, 0.968\\
&            &  5 & [435, 5065] & 0.980, 0.972\\
&            & 10~~ & [470, 4750] & 0.936, 0.967\\\hline
$m_X$ & LeadB &  3 & [420, 3050] & 0.930, 0.949\\
&             & 5 & [430, 2960] & 0.926, 0.948\\
&             & 10~~ & [750, 2750] & 0.919, 0.941\\\hline
$m_X$ & LeadFatJ & 3 & [730, 5580]  & 0.964, 0.976\\
&                &  5 & [745, 5540] & 0.903, 0.981\\
&                & 10~~ & [1100, 5070]~~ & 0.917, 0.959\\\hline
$m_X$ & LeadP & 3 & [185, 1620]  & 0.984, 0.960\\
&             & 5 & [190, 1580] & 0.990, 0.955\\
&             & 10~~ & [205, 1475] & 0.996, 0.951\\\hline
$m_X$ & LeadE & 3 & [180, 1350]  & 0.946, 0.953\\
&            & 5 & [180, 1315] & 0.946, 0.947\\
&            & 10~~ & [200, 1230] & 0.948, 0.945\\\hline
$m_X$ & LeadM & 3 & [175, 1065]  & 0.937, 0.944\\
&             &  5 & [180, 1045] & 0.956, 0.944\\
&             & 10~~ & [195, 980]~~ & 0.961, 0.937\\\hline
$m_X$ & Inclusive & 3 & [650, 5940]  & 0.923, 0.982\\
&                   & 5 & [665, 5795] & 0.936, 0.978\\
&                   & 10~~ & [900, 5375] & 0.906, 0.956\\\hline\hline
 
$m_{ZX}$ & LeadJ & 3 & [840, 5895]  & 1.000, 0.971\\
&           & 5 & [860, 5750] & 0.997, 0.974\\
&           & 10~~ & [915, 5360] & 0.956, 0.955\\\hline
$m_{ZX}$ & LeadB & 3 & [665, 3815]  & 0.932, 0.953\\
&            & 5 & [680, 3705] & 0.901, 0.931\\
&            & 10~~ & [1000, 3460]~~ & 0.921, 0.926\\\hline
$m_{ZX}$ & LeadFatJ & 3 & [730, 6505]  & 0.964, 0.969\\
&                 & 5 & [750, 6345] & 0.957, 0.971\\
&                 & 10~~ & [815, 5910] & 0.953, 0.966\\\hline
$m_{ZX}$ & LeadP & 3 & [300, 2255] & 0.960, 0.962\\
&            & 5 & [310, 2190] & 0.958, 0.954\\
&            & 10~~ & [350, 2035] & 0.991, 0.942\\\hline
$m_{ZX}$ & LeadE & 3 & [360, 1960]  & 0.967, 0.947\\
&            & 5 & [370, 1910] & 0.934, 0.948\\
&            & 10~~ & [410, 1785] & 0.953, 0.930\\\hline
$m_{ZX}$ & LeadM & 3 & [320, 1680]  & 0.957, 0.949\\
&             & 5 & [325, 1645] & 0.941, 0.956\\
&             & 10~~ & [365, 1530] & 0.940, 0.960\\\hline
$m_{ZX}$ & Inclusive & 3 & [740, 7230] & 0.954, 0.983\\
&                  & 5 & [755, 7040] & 0.943, 0.980\\
&                  & 10~~ & [815, 6550] & 0.935, 0.982\\\hline
\end{tabular}
\end{center}
\end{table}


\clearpage
\printbibliography
 
\clearpage
 
\begin{flushleft}
\hypersetup{urlcolor=black}
{\Large The ATLAS Collaboration}

\bigskip

\AtlasOrcid[0000-0002-6665-4934]{G.~Aad}$^\textrm{\scriptsize 101}$,
\AtlasOrcid[0000-0002-5888-2734]{B.~Abbott}$^\textrm{\scriptsize 119}$,
\AtlasOrcid[0000-0002-7248-3203]{D.C.~Abbott}$^\textrm{\scriptsize 102}$,
\AtlasOrcid[0000-0002-1002-1652]{K.~Abeling}$^\textrm{\scriptsize 55}$,
\AtlasOrcid[0000-0002-8496-9294]{S.H.~Abidi}$^\textrm{\scriptsize 29}$,
\AtlasOrcid[0000-0002-9987-2292]{A.~Aboulhorma}$^\textrm{\scriptsize 35e}$,
\AtlasOrcid[0000-0001-5329-6640]{H.~Abramowicz}$^\textrm{\scriptsize 150}$,
\AtlasOrcid[0000-0002-1599-2896]{H.~Abreu}$^\textrm{\scriptsize 149}$,
\AtlasOrcid[0000-0003-0403-3697]{Y.~Abulaiti}$^\textrm{\scriptsize 116}$,
\AtlasOrcid[0000-0003-0762-7204]{A.C.~Abusleme~Hoffman}$^\textrm{\scriptsize 136a}$
% ...
\bigskip
\\

% ...
$^{29}$Physics Department, Brookhaven National Laboratory, Upton NY; United States of America.\\
% ...
$^{35}$$^{(a)}$Facult\'e des Sciences Ain Chock, R\'eseau Universitaire de Physique des Hautes Energies - Universit\'e Hassan II, Casablanca;$^{(b)}$Facult\'{e} des Sciences, Universit\'{e} Ibn-Tofail, K\'{e}nitra;$^{(c)}$Facult\'e des Sciences Semlalia, Universit\'e Cadi Ayyad, LPHEA-Marrakech;$^{(d)}$LPMR, Facult\'e des Sciences, Universit\'e Mohamed Premier, Oujda;$^{(e)}$Facult\'e des sciences, Universit\'e Mohammed V, Rabat;$^{(f)}$Institute of Applied Physics, Mohammed VI Polytechnic University, Ben Guerir; Morocco.\\
% ...
$^{55}$II. Physikalisches Institut, Georg-August-Universit\"{a}t G\"ottingen, G\"ottingen; Germany.\\
% ...
$^{101}$CPPM, Aix-Marseille Universit\'e, CNRS/IN2P3, Marseille; France.\\
$^{102}$Department of Physics, University of Massachusetts, Amherst MA; United States of America.\\
% ...
$^{116}$Department of Physics, New York University, New York NY; United States of America.\\
% ...
$^{119}$Homer L. Dodge Department of Physics and Astronomy, University of Oklahoma, Norman OK; United States of America.\\
% ...
$^{136}$$^{(a)}$Departamento de F\'isica, Pontificia Universidad Cat\'olica de Chile, Santiago;$^{(b)}$Millennium Institute for Subatomic physics at high energy frontier (SAPHIR), Santiago;$^{(c)}$Instituto de Investigaci\'on Multidisciplinario en Ciencia y Tecnolog\'ia, y Departamento de F\'isica, Universidad de La Serena;$^{(d)}$Universidad Andres Bello, Department of Physics, Santiago;$^{(e)}$Instituto de Alta Investigaci\'on, Universidad de Tarapac\'a, Arica;$^{(f)}$Departamento de F\'isica, Universidad T\'ecnica Federico Santa Mar\'ia, Valpara\'iso; Chile.\\
% ...
$^{149}$Department of Physics, Technion, Israel Institute of Technology, Haifa; Israel.\\
$^{150}$Raymond and Beverly Sackler School of Physics and Astronomy, Tel Aviv University, Tel Aviv; Israel.\\
% ...
$^{a}$ Also Affiliated with an institute covered by a cooperation agreement with CERN.\\
$^{b}$ Also at An-Najah National University, Nablus; Palestine.\\
$^{c}$ Also at Borough of Manhattan Community College, City University of New York, New York NY; United States of America.\\
$^{d}$ Also at Bruno Kessler Foundation, Trento; Italy.\\
$^{e}$ Also at Center for High Energy Physics, Peking University; China.\\
$^{f}$ Also at Centro Studi e Ricerche Enrico Fermi; Italy.\\
% ...

\end{flushleft}


\end{document}